\documentclass[12pt]{article}
\RequirePackage[OT1]{fontenc}

\usepackage{amsmath,amsfonts,amsthm,amssymb,amsbsy}
\usepackage{array,epsfig,fancyhdr,rotating}
\usepackage{booktabs}
\usepackage{color, dsfont}
\usepackage{enumerate}
\usepackage{float}
\usepackage{graphicx,psfrag,epsf}
\usepackage[]{hyperref}  
\usepackage{ltxtable}
\usepackage{multirow}
\usepackage{pdflscape}
\usepackage[sectionbib]{natbib}
\usepackage{sectsty, secdot}
\usepackage{subcaption}
\usepackage{threeparttable}
\usepackage{url} 

\renewcommand{\theequation}{\arabic{equation}}
\def\thefigure{\arabic{figure}}
\def\thetable{\arabic{table}}

\newtheorem{algorithm}{Algorithm}

\newtheorem{corollary}{Corollary}

\newtheorem{example}{Example}
\newtheorem{lemma}{Lemma}

\newtheorem{remark}{Remark}
\newtheorem{theorem}{Theorem}

\begin{document}


\title{\Large\bf A-optimal Designs under Generalized Linear Models}
\author{Yingying Yang$^1$, Xiaotian Chen$^2$, and Jie Yang$^1$\\ 
	$^1$University of Illinois Chicago and $^2$University of Hong Kong
}

\maketitle


\begin{abstract}
Designing efficient experiments under practical constraints is critical in both scientific research and industrial practice. Focusing on minimizing the average variance of the parameter estimates, A-optimal designs show advantages in screening factors and reducing prediction errors. Compared with other criteria, however, algorithms and software for generating A-optimal designs are scarce. In this paper, we characterize A-optimal designs under generalized linear models theoretically and develop efficient algorithms for identifying them. When a predetermined finite set of experimental settings is given, we derive analytic solutions or establish necessary and sufficient conditions for obtaining A-optimal approximate allocations. We show that a lift-one algorithm based on our formulae outperforms commonly used algorithms for finding A-optimal allocations. When continuous factors or design regions get involved, we develop a ForLion algorithm to find A-optimal designs with mixed factors. Numerical studies on real-data experiments show that our algorithms can find highly efficient designs with reduced numbers of distinct experimental settings, which may save both experimental time and cost significantly. Along with a rounding-off algorithm that converts approximate allocations to exact ones, we demonstrate that A-optimal designs may provide more accurate parameter estimates than commonly used alternatives.
\end{abstract}


\section{Introduction}\label{sec:introduction}

Designing efficient experiments or clinical trials under practical constraints, such as budget, logistics, or ethical considerations, is a fundamental problem in many applied studies. In practice, researchers often face a large pool of heterogeneous experimental units or individuals but can only collect responses from a limited number of subjects, making the choice of sampling allocation critical for more accurate estimation and prediction.

As an illustration, we consider a real-world potato packing experiment \citep{woods2006, li2021}, where a binary response indicates whether liquid appears after seven days of storage, and three continuous factors (a vitamin concentration and two gas levels) are controlled under a logistic regression model. Each support point of a design corresponds to a distinct factor-level combination, which requires preparing and monitoring a separate batch of packages over the seven-day period. A design with fewer support points therefore lowers the material and labor cost of the study. Using our proposed ForLion algorithm (Algorithm~\ref{algo:A_opt_Forlion}), the A-optimal design has $21$ support points, compared with $29$ for the design of \cite{li2021}, while matching its A-efficiency. Across $100$ simulated parameter settings, our A-optimal designs have fewer support points in $96$ cases ($18.50$ versus $26.29$ on average). Moreover, at the assumed parameter values, the A-optimal design attains a smaller average root mean squared error (RMSE) in estimating the model coefficients than the I- and D-optimal designs of \cite{li2021} at every sample size considered, with a statistically significant improvement over the D-optimal design (see Example~\ref{ex:potato_example}). This is consistent with the A-criterion directly targeting the average variance of the coefficient estimates and illustrates the practical benefits of A-optimal designs.

In this paper, we consider experiments with a univariate response $Y$ whose probability distribution belongs to the exponential family with a single parameter $\theta \in \mathbb{R}$ in its canonical form
$f(y;\theta) = \exp\{y b(\theta) + c(\theta) + d(y)\}$. Examples, along with $\log f(y; \theta)$ listed in Table~\ref{tab:exponential_list} of Section~\ref{sec:tables} in the Supplementary Material, include Bernoulli and binomial distributions for binary responses, Poisson distribution for count responses, Gamma and Inverse Gaussian (IG) distributions for positive responses, and normal distribution for general responses. Suppose $Y_1, \ldots, Y_n \in \mathbb{R}$ are collected independently with covariates ${\mathbf x}_1, \ldots, {\mathbf x}_n$~, respectively, where ${\mathbf x}_i = (x_{i1}, \ldots, x_{id})^T \in \mathbb{R}^d$. A generalized linear model \citep{pmcc1989, dobson2018} consists of a link function $g$, a list of predictor functions $q_1, \ldots, q_p$~, and the corresponding parameters of interest $\boldsymbol{\beta} = (\beta_1, \ldots, \beta_p)^T$, such that
\begin{equation}\label{eq:glm}
E(Y_i) = \mu_i\mbox{ and } \eta_i = g(\mu_i)= {\mathbf X}_i^T\boldsymbol{\beta}\ ,
\end{equation}
where ${\mathbf X}_i = {\mathbf q}({\mathbf x}_i) = (q_1({\mathbf x}_i), \ldots, q_p({\mathbf x}_i))^T$. For many applications, $q_1({\mathbf x}_i)\equiv 1$ represents the intercept, and ${\mathbf q}({\mathbf x}_i) = (1, {\mathbf x}_i^T)^T \in \mathbb{R}^{d+1}$, known as a main-effects model. 
If there are only $m$ distinct experimental settings ${\bf x}_1, \ldots, {\bf x}_m$
with the numbers of replicates $n_1, \ldots, n_m$~, respectively, the Fisher information matrix can be written as \citep{dobson2018, ym2015} ${\mathbf F} = n{\mathbf X}^T{\mathbf W}{\mathbf X} = n \sum_{i=1}^m w_i \nu_i {\mathbf q}({\mathbf x}_i) {\mathbf q}({\mathbf x}_i)^T\ $, where $n=\sum_{i=1}^m n_i$ is the sample size, ${\mathbf X} = ({\mathbf X}_1, \ldots, {\mathbf X}_m)^T$ is the model matrix, and ${\mathbf W}={\rm diag}\{w_1\nu_1,\ldots, w_m\nu_m\}$ with
$\nu_i = (\partial \mu_i/\partial \eta_i)^2/{\rm Var}(Y_i)$, and $w_i = n_i/n$, $i=1,\ldots, m$.

To estimate the parameters $\boldsymbol{\beta}$ more accurately, we aim to find either the ``optimal'' ${\mathbf w} = (w_1, \ldots, w_m)^T$ for a given set $\{{\mathbf x}_1, \ldots, {\mathbf x}_m\}$ of experimental settings (see Section~\ref{sec:A_optimal_designs}), or the ``optimal'' $\boldsymbol{\xi} = \{({\mathbf x}_i, w_i), i=1, \ldots, m\}$ for a given design space ${\cal X}\subset \mathbb{R}^d$, such that ${\mathbf x}_i \in {\cal X}$ and $m\geq 1$ (see Section~\ref{sec:Aopt_mixed_factors}).
Since the maximum likelihood estimator of $\boldsymbol\beta $ has an asymptotic covariance matrix \citep{pmcc1989, dobson2018, khuri2006} that is the inverse of ${\mathbf F}$,  we look for an A-optimal design ${\mathbf w}$ or $\boldsymbol{\xi}$ maximizing 
$[{\rm tr}(({\mathbf X}^T{\mathbf W}{\mathbf X})^{-1})]^{-1}$, which is the inverse (up to a constant) of the average variance
of parameter estimators \citep{elfving1952optimum}. Since ${\mathbf F}$ depends on $\boldsymbol{\beta}$ for most generalized linear models (GLMs), we adopt the local optimality approach of \cite{chernoff1953} with an assumed $\boldsymbol{\beta}$, which may be estimated from a pilot or previous study (see \cite{ford1992} for more justifications on the significance of locally optimal designs). The ordinary linear model is a notable special case in which ${\mathbf F}$ is independent of $\boldsymbol{\beta}$, so that the optimal design is global. In the general GLM setting, the parameter dependence of ${\mathbf F}$, arising through the link and variance functions, constitutes a structural distinction, in that classical A-optimality results for linear models do not directly extend to the GLM setting, which necessitates the development of a dedicated design theory.

Compared with D-optimal designs maximizing $|{\mathbf X}^T{\mathbf W}{\mathbf X}|$ \citep{wald1943efficient}, A-optimal designs are much more difficult to construct and study \citep{chen2023}, and the corresponding algorithms and software are scarce \citep{jones2021optimal}. The theoretical results on A-optimal designs often either require additional technical conditions \citep{sitter1993optimal, mathew2001optimal, yangmin2008, chen2023}, or focus on special cases such as binary responses \citep{sitter1993optimal, mathew2001optimal, yangmin2008}, Gamma models \citep{gaffke2019, chen2023}, Inverse Gaussian models \citep{chen2023}, with one factor  \citep{sitter1993optimal, yangmin2008, chen2023} or two factors only \citep{gaffke2019}.

On the other hand, \cite{jones2021optimal} showed that A-optimal designs can do better than D-optimal ones in screening experiments under linear models (i.e., normal models in GLMs), and \cite{stallrich2023optimal} identified cases where D-optimal screening designs have undesirable estimation variance properties, whereas A-optimal designs drive the variances closer to their minimum attainable values. Moreover, \cite{chen2026optimal}'s recent study implied that A-optimal designs are more efficient than D-optimal ones (99.64\% versus 84.98\%) in terms of their generalized integrated variance for differences criterion under linear models when the experimental region is symmetric. 

In this paper, we characterize A-optimal designs under a fairly general GLM, including but not limited to the models listed in Table~\ref{tab:exponential_list}. Given a predetermined finite set $\{{\mathbf x}_1, \ldots, {\mathbf x}_m\}$ of experimental settings, we provide analytic solutions for A-optimal approximate allocation ${\mathbf w}$ if $m=p$ (Theorem~\ref{thm:m=p}) , and analytic necessary and sufficient conditions for identifying A-optimal allocations when $m>p$ (Theorems~\ref{thm:for_lift_one} and \ref{thm:max_h_i(x)}). Similarly to the lift-one algorithm for D-optimality \citep{ymm2016, ym2015}, we develop a lift-one A-optimality algorithm (Algorithm~\ref{algo:A_opt_lift_one}), which outperforms commonly used algorithms for finding A-optimal allocations (see Example~\ref{ex:main_effects}). When some factors are continuous, we follow the ForLion algorithm for D-optimality \citep{huang2024forlion} and develop a ForLion A-optimality algorithm (Algorithm~\ref{algo:A_opt_Forlion}) that is guaranteed to find A-optimal designs with mixed factors when the converging condition is met (Theorem~\ref{thm:GLM_A_optimality} and Corollary~\ref{cor:ForLion_Aopt}). Numerical studies show that our algorithms are able to find highly efficient designs with reduced numbers of design points (see Examples~\ref{ex:main_effects}, \ref{ex:compare_ForLion_OptimalDesign} and \ref{ex:potato_example}). Together with our rounding-off algorithm (Algorithm~\ref{algo:exact_A}) that converts an approximate allocation into an exact one, these algorithms are presented in detail in the Supplementary Material (Section~\ref{sec:algorithms}). Numerical studies show that stratified samplers based on A-optimal allocations may provide more accurate estimates on factor effects than commonly used samplers, including D-optimal ones (Example~\ref{ex:RMSE} in Section~\ref{sec:examples_supp} of the Supplementary Material).

\section{A-optimal Designs for Discrete Factors}\label{sec:A_optimal_designs}

In this section, we consider ${\mathbf w} \in S_m = \{(w_1, \ldots, w_m)^T \in \mathbb{R}^m \mid w_i\geq 0, \sum_{i=1}^m w_i=1\}$, known as an {\it approximate allocation}, given a finite set $\{{\mathbf x}_1, \ldots, {\mathbf x}_m\}$ of experimental settings. It covers the cases when each factor has only a finite number of levels. 

If $m=1$, the A-optimal allocation is ${\mathbf w} = w_1 = 1$, which is trivial. If $m\geq 2$ and $p=1$, then ${\mathbf F} = n\sum_{i=1}^m w_i \nu_i q_1({\mathbf x}_i)^2$ and ${\mathbf e}_{i_*}$ is an A-optimal allocation, whose $i_*$th coordinate is $1$ and others are zeros, where $i_* = {\rm argmax}_i \nu_i q_1({\mathbf x}_i)^2$. 
Note that $\nu_i$ depends on both  $\boldsymbol\beta$ and the link function $g$. For commonly used GLMs (see Table~5 in the Supplementary Material of \cite{huang2025constrained}), $g$ is one-to-one and differentiable, and $\nu_i = \nu(\eta_i) = \nu({\mathbf X}_i^T\boldsymbol\beta) \geq 0$, where $\nu(\eta_i) =  [(g^{-1})'(\eta_i)]^2/s(\eta_i)$ and $s(\eta_i)={\rm Var}(Y_i)$.

From now on, we consider the cases with $m\geq 2$ and $p\geq 2$. For given $\{{\mathbf x}_1, \ldots, {\mathbf x}_m\}$ and assumed $\boldsymbol{\beta}$, we let  $f({\mathbf w}) = |{\mathbf X}^T{\mathbf W}{\mathbf X}| \geq 0$ and 
\begin{equation}\label{eq:h(w)}
h({\mathbf w}) = \left\{
\begin{array}{cl}
\left[{\rm tr}\left(({\mathbf X}^T{\mathbf W}{\mathbf X})^{-1}\right)\right]^{-1} & ,\mbox{ if }f({\mathbf w}) > 0;\\
0 & ,\mbox{ if } f({\mathbf w})=0.
\end{array}\right.
\end{equation}
A (locally) D-optimal allocation is a ${\mathbf w}$ maximizing $f({\mathbf w})$, while a (locally) A-optimal allocation is a ${\mathbf w}$ maximizing $h({\mathbf w})$. Both D- and A-criteria can be regarded as special kinds of matrix means (see, e.g., Section~6.1 in \cite{pukelsheim1993}). We further denote by $S_m^+ = \{{\mathbf w}\in S_m \mid f({\mathbf w})>0\}$, the collection of all feasible approximate allocations. To avoid trivial maximization problems, we assume $S_m^+ \neq \emptyset$, that is, there exists a ${\mathbf w}\in S_m$~, such that, $f({\mathbf w})>0$. Since ${\mathbf X}^T{\mathbf W}{\mathbf X}$ is a $p\times p$ matrix with a rank no more than $\min\{m,p\}$, $m\geq p$ is required for the existence of $({\mathbf X}^T{\mathbf W}{\mathbf X})^{-1}$. For the same reason, we assume ${\rm rank}({\mathbf X}) = p$ to avoid trivial cases.

\subsection{A-optimal designs when $m=p\geq 2$}
\label{sec:m=p}

When $m=p\geq 2$, the design is saturated and can be obtained analytically. 

\begin{theorem}\label{thm:m=p}
    Consider the GLM~\eqref{eq:glm} with $m=p\geq 2$, ${\rm rank}({\mathbf X})=p$, and $\nu_i>0$ for each $i$. Then an allocation ${\mathbf w}_* = (w_1^*, \ldots, w_m^*)^T$ is A-optimal if and only if
    \[
    w_i^* = \frac{\sqrt{c_i/\nu_i}}{\sum_{j=1}^m \sqrt{c_j/\nu_j}}\ , 
    \]
    $i=1, \ldots, m$, where $c_i > 0$ is the $i$th diagonal element of $({\mathbf X}{\mathbf X}^T)^{-1}$.
\end{theorem}

The proof of Theorem~\ref{thm:m=p}, as well as other proofs, is relegated to Section~\ref{sec:proofs} in the Supplementary Material. According to Theorem~\ref{thm:m=p}, an A-optimal design with $m=p$ must satisfy $w_i^* \propto \sqrt{c_i/\nu_i}$~. In other words, a uniform design with $w^*_i=1/m$ is not A-optimal in general, which is different from D-optimality (see, e.g., Theorem~3.2 in \cite{ym2015}).

The A-optimal allocation in Theorem~\ref{thm:m=p} is analogous to the classical result for linear models (see, e.g., Section~8.8 in \cite{pukelsheim1993}). However, in the GLM setting, the presence of the factor $\nu_i$ arising from the mean–variance relationship leads to modified optimal weights. In particular, when $\nu_i \equiv 1$, the result reduces to the classical linear model case.

\subsection{A-optimal designs when $m>p\geq 2$}
\label{sec:m>p}

It is known that $f({\mathbf w})$ is an order-$p$ homogeneous polynomial of $w_1, \ldots,$ $w_m$ (see, e.g., Lemma~3.1 in \cite{ym2015}). More specifically, 
\begin{equation}\label{eq:|xwx|}
f({\mathbf w})=|{\mathbf X}^T{\mathbf W}{\mathbf X}|=\sum_{1\leq i_1<\cdots<i_p\leq m} |{\mathbf X}[i_1,\ldots,i_p]|^2 \cdot w_{i_1}\nu_{i_1}\cdots w_{i_p}\nu_{i_p}\ ,
\end{equation}
where ${\mathbf X}[i_1,\ldots,i_p]$ is the $p \times p$ submatrix consisting of the $i_1\mbox{th}$, $\ldots$, $i_p\mbox{th}$ rows of the model matrix ${\mathbf X}$. In Lemma~\ref{lem:aobjective} of the Supplementary Material, we show that the objective function $h({\mathbf w})$ (see \eqref{eq:h(w)}) of A-optimality can be written as the ratio of an order-$p$ homogeneous polynomial and an order-$(p-1)$ homogeneous polynomial of $w_1, \ldots, w_m$~.

We let ${\mathbf X}_{-j}$ be the $m\times (p-1)$ submatrix of ${\mathbf X}$ after removing its $j$th column. Then $f_{-j}({\mathbf w}) = |{\mathbf X}_{-j}^T{\mathbf W}{\mathbf X}_{-j}|$ is an order-$(p-1)$ homogeneous polynomial of $w_1, \ldots, w_m$~. Following \cite{ymm2016} and \cite{ym2015}, for $i=1, \ldots, m$, if $0\leq w_i<1$, we define for $z \in [0, 1]$, ${\mathbf w_i(z)} = \left(\frac{1-z}{1-w_i}w_1,\ldots,\frac{1-z}{1-w_i}w_{i-1},z, \frac{1-z}{1-w_i}w_{i+1},\ldots, \frac{1-z}{1-w_i}w_{m}\right)^T\ = \frac{1-z}{1-w_i}{\mathbf w} + \frac{z-w_i}{1-w_i}{\mathbf e_{i}}$~, which is associated with ${\mathbf w} = (w_1, \ldots, w_m)^T$, ${\mathbf e_{i}} = (0, \ldots, 0, 1, 0, \ldots, 0)^T \in \mathbb{R}^m$, whose $i$th coordinate is 1, and the set of distinct design points $\{{\mathbf x}_1, \ldots, {\mathbf x}_m\}$. We also define $f_i(z) = f\left({\mathbf w_i(z)}\right)$, $f_i^{(-j)}(z) = f_{-j}\left({\mathbf w_i(z)}\right)$, and $h_i(z) = h\left({\mathbf w_i(z)}\right)$. Due to Lemma~\ref{lem:aobjective},
\begin{equation}\label{eq:h_i(x)_1}
h_i(z) = \frac{f_i(z)}{\sum_{j=1}^p f_i^{(-j)}(z)}\ .
\end{equation}

The maximum of $h_i(z)$ over $z \in [0,1]$ admits an analytic solution. As shown in Theorem~\ref{thm:max_h_i(x)} of the Supplementary Material (Section~\ref{sec:additional_theorems}), the maximizer $z_*$ and its corresponding maximum admit closed-form expressions, categorized into four cases dictated by coefficients derived from $f_i(z)$ and $f_i^{(-j)}(z)$.

As a special case of the general equivalence theorem \citep{kiefer1974, pukelsheim1993, atkinson2007, stufken2012, fedorov2014}, along with Theorem~\ref{thm:max_h_i(x)} in the Supplementary Material, the following theorem provides analytic necessary and sufficient conditions on A-optimality within $S_m$~.

\begin{theorem}\label{thm:for_lift_one}
Consider the GLM~\eqref{eq:glm} with a predetermined set of distinct design points $\{{\mathbf x}_1, \ldots, {\mathbf x}_m\} \subset \mathbb{R}^d$, $d\geq 1$, and $m > p \geq 2$. Then (i) an A-optimal allocation that maximizes $h({\mathbf w}), {\mathbf w}\in S_m$ must exist; (ii) the set of A-optimal allocations is convex; and (iii) if there is a ${\mathbf w}_* = (w_1^*, \ldots, w_m^*)^T \in S_m$ satisfying $f({\mathbf w}_*) > 0$, then ${\mathbf w}_*$ is A-optimal among $S_m$ if and only if $w_i^*$ maximizes $h_i(z), z\in [0,1]$ associated with ${\mathbf w}_*$ for each $i$.    
\end{theorem}

\subsection{Lift-one algorithm for A-optimal designs}
\label{sec:lift_one_discrete_factor}

Inspired by the lift-one algorithm for D-optimality \citep{ymm2016, ym2015}, we propose a lift-one algorithm (Algorithm~\ref{algo:A_opt_lift_one} in the Supplementary Material) for finding an A-optimal allocation ${\mathbf w}\in S_m$ under a general GLM with given $\{{\mathbf x}_1, \ldots, {\mathbf x}_m\}$ and $\boldsymbol{\beta}$, based on Theorem~\ref{thm:for_lift_one} together with Theorem~\ref{thm:max_h_i(x)} in the Supplementary Material.

As a direct conclusion of Theorem~\ref{thm:for_lift_one} and Lemma~\ref{lem:lift_one_initial} in the Supplementary Material (Section~\ref{sec:lemmas_remark}), we have the following corollary to guarantee that a converged approximate allocation in Algorithm~\ref{algo:A_opt_lift_one} is A-optimal.

\begin{corollary}\label{cor:lift_one}
Under the conditions of Theorem~\ref{thm:for_lift_one}, if there exists a ${\mathbf w}\in S_m$ such that $f({\mathbf w})>0$, then Algorithm~\ref{algo:A_opt_lift_one} converges at an approximate allocation ${\mathbf w}_*$ if and only if ${\mathbf w}_*$ is A-optimal among $S_m$~.    
\end{corollary}

As demonstrated in the following example, a comprehensive simulation study shows that Algorithm~\ref{algo:A_opt_lift_one} outperforms commonly used methods for finding A-optimal allocations in $2^k$ factorial experiments.

\begin{example}\label{ex:main_effects} {\bf Comparison of algorithms for finding A-optimal allocations}\quad {\rm  Following Table~2 in \cite{ymm2016}, to compare different algorithms, we adopt a $2^k$ experiment under a main-effects logistic regression model, ${\rm logit}(P(Y_{i} = 1|x_{i1}, \ldots, x_{ik})) = \beta_0 + \beta_1 x_{i1} + \cdots + \beta_{k} x_{ik}$~,
where $i = 1, \ldots, 2^k$, $Y_i \in \{0, 1\}$, $x_{ij} \in \{-1, 1\}$, and ${\rm logit}(\mu) = \log(\mu/(1 - \mu))$.

Similarly to \cite{ymm2016}, for each $k \in \{2, 3, \ldots, 7\}$, we generate 100 sets of parameters $\boldsymbol{\beta} = (\beta_0, \dots, \beta_k)^T$, with each $\beta_j$ drawn independently from a uniform distribution $U(-3,3)$. For each simulated $\boldsymbol{\beta}$, we treat it as the true parameter vector and determine A-optimal allocations ${\mathbf w} = (w_1, \ldots, w_{2^k})^T$ using one of the following algorithms: {\it (i)} Algorithm~\ref{algo:A_opt_lift_one}; {\it (ii)} classical optimization methods based on R function \texttt{constrOptim}, including Nelder-Mead, quasi-Newton, conjugate-gradient, and simulated annealing \citep{nocedal1999}; {\it (iii)} algorithms implemented by R package \texttt{OptimalDesign} \citep{harman2025package} including their randomized exchange algorithm (REX, see also \cite{harman2020randomized}), a vertex-direction method (VDM, see also \cite{fedorov2025model}), and a multiplicative algorithm (MUL, see also \cite{silvey1978}).

\begin{figure}[ht]
    \centering
    \includegraphics[width=\textwidth]{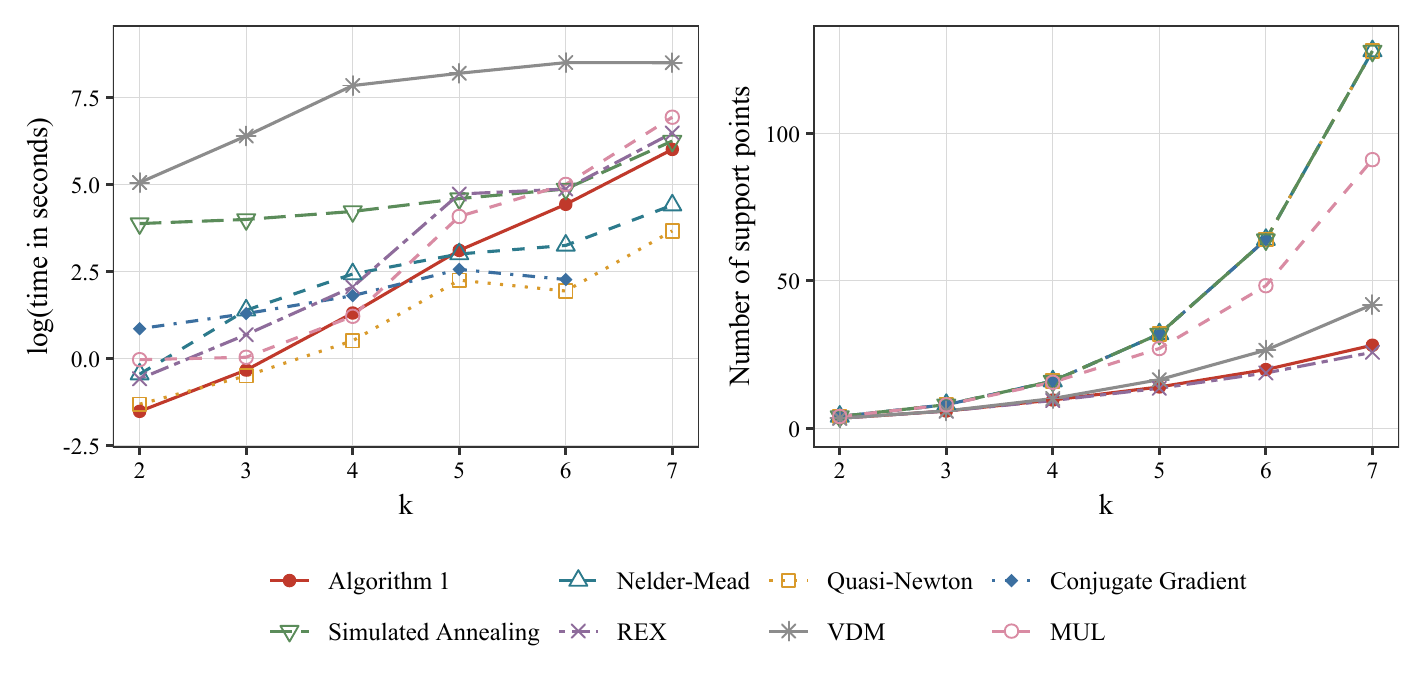}
    \caption{A-optimal designs for the $2^k$ main-effects model (Example~\ref{ex:main_effects}) as a function of $k$: (left) log total computation time; (right) average number of support points, each averaged across 100 simulations.}
    \label{fig:2^k_main_fig}
    \caption*{\footnotesize{Note: Lines ending before $k = 7$ indicate algorithmic failure for those settings (corresponding to ``NA'' in Table~\ref{tab:2^k_main_tab} of the Supplementary Material).}}
\end{figure}

The total computation time in seconds (shown in Figure~\ref{fig:2^k_main_fig} and Table~\ref{tab:2^k_main_tab} in the Supplementary Material) is obtained on a Windows-based laptop with 12th Gen Intel Core i7-1260P processor (2.10 GHz) and 8 GB memory (5200 MT/s). Figure~\ref{fig:2^k_main_fig} displays the total computation time (on the natural log scale) and the average number of support points (i.e., design points with positive allocations) for each algorithm. The underlying values, together with the average relative efficiency with respect to ${\mathbf w}_a$ obtained by Algorithm~\ref{algo:A_opt_lift_one} (i.e., $h({\mathbf w})/h({\mathbf w}_a)$), are reported in Table~\ref{tab:2^k_main_tab} of the Supplementary Material. Overall, Algorithm~\ref{algo:A_opt_lift_one} demonstrates notable advantages across multiple evaluation metrics. It is consistently faster than simulated annealing, REX, VDM, and MUL. Regarding design sparsity, it needs much less support points than other algorithms, except REX, which produces comparable sparsity. As pointed out by \cite{huang2024forlion}, in many applications, a smaller number of support points typically indicates reduced time and cost of the experiment. In terms of relative efficiency, Algorithm~\ref{algo:A_opt_lift_one} is comparable with REX, VDM, and MUL, and outperforms the classical optimization algorithms in finding A-optimal designs.
}\hfill{$\Box$}
\end{example}

\section{A-optimal Designs with Mixed Factors}\label{sec:Aopt_mixed_factors}

In this section, we consider experiments under a GLM with mixed factors. We denote $I_j$ as the collection of numerical levels of the $j$th factor, for $1\leq j\leq d$. For typical experiments, $I_j = [l_j, r_j]$ is a finite closed interval for a continuous factor, or a finite set for a discrete factor. To simplify the notations, we assume that the first $s \in \{0, 1, \ldots, d\}$ factors are continuous. Following \cite{lin2025ew}, if $s=d$, i.e., all factors are continuous, we consider a design region ${\cal X} = \prod_{j=1}^d I_j$~; if $1\leq s\leq d-1$, i.e., the factors are mixed, then ${\cal X} = \prod_{j=1}^s I_j \times {\cal D}$ for some ${\cal D} \subseteq \prod_{j=s+1}^d I_j$ (see Examples~1 \& 2 in \cite{lin2025ew}); if $s=0$, i.e., all factors are discrete, then ${\cal X}={\cal D}$, which has been discussed in Section~\ref{sec:A_optimal_designs}. Note that ${\cal X}$ considered in this paper is always closed and bounded, and thus compact.

In this section, we assume $1 \leq s \leq d$, i.e., there is at least one continuous factor. A feasible design is denoted as $\boldsymbol{\xi} = \{({\mathbf x}_i, w_i), i=1, \ldots, m\}$, such that $m\geq 1$, ${\mathbf x}_i \in {\cal X}$, $w_i\geq 0$, and $\sum_{i=1}^m w_i = 1$. We denote $\boldsymbol{\Xi}$ as the collection of all feasible designs. To avoid trivial optimization problems, we assume that there exists a $\boldsymbol{\xi} \in \boldsymbol{\Xi}$, such that, $f(\boldsymbol{\xi}) = |{\mathbf X}_{\boldsymbol{\xi}}^T {\mathbf W}_{\boldsymbol{\xi}}  {\mathbf X}_{\boldsymbol{\xi}}| > 0$, where ${\mathbf X}_{\boldsymbol{\xi}} = ({\mathbf q}({\mathbf x}_1), \ldots, {\mathbf q}({\mathbf x}_m))^T$, ${\mathbf W}_{\boldsymbol{\xi}} = {\rm diag}\{w_1 \nu(\boldsymbol{\beta}^T {\mathbf q}({\mathbf x}_1)), \ldots, w_m \nu(\boldsymbol{\beta}^T {\mathbf q}({\mathbf x}_m))\}$. According to \eqref{eq:h(w)} and Lemma~\ref{lem:f_-j(w)_>0} in the Supplementary Material, $f(\boldsymbol{\xi})>0$ implies $h(\boldsymbol{\xi}) > 0$ as well, where $h(\boldsymbol{\xi}) =  [{\rm tr}(({\mathbf X}_{\boldsymbol{\xi}}^T {\mathbf W}_{\boldsymbol{\xi}}  {\mathbf X}_{\boldsymbol{\xi}})^{-1})]^{-1}$.

\subsection{Characterization of A-optimal designs under a GLM}\label{sec:GLM_forlion_formula}

For a feasible design $\boldsymbol{\xi} \in \boldsymbol{\Xi}$, it contains only a finite number of design points, and can be regarded as a discrete probability measure $\{w_i, \ldots, w_m\}$ defined on $\{{\mathbf x}_1, \ldots, {\mathbf x}_m\}$. We let ${\mathbf F}_{\mathbf x} = \nu(\boldsymbol{\beta}^T{\mathbf q}({\mathbf x})) {\mathbf q}({\mathbf x}) {\mathbf q}({\mathbf x})^T \in \mathbb{R}^{p\times p}$ stand for the Fisher information matrix at ${\mathbf x} \in {\cal X}$. Then the Fisher information matrix (up to a multiplicative constant $n$) associated with $\boldsymbol{\xi}$ is ${\mathbf F}(\boldsymbol{\xi}) = {\mathbf X}_{\boldsymbol{\xi}}^T {\mathbf W}_{\boldsymbol{\xi}}  {\mathbf X}_{\boldsymbol{\xi}} = \sum_{i=1}^m w_i {\mathbf F}_{{\mathbf x}_i}$~. Theoretically, as the limit of a sequence of feasible designs, a general design $\boldsymbol{\xi}$ can be defined as a probability measure on the design space ${\cal X}$, which may contain infinitely many support points. Following \cite{fedorov2014} and \cite{lin2025ew}, we denote the collection of general designs as $\boldsymbol{\Xi}({\cal X})$, which is also the collection of probability measures on ${\cal X}$. Then for each $\boldsymbol{\xi}\in \boldsymbol{\Xi}({\cal X})$, the corresponding ${\mathbf F}(\boldsymbol{\xi}) = \int_{\cal X} {\mathbf F}_{\mathbf x} \boldsymbol{\xi}(d{\mathbf x})$ under the GLM \eqref{eq:glm}. 
Based on Theorems~2.1 and 2.2 in \cite{fedorov2014}, we obtain the following theorem assuring that we can find a design with only a finite number of design points that is A-optimal among $\boldsymbol{\Xi}({\cal X})$.

\begin{theorem}\label{thm:GLM_A_optimality} For GLM~\eqref{eq:glm}, suppose ${\cal X}$ is compact, $\nu(\eta)$ is continuous, $q_1, \ldots, q_p$ are continuous with respect to all continuous factors of $\mathbf{x}\in {\cal X}$, and there exists a $\boldsymbol{\xi}\in \boldsymbol{\Xi}({\cal X})$ satisfying $|{\mathbf F}(\boldsymbol{\xi})|>0$. Then (i) there exists an A-optimal design $\boldsymbol{\xi}^*$ that contains no more than $p(p+1)/2$ design points;
(ii) the set of A-optimal designs is convex;
and (iii) a design $\boldsymbol\xi = \{({\mathbf x}_i, w_i), i=1, \ldots, m\}$ is A-optimal among $\boldsymbol{\Xi}({\cal X})$ if and only if
\begin{equation}\label{eq:A_opt_condition}
\max_{{\mathbf x} \in {\mathcal X}} \varphi({\mathbf x}, \boldsymbol{\xi})  \leq {\rm tr}(({\mathbf X}_{\boldsymbol{\xi}}^T {\mathbf W}_{\boldsymbol{\xi}}  {\mathbf X}_{\boldsymbol{\xi}})^{-1})\ ,
\end{equation}
where $\varphi({\mathbf x}, \boldsymbol{\xi}) = \nu\left(\boldsymbol\beta^T {\mathbf q}({\mathbf x})\right) \cdot {\mathbf q}({\mathbf x})^T({\mathbf X}^T_{\boldsymbol\xi} {\mathbf W}_{\boldsymbol\xi} {\mathbf X}_{\boldsymbol\xi})^{-2}{\mathbf q}({\mathbf x})$.
\end{theorem}

The function $\varphi({\mathbf x}, \boldsymbol{\xi})$ in \eqref{eq:A_opt_condition} is known as the sensitivity function at $\boldsymbol{\xi}$ along the direction of ${\mathbf x}$ \citep{fedorov2014, huang2024forlion}. For computational purposes, when verifying the condition \eqref{eq:A_opt_condition}, we may first find the eigen decomposition ${\mathbf O}\boldsymbol{\Lambda}{\mathbf O}^T$ of ${\mathbf X}_{\boldsymbol{\xi}}^T {\mathbf W}_{\boldsymbol{\xi}}  {\mathbf X}_{\boldsymbol{\xi}}$~, where ${\mathbf O}$ is an orthogonal matrix, and $\boldsymbol{\Lambda} = {\rm diag}\{\lambda_1, \ldots, \lambda_p\}$. Then $({\mathbf X}^T_{\boldsymbol\xi} {\mathbf W}_{\boldsymbol\xi} {\mathbf X}_{\boldsymbol\xi})^{-2} = {\mathbf O}\boldsymbol{\Lambda}^{-2} {\mathbf O}^T$, and  ${\rm tr}(({\mathbf X}_{\boldsymbol{\xi}}^T {\mathbf W}_{\boldsymbol{\xi}}  {\mathbf X}_{\boldsymbol{\xi}})^{-1})= \sum_{i=1}^p \lambda_i^{-1}$. 

Theorem~\ref{thm:GLM_A_optimality} is also closely related to the equivalence theorem for matrix means under linear models (see Section~7.20 of \cite{pukelsheim1993}). In terms of A-criterion, Pukelsheim's equivalence theorem states that a positive definite moment matrix ${\mathbf M}$ is A-optimal if and only if ${\mathbf q}({\mathbf x})^T {\mathbf M}^{-2} {\mathbf q}({\mathbf x}) \leq \mathrm{tr}({\mathbf M}^{-1})$ for all ${\mathbf x} \in \mathcal{X}$. Theorem~\ref{thm:GLM_A_optimality} is the GLM counterpart of this result, where the Fisher information matrix takes the form ${\mathbf F}(\boldsymbol{\xi})$ with the weight matrix ${\mathbf W}_{\boldsymbol{\xi}}$ reflecting the heteroscedastic nature of GLMs, and the sensitivity function in \eqref{eq:A_opt_condition} is the GLM analogue of ${\mathbf q}({\mathbf x})^T {\mathbf M}^{-2} {\mathbf q}({\mathbf x})$ with an additional weight $\nu(\boldsymbol{\beta}^T {\mathbf q}({\mathbf x}))$ arising from the non-constant variance structure. To accommodate the heteroscedastic structure of GLMs, we rely on the framework of \cite{fedorov2014}, deriving our Theorem~\ref{thm:GLM_A_optimality} from their Theorems~2.1 and 2.2.

\subsection{ForLion algorithm for A-optimal designs with mixed factors}
\label{sec:forlion_mixed_factor}

When there exists at least one continuous factor, we follow \cite{huang2024forlion} and propose a ForLion algorithm for finding A-optimal designs under a GLM with mixed factors.
Given GLM~\eqref{eq:glm}, our objective is to determine an A-optimal design $\boldsymbol\xi = \{({\mathbf x}_i, w_i), i=1, \ldots, m\}$ that maximizes $h(\boldsymbol{\xi})$ on $\boldsymbol{\Xi}$ (or equivalently, $\boldsymbol{\Xi}({\cal X})$, according to Theorem~\ref{thm:GLM_A_optimality}). 

Similarly to the original ForLion algorithm \citep{huang2024forlion} for D-optimality, given a design ${\boldsymbol\xi}_t = \{({\mathbf x}_i^{(t)}, w_i^{(t)}), i=1, \ldots, m_t\}$ at the $t$th iteration, we need to identify a design point ${\mathbf x}^*  \in {\mathcal X}$ maximizing $\varphi({\mathbf x}, {\boldsymbol\xi}_t)$ as established by Theorem~\ref{thm:GLM_A_optimality}. If $\varphi({\mathbf x}^*, {\boldsymbol\xi}_t) \leq {\rm tr}(({\mathbf X}_{\boldsymbol{\xi_t}}^T {\mathbf W}_{\boldsymbol{\xi_t}}  {\mathbf X}_{\boldsymbol{\xi_t}})^{-1})$, the current design ${\boldsymbol\xi}_t$ is reported as A-optimal, and the iterative process is terminated. If $\varphi({\mathbf x}^*, {\boldsymbol\xi}_t) > {\rm tr}(({\mathbf X}_{\boldsymbol{\xi_t}}^T {\mathbf W}_{\boldsymbol{\xi_t}}  {\mathbf X}_{\boldsymbol{\xi_t}})^{-1})$, the new design point ${\mathbf x}^*$ is incorporated into the updated design ${\boldsymbol{\xi}}_{t+1}$~. Similarly to Theorem~5 in \cite{huang2024forlion} for D-optimality, we let $\boldsymbol{\xi}_{t+1} = \{({\mathbf x}_i^{(t)}, (1-\alpha_t)w_i^{(t)}), i=1, \ldots, m_t\} \bigcup \{({\mathbf x}^*, \alpha_t)\}$, denoted as $(1-\alpha_t) {\boldsymbol\xi}_t \bigcup \{({\mathbf x}^*, \alpha_t)\}$ for simplicity, where $\alpha_t \in [0,1]$ is determined analytically by Theorem~\ref{thm:alphat} in the Supplementary Material (Section~\ref{sec:additional_theorems}). The complete procedure is summarized in Algorithm~\ref{algo:A_opt_Forlion} in the Supplementary Material (Section~\ref{sec:algorithms}) for finding A-optimal designs under a GLM when there exists at least one continuous factor (i.e., $s\geq 1$).

As a direct conclusion of Theorem~\ref{thm:GLM_A_optimality}, the design reported by Algorithm~\ref{algo:A_opt_Forlion} is A-optimal when the stopping condition \eqref{eq:A_opt_condition} is met.

\begin{corollary}\label{cor:ForLion_Aopt}
Under the conditions of Theorem~\ref{thm:GLM_A_optimality}, a design $\boldsymbol{\xi}_t \in \boldsymbol{\Xi}$ is reported by Algorithm~\ref{algo:A_opt_Forlion} if and only if it is A-optimal among $\boldsymbol{\Xi}({\cal X})$.    
\end{corollary}

Conclusions like Corollary~\ref{cor:ForLion_Aopt} (see also Theorem~1 and Remark~2 in \cite{huang2024forlion}) make ForLion algorithms different from stochastic optimization algorithms such as particle swarm optimization (PSO), which do not guarantee an optimal solution to be found \citep{kennedy1995particle, poli2007particle, huang2024forlion}. Compared with commonly used algorithms that discretize the continuous factors first (see, e.g., \cite{yangmin2013} and \cite{harman2020randomized}),  Algorithm~\ref{algo:A_opt_Forlion} is able to find highly efficient designs with reduced numbers of design points, as illustrated by the following example.

\begin{example}\label{ex:compare_ForLion_OptimalDesign} {\bf Comparison of Algorithm~\ref{algo:A_opt_Forlion} and the REX algorithm}\quad {\rm  
We revisit Example~4.7 in \cite{stufken2012} (see also Example~2 in \cite{huang2024forlion}) under a logistic regression model, ${\rm logit}(\mu_i) = \beta_0 + \beta_1 x_{i1} + \beta_2 x_{i2} + \beta_3 x_{i3}$ with $(\beta_0, \beta_1, \beta_2, \beta_3) = (1, -0.5, 0.5, 1)$ and three continuous factors $x_{i1} \in [-2, 2]$, $x_{i2} \in [-1,1]$, and $x_{i3} \in [-3, 3]$. We compare Algorithm~\ref{algo:A_opt_Forlion} with the REX algorithm (see also Example~\ref{ex:main_effects}) in R package \texttt{OptimalDesign} \citep{harman2025package}, which can find an A-optimal design in this case by discretizing the continuous factors first.

\begin{table}[ht] 
\caption{A-optimal designs by REX for Example~\ref{ex:compare_ForLion_OptimalDesign}} 
\label{tab:A_REX_tab_-3_3} 
\centering
{\renewcommand{\arraystretch}{0.5}
\resizebox{\textwidth}{!}{%
\begin{tabular}{ccccc} 
\toprule 
Grid level & $\#$ of levels of factors & $\#$ of support points & Time cost in seconds & Relative efficiency w.r.t. $\boldsymbol{\xi}_a$ \\ 
\midrule 
0.05 & (81,41,121) & 13 & 2.99 & 99.9951\% \\ 
0.02 & (201,101,301) & 13 & 59.17 & 99.9992\% \\ 
0.0125 & (321,161,481) & 10 & 247.92 & 96.0565\% \\ 
0.01 & (401,201,601) & - & - & - \\ 
\bottomrule 
\end{tabular}
}
}\end{table}

Algorithm~\ref{algo:A_opt_Forlion} yields an A-optimal design $\boldsymbol{\xi}_a$ with $8$ support points in $200.02$ seconds. Table~\ref{tab:A_REX_tab_-3_3} summarizes the A-optimal designs obtained by REX under different grid levels, where each continuous factor is first discretized into equally spaced grid points with spacing $\Delta$ over its design region. With relatively coarse grids, REX achieves high relative efficiency but requires more support points, whereas finer grids substantially increase computational cost, degrade efficiency, or fail to return a solution ($\Delta = 0.01$). Overall, by directly searching the continuous design space, Algorithm~\ref{algo:A_opt_Forlion} produces highly efficient designs with fewer support points and remains computationally feasible with multiple continuous factors.
}\hfill{$\Box$}
\end{example}

\section{Applications to Real Experiments}
\label{sec:examples}

In this section, we first illustrate the lift-one algorithm (Algorithm~\ref{algo:A_opt_lift_one} in the Supplementary Material, Section~\ref{sec:algorithms}) for A-optimal designs with discrete factors. Using real data from a wave-solder experiment under a Poisson regression model (see Example~\ref{ex:wavesolder_example}), we show that the A-optimal design yields more accurate coefficient estimates than the corresponding D-optimal design and the original fractional factorial deign. Further comparisons with commonly used sampling schemes and the D-optimal one, showing that the A-optimal sampling scheme may lead to more accurate parameter estimates and predictions, are provided in the Supplementary Material (see Examples~\ref{ex:RMSE}, ~\ref{ex:saturated}, and~\ref{ex:compare_A_D} in Section~\ref{sec:examples_supp}).

We then illustrate the ForLion algorithm (Algorithm~\ref{algo:A_opt_Forlion} in the Supplementary Material, Section~\ref{sec:algorithms}) for A-optimal designs with mixed factors. In Example~\ref{ex:potato_example} about a real potato packing study, we compare the A-optimal design found by Algorithm~\ref{algo:A_opt_Forlion} with the A-, I-, and D-optimal designs obtained by the sequential approach of \cite{li2021}. More examples about Algorithm~\ref{algo:A_opt_Forlion} can be found in the Supplementary Material (see Example~\ref{ex:2_paramrters_logit_continuous} for a logistic regression model and Example~\ref{ex:3_paramrters_gamma} under a Gamma model in Section~\ref{sec:examples_supp}).

\begin{example}\label{ex:wavesolder_example} {\bf Wave-soldering defects}\quad {\rm We consider a wave-solder experiment analyzed by \citet{hamada1997glmquality}, whose data are also distributed as \texttt{wavesolder} in the \textsf{R} package \textsf{faraway} \citep{faraway2025package}. The response is the count of defects in a wave-soldering process for attaching components to an electronic circuit card.
The original experiment used a $2^{7-3}$ fractional factorial design with seven two-level (0 or 1) factors: (A)~prebake condition, (B)~flux density, (C)~conveyer speed, (D)~preheat condition, (E)~cooling time, (F)~ultrasonic solder agitator, and (G)~solder temperature, with three boards at each of the $16$ runs, giving $16\times 3 = 48$ observations. \citet{hamada1997glmquality} analyzed the data using a Poisson regression model with log link and concluded that four out of six interactions under consideration were not significant (see Section~\ref{sec:examples_supp} in the Supplementary Material). For illustration purposes, we fit the data with a main-effects Poisson model with log link and obtain the fitted coefficients  $\boldsymbol{\beta} = (3.043,\, 0.651,\, -0.577,\, 1.229,\, -0.166,\, -0.162,\, -0.090,\, -0.698)^T$, ordered as the intercept followed by factors (A)--(G). Treating $\boldsymbol{\beta}$ as the true parameter values, we compare D- and A-optimal designs restricted to the same sixteen runs considered by the original design (see Table~5 in \cite{hamada1997glmquality} with levels 1 and 2).

More specifically, the D-optimal design is obtained via the lift-one algorithm for D-optimality \citep{ymm2016, ym2015}, the A-optimal design is calculated by Algorithm~\ref{algo:A_opt_lift_one}, and the original design assigns equal weight $1/16$ to each of the 16 runs. The D- and A-optimal designs are supported on $11$ and $12$ of the $16$ runs, respectively, while the original design uses all $16$. For each design and a sample size $n\in\{48,100,200,500,1000\}$, where $n=48$ matches the original experiment, we convert the approximate design into an exact design using a round-off algorithm (see Algorithm~\ref{algo:exact_A} in the Supplementary Material, Section~\ref{sec:algorithms}), simulate Poisson responses under the true coefficients $\boldsymbol\beta$, compute the maximum likelihood estimate $\hat{\boldsymbol\beta}$ based on the simulated data, and record the root mean squared error ${\rm RMSE} = \big[\sum_{i \in I} (\hat{\beta}_i - \beta_i)^2/|I|\big]^{1/2}$ over the index set $I$ of all eight coefficients.

\begin{figure}[t]
  \centering
  \includegraphics[width=0.5\textwidth]{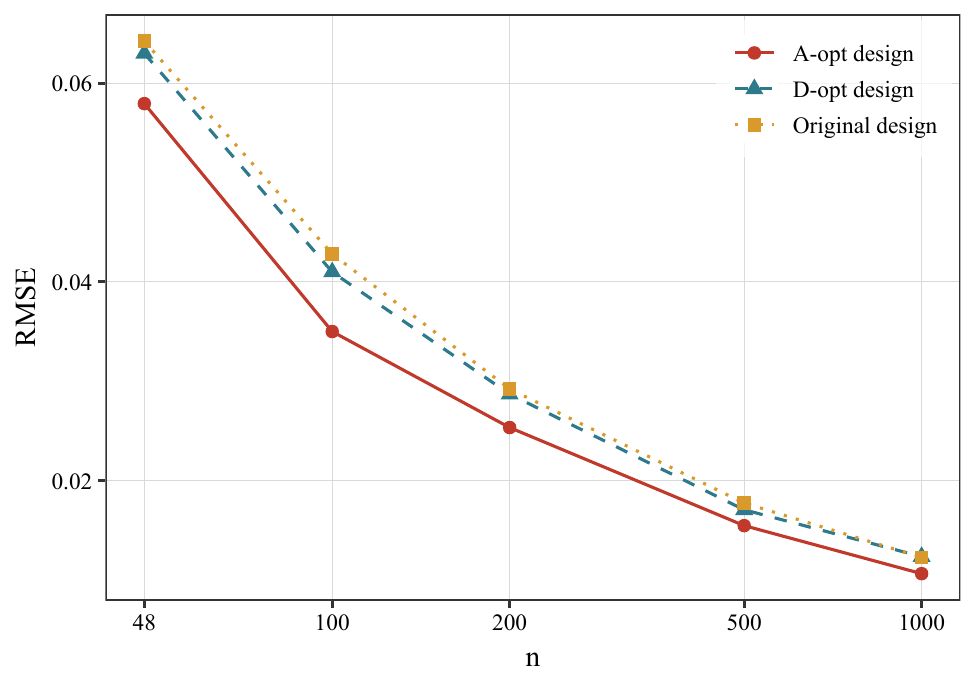}
  \caption{Mean RMSE versus sample size $n$ over 100 simulations in Example~\ref{ex:wavesolder_example}, with the horizontal axis on a logarithmic scale.}
  \label{fig:rmse_ws}
\end{figure}

Figure~\ref{fig:rmse_ws} plots the mean RMSE averaged over $100$ simulated data sets against the sample size. At every sample size the A-optimal design attains the smallest RMSE, while the D-optimal design and the original one are comparable to each other, both noticeably less accurate. Two-sided paired $t$-tests confirm that A-optimal designs are typically more accurate ($p<0.05$) than D-optimal designs and the original one. At $n=48$, the size of the original experiment, the A-optimal design still reduces the mean RMSE by about $10\%$ relative to the original one. Exact values of the mean RMSE, standard deviations, and significance markings, are given in Table~\ref{tab:rmse_ws} in the Supplementary Material.
}\hfill{$\Box$}
\end{example}

\begin{example}\label{ex:potato_example} {\bf A-optimal design for the potato packing experiment}\quad {\rm  We consider a real-world example from a potato packing study analyzed by \cite{woods2006} and revisited later by \cite{li2021}. A binary response indicates the presence of liquid after seven days of storage. Three standardized continuous factors are involved, namely the vitamin concentration in the prepackaging dip and the levels of two gases in the packaging atmosphere, with the design space $\mathcal{X} = [-1,1]^3$. For comparison purposes, we adopt the same model and parameter setting considered in \cite{li2021}, namely ${\rm logit}(\mu_i) = \boldsymbol\beta^T{\mathbf q}({\mathbf x}_i)$ with ${\mathbf q}({\mathbf x}_i) = (1, x_{i2}, x_{i3}, x_{i2}x_{i3}, x_{i1}^2, x_{i2}^2, x_{i3}^2)^T$ and $\boldsymbol\beta = (-2.93, -0.52, -0.79, -0.66, 0.94, 0.79, 1.82)^T$. In this example, we compare the A-optimal designs obtained by our Algorithm~\ref{algo:A_opt_Forlion} and the sequential algorithm proposed by \citet{li2021}, which combines the Fedorov-Wynn algorithm \citep{wynn1970} and the multiplicative algorithm \citep{titterington1978} and is implemented in \textsc{Matlab}. Using Algorithm~\ref{algo:A_opt_Forlion}, the resulting A-optimal design has $21$ support points, compared with $29$ for \cite{li2021}. In terms of A-efficiency, Algorithm~\ref{algo:A_opt_Forlion} achieves about $0.16\%$ higher.

To compare the algorithms across different parameter settings, we conduct a simulation study with 100 parameter vectors sampled from a multivariate normal distribution centered at $\boldsymbol{\beta}$. For each simulated parameter vector, we obtain A-optimal designs using Algorithm~\ref{algo:A_opt_Forlion} and  \cite{li2021}'s algorithm, respectively. Overall, designs from Algorithm~\ref{algo:A_opt_Forlion} require significantly fewer support points (an average of 18.50 versus 26.29; fewer in 96 cases; Figure~\ref{fig:boxplot_points}). While our algorithm achieved higher A-efficiency in 69 cases, a paired $t$-test shows this difference is not statistically significant. Thus, Algorithm~\ref{algo:A_opt_Forlion} matches the A-efficiency of \cite{li2021} while substantially reducing the number of support points. Each support point requires a separate batch of packages stored under specific conditions for seven days. Having fewer support points cuts down on these batches, which lowers material and labor costs even if A-efficiency stays the same.

\begin{figure}[htbp]
    \centering
    \begin{minipage}[t]{0.45\textwidth}
        \centering
        \includegraphics[width=\textwidth]{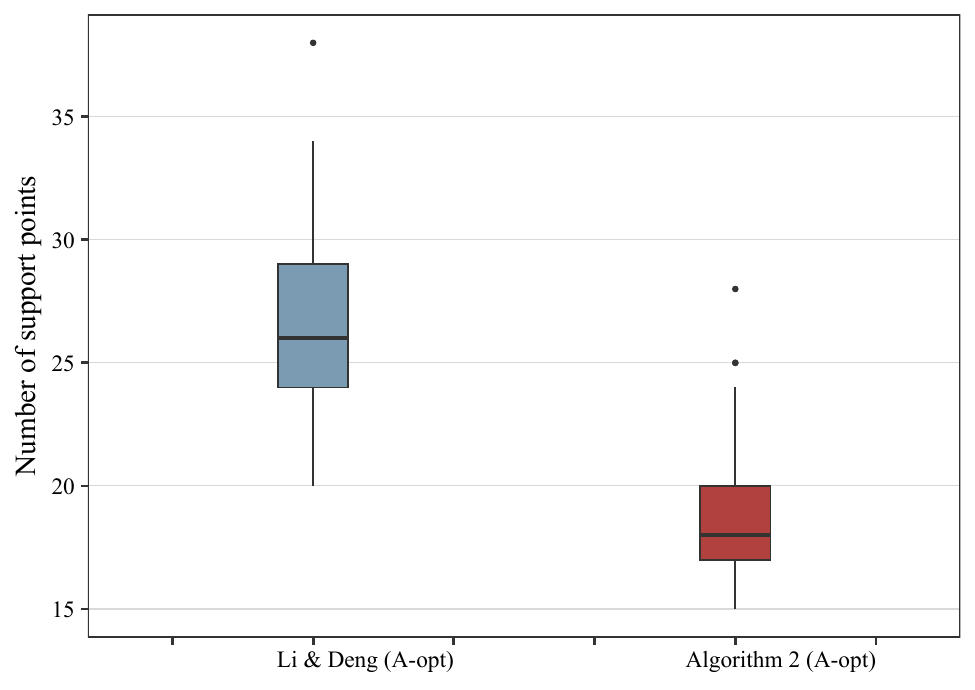}
    \caption{Boxplots of numbers of support points over 100 simulations for Example~\ref{ex:potato_example}.}
    \label{fig:boxplot_points}
    \end{minipage}
    \hfill
    \begin{minipage}[t]{0.45\textwidth}
        \centering
        \includegraphics[width=\textwidth]{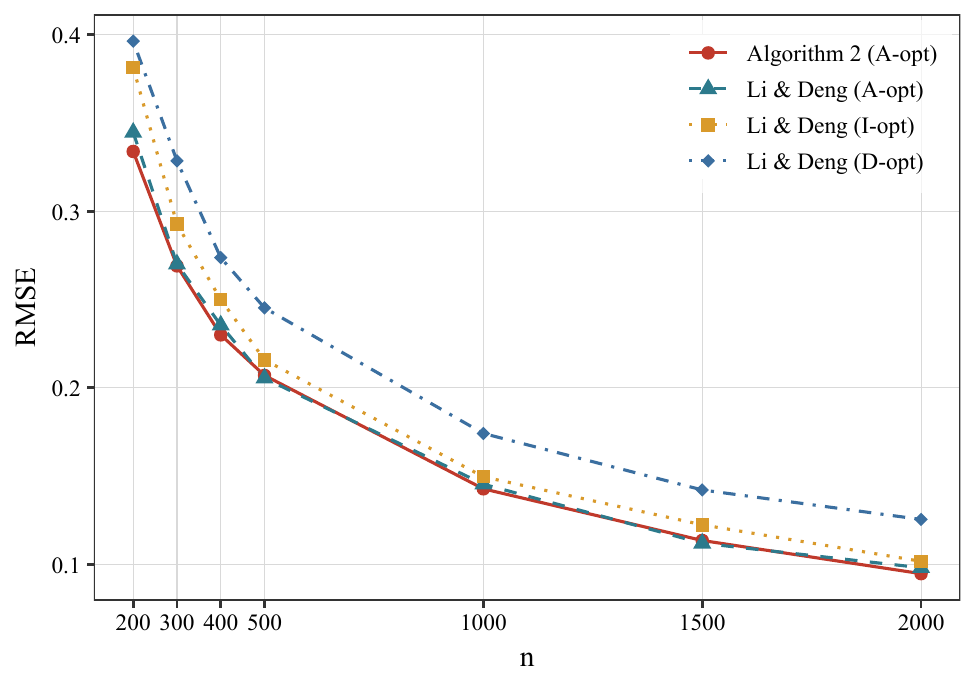}
        \caption{RMSE against the sample size $n$ over 100 simulations in Example~\ref{ex:potato_example}.}
    \label{fig:rmse_potato}
    \end{minipage}
\end{figure}

Beyond the comparison of A-efficiency and support-point counts above, we further examine how the choice of optimality criterion affects estimation accuracy. We restrict attention to the A-, I-, and D-criteria supported by \cite{li2021}'s sequential algorithm. Among the three, the A-criterion minimizes ${\rm tr}(({\mathbf X}^T{\mathbf W}{\mathbf X})^{-1})$, the average variances of the parameter estimates. Here the I-criterion is the elastic I-optimality of \cite{li2021} under the uniform measure, which coincides with classical I-optimality. Because the average variance approximates the mean squared error of $\hat{\boldsymbol\beta}$ in large samples, the A-criterion is the one among the three most directly aligned with coefficient-estimation accuracy. To assess whether this yields a measurable advantage, we compare the A-optimal design against the I- and D-optimal designs using the RMSE of $\hat{\boldsymbol\beta}$. Following the same simulation protocol as in Example~\ref{ex:wavesolder_example}, we compare the designs by the RMSE of $\hat{\boldsymbol\beta}$ over a range of sample sizes $n$, averaged over $100$ simulated data sets at each $n$.

Figure~\ref{fig:rmse_potato} shows the RMSE curves against the sample size. The underlying averages and standard deviations, together with the corresponding pairwise significance tests, are reported in Tables~\ref{tab:rmse_potato} and~\ref{tab:potato_rmse_pairwise} of the Supplementary Material. Across all sample sizes, the two A-optimal designs, obtained by Algorithm~\ref{algo:A_opt_Forlion} and by \cite{li2021}, respectively, attain the smallest RMSE and are statistically indistinguishable, consistent with their nearly identical A-optimality and confirms that Algorithm~\ref{algo:A_opt_Forlion} recovers the A-optimal design. The D-optimal design is uniformly the least accurate, while the I-optimal design lies in between and is only marginally inferior to the A-optimal designs. The advantage of the A-optimal designs over the D-optimal design is statistically significant from $n=200$ onward. This ordering of RMSE matches the A-efficiencies of the I- and D-optimal designs relative to the A-optimal design ($81\%$ and $57\%$). The empirical RMSE ratios at large $n$ agree closely with these efficiency-based predictions. 
}\hfill{$\Box$}
\end{example}

\section{Discussion}
\label{sec:discussion}

Constructing A-optimal designs is computationally more challenging than constructing D-optimal designs. Under generalized linear models (GLMs), this difficulty is compounded because the Fisher information matrix depends on unknown parameters through the link and variance functions, preventing classical linear model results from directly carrying over. Consequently, existing theory has largely been confined to specific distributions, one or two factors, or designs that satisfy restrictive technical assumptions. Our goal is to make A-optimal designs computationally feasible without these limitations. When $m = p$, the A-optimal weights have a closed form. When $m > p$, the maximum of the objective function along the lift-one direction, which raises a single weight and rescales the remaining ones proportionally, falls into one of four analytical cases. That result drives both the lift-one algorithm for a fixed set of experimental settings and the ForLion algorithm for design regions with continuous or mixed factors. A rounding-off algorithm converts an approximate allocation into an exact one, assigning each leftover unit where it does most for the A-criterion.

A-optimality minimizes the average variance of the parameter estimates, which makes it a natural criterion when a study is meant to estimate those parameters accurately. In our examples, A-optimal designs tend to attain smaller RMSE than the D-optimal and standard alternatives. Where continuous factors are involved, our algorithms further return designs with fewer support points than existing algorithms at comparable efficiency, and since each support point is a distinct experimental setting to prepare and maintain, this can save considerable experimental time and cost. Two features of the criterion deserve attention before it is adopted. Because the A-criterion adds the variances of the individual parameter estimates, the factors should be coded on comparable scales. The criterion also weighs every parameter equally, so a weighted version is preferable when some parameters matter more than others. If prediction rather than estimation is the goal, the I-criterion may be an appropriate choice.

Our designs are locally optimal, so an assumed $\boldsymbol{\beta}$ is required and efficiency falls away as that guess departs from the truth. 
When little information about model parameters is available, robust designs are needed to protect against misspecified parameters in GLMs. The analytical results and algorithms developed in this paper may help construct these robust designs.


\subsection*{Supplementary Material}

The Supplementary Material consists of seven sections: S1 provides a list of notations to facilitate users' reading; S2 summarizes commonly used exponential family distributions; S3 provides the lift-one (Algorithm~\ref{algo:A_opt_lift_one}), ForLion (Algorithm~\ref{algo:A_opt_Forlion}), and round-off (Algorithm~\ref{algo:exact_A}) algorithms; S4 provides consequences of Theorem~\ref{thm:m=p} for special designs and technical results for Algorithms~\ref{algo:A_opt_lift_one} and~\ref{algo:A_opt_Forlion} (Theorems~\ref{thm:max_h_i(x)} and~\ref{thm:alphat}); S5 contains lemmas for Sections~\ref{sec:m>p} \& \ref{sec:lift_one_discrete_factor} and two remarks; S6 contains proofs; S7 provides additional examples and tables.

\subsection*{Acknowledgments}

The authors gratefully acknowledge the authors of \cite{li2021} for kindly sharing their source codes, which we used to implement and compare their methods with ours.

\subsection*{Declaration of Generative AI Use}

During the preparation of this work, the authors used Google Gemini (version 1.5 Pro) solely to check and improve the grammar, spelling, and sentence flow of the manuscript. The tool was not used to generate, analyze, or interpret data, nor to author original scientific content. After using this tool, the authors reviewed and edited the text as needed and takes full responsibility for the content of the published article.

\subsection*{Data Availability Statement}

The data that support the findings of this study are available from the corresponding author, J.Y., upon reasonable request.

\clearpage



\clearpage
\setcounter{page}{1}
\def\thepage{S\arabic{page}}



\centerline{\Large\bf A-optimal Designs under Generalized Linear Models}
\vspace{.25cm}
\centerline{Yingying Yang$^1$, Xiaotian Chen$^2$, and Jie Yang$^1$}
\vspace{.4cm}
\centerline{\it $^1$University of Illinois Chicago and $^2$University of Hong Kong}
\vspace{.55cm}
\centerline{\bf Supplementary Material}
\vspace{.55cm}
\par

\begin{description}
  \item[\textbf{S1}] List of notations
  \item[\textbf{S2}] Commonly used exponential family distributions
  \item[\textbf{S3}] Lift-one, ForLion, and Round-off algorithms
  \item[\textbf{S4}] Additional theorems and technical details
  \item[\textbf{S5}] Lemmas and remarks 
  \item[\textbf{S6}] Proofs
  \item[\textbf{S7}] Additional examples and tables
\end{description}

\renewcommand{\thesection}{S\arabic{section}}
\setcounter{section}{0}
\setcounter{equation}{0}
\def\theequation{S\arabic{section}.\arabic{equation}}
\setcounter{table}{0}
\def\thetable{S.\arabic{table}}
\setcounter{figure}{0}
\def\thefigure{S.\arabic{figure}}

\fontsize{12}{14pt plus.8pt minus .6pt}\selectfont

\section{List of notations}\label{sec:list}

\begin{list}{}{
		\setlength{\labelwidth}{0.5in}
		\setlength{\leftmargin}{0.6in}
		\setlength{\labelsep}{.3in}
		\setlength{\rightmargin}{\leftmargin}
	}
\item[$a_i$\hfill] Nonnegative constant in $f_i(z) = a_i z(1-z)^{p-1} + b_i(1-z)^p$, note that $a_i$ depends on $i$, ${\mathbf X}$, and ${\mathbf W}$, see the context of equation~\eqref{eq:f_i_a_b}
\item[$A_i$\hfill] Nonnegative constant $\sum_{j=1}^p a_{ij}$~, note that it depends on $i$, ${\mathbf X}$, and ${\mathbf W}$, see the context of equation~\eqref{eq:h_i(x)}
\item[${\mathbf A}$\hfill] A $p\times p$ matrix, ${\mathbf A} = ({\mathbf X}^T_{\boldsymbol{\xi}_t} {\mathbf W}_{\boldsymbol{\xi}_t} {\mathbf X}_{\boldsymbol{\xi}_t})^{-2}$, see Remark~\ref{rem:partial_gradient}
\item[$a_{ij}$\hfill] Nonnegative constant in $f_i^{(-j)}(z) = a_{ij} z(1-z)^{p-2} + b_{ij}(1-z)^{p-1}$, $j=1, \ldots, p$, note that $a_{ij}$ also depends on $i$, ${\mathbf X}$, and ${\mathbf W}$, see Lemma~\ref{lem:f(w)>0_a_jb_j_>0}
\item[$b_i$\hfill] Nonnegative constant in $f_i(z) = a_i z(1-z)^{p-1} + b_i(1-z)^p$, note that $b_i$ depends on $i$, ${\mathbf X}$, and ${\mathbf W}$, see the context of equation~\eqref{eq:f_i_a_b}
\item[$B_i$\hfill] Nonnegative constant $\sum_{j=1}^p b_{ij}$~, note that it depends on $i$, ${\mathbf X}$, and ${\mathbf W}$, see the context of equation~\eqref{eq:h_i(x)}
\item[$b(\theta)$\hfill] Function of $\theta$ in the canonical form of $f(y;\theta)$ 
\item[$b_{ij}$\hfill] Nonnegative constant in $f_i^{(-j)}(z) = a_{ij} z(1-z)^{p-2} + b_{ij}(1-z)^{p-1}$, $j=1, \ldots, p$, note that $b_{ij}$ also depends on $i$, ${\mathbf X}$, and ${\mathbf W}$, see Lemma~\ref{lem:f(w)>0_a_jb_j_>0}
\item[$c(\theta)$\hfill] Function of $\theta$ in the canonical form of $f(y;\theta)$ 
\item[$c_i$\hfill] The $i$th diagonal element of $({\mathbf X}{\mathbf X}^T)^{-1}$
\item[$d$\hfill] Number of covariates or experimental factors under consideration, $d\geq 1$ 
\item[${\cal D}$\hfill] Collection of level combinations of discrete factors under consideration, ${\cal D} \subseteq \prod_{j=s+1}^d I_j$ if $s<d$ 
\item[$d(y)$\hfill] Function of $y$ in the canonical form of $f(y;\theta)$ 
\item[${\mathbf e}_i$\hfill] Vector in $\mathbb{R}^m$ whose $i$th coordinate is $1$ and others are zeros
\item[${\mathbf F}$\hfill] Fisher information matrix in $\mathbb{R}^{p\times p}$, ${\mathbf F} = n{\mathbf X}^T{\mathbf W}{\mathbf X} = n \sum_{i=1}^m w_i \nu_i {\mathbf q}({\mathbf x}_i) {\mathbf q}({\mathbf x}_i)^T$ under a GLM
\item[${\mathbf F}_{\mathbf x}$\hfill] Fisher information matrix at ${\mathbf x} \in {\cal X}$, i.e.,\\ ${\mathbf F}_{\mathbf x} = \nu(\boldsymbol{\beta}^T {\mathbf q}({\mathbf x})) {\mathbf q}({\mathbf x}) {\mathbf q}({\mathbf x})^T \in \mathbb{R}^{p\times p}$
\item[$f(y; \theta)$\hfill] Probability density function or probability mass function of $Y$  
\item[$f({\mathbf w})$\hfill] Objective function for D-optimality of ${\mathbf w}$, under a GLM, $f({\mathbf w}) = |{\mathbf X}^T{\mathbf W}{\mathbf X}|$ is an order-$p$ homogeneous polynomial of $w_1, \ldots, w_m$
\item[${\mathbf F}({\mathbf w})$\hfill] Fisher information matrix (up to a multiplicative constant $n$) associated with ${\mathbf w}$, ${\mathbf F}({\mathbf w})  = {\mathbf X}^T {\mathbf W}  {\mathbf X}$ under a GLM, see the proof of Theorem~\ref{thm:for_lift_one} in Section~\ref{sec:proofs} 
\item[$f(\boldsymbol{\xi})$\hfill] Objective function for D-optimality of $\boldsymbol{\xi}$, $f(\boldsymbol{\xi}) = |{\mathbf F}(\boldsymbol{\xi})|$
\item[${\mathbf F}(\boldsymbol{\xi})$\hfill] Fisher information matrix (up to a multiplicative constant $n$) of $\boldsymbol{\xi}$, under a GLM, ${\mathbf F}(\boldsymbol{\xi})  = {\mathbf X}_{\boldsymbol{\xi}}^T {\mathbf W}_{\boldsymbol{\xi}}  {\mathbf X}_{\boldsymbol{\xi}}$ for $\boldsymbol{\xi} \in \boldsymbol{\Xi}$, or ${\mathbf F}(\boldsymbol{\xi}) = \int_{\cal X} {\mathbf F}_{\mathbf x} \boldsymbol{\xi}(d{\mathbf x})$ for $\boldsymbol{\xi}\in \boldsymbol{\Xi}({\cal X})$
\item[$f_i(z)$\hfill] $f({\mathbf w}_i(z)) = f\left(\frac{1-z}{1-w_i}w_1,\ldots,\frac{1-z}{1-w_i}w_{i-1},z, \frac{1-z}{1-w_i}w_{i+1},\ldots, \frac{1-z}{1-w_i}w_{m}\right)$, given \\ $0\leq w_i<1$, $i=1, \ldots, m$, $z\in [0,1]$
\item[$f_i^{(-j)}(z)$\hfill] $f_{-j}({\mathbf w}_i(z)) = f_{-j}\left(\frac{1-z}{1-w_i}w_1,\ldots,\frac{1-z}{1-w_i}w_{i-1},z, \frac{1-z}{1-w_i}w_{i+1},\ldots, \frac{1-z}{1-w_i}w_{m}\right)$, \\ given $0\leq w_i<1$, $i=1, \ldots, m$, $j=1, \ldots, p$, $z\in [0,1]$
\item[$f_{-j}({\mathbf w})$\hfill] $f_{-j}({\mathbf w}) = |{\mathbf X}_{-j}^T{\mathbf W}{\mathbf X}_{-j}|$, an order-$(p-1)$ homogeneous polynomial of $w_1, \ldots, w_m$~, $j=1, \ldots, p$
\item[$f_{-j}(\boldsymbol{\xi})$\hfill] $f_{-j}(\boldsymbol\xi) = |{\mathbf X}_{\boldsymbol\xi, -j}^T {\mathbf W}_{\boldsymbol\xi} {\mathbf X}_{\boldsymbol\xi, -j}|$, an order-$(p-1)$ homogeneous polynomial of $w_1, \ldots, w_m$~, $j=1, \ldots, p$
\item[$g$\hfill] Link function for a generalized linear model, such that, $\eta_i = g(\mu_i)$ 
\item[$h({\mathbf w})$\hfill] Objective function for A-optimality of ${\mathbf w}$, under a GLM, $h({\mathbf w}) = \left[{\rm tr}\left(({\mathbf X}^T{\mathbf W}{\mathbf X})^{-1}\right)\right]^{-1}$
\item[$h(\boldsymbol{\xi})$\hfill] Objective function for A-optimality of $\boldsymbol{\xi}$, under a GLM, $h(\boldsymbol{\xi}) =  [{\rm tr}(({\mathbf X}_{\boldsymbol{\xi}}^T {\mathbf W}_{\boldsymbol{\xi}}  {\mathbf X}_{\boldsymbol{\xi}})^{-1})]^{-1}$ 
\item[$h_i(z)$\hfill] $h({\mathbf w}_i(z)) = h\left(\frac{1-z}{1-w_i}w_1,\ldots,\frac{1-z}{1-w_i}w_{i-1},z, \frac{1-z}{1-w_i}w_{i+1},\ldots, \frac{1-z}{1-w_i}w_{m}\right)$, which also equals $\frac{f_i(z)}{\sum_{j=1}^p f_i^{(-j)}(z)}$ and $\frac{(b-a)z^2+(a-2b)z+b}{(A - B)z + B}$, given $0\leq w_i<1$, $i=1, \ldots, m$, $z\in [0,1]$
\item[$I_j$\hfill] Range of numerical levels of the $j$th factor, $j=1, \ldots, d$, $I_j$ is a finite closed interval $[l_j~,\ r_j]$ for continuous factors, or a finite set for discrete factors
\item[$I_{\bf w}$\hfill] Index set, $I_{\bf w} = \{i\in \{1, \ldots, m\}\mid w_i>0\}$, with $|I_{\bf w}| = \#\{i\mid w_i>0\} \geq p$ if $f({\mathbf w})>0$, see Proof of Theorem~\ref{thm:max_h_i(x)} in Section~\ref{sec:proofs}
\item[$l_j$\hfill] Left end of the range of numerical levels of the $j$th factor, $j=1, \ldots, d$ 
\item[$m$\hfill] Number of distinct design points or support points in a design
\item[$n$\hfill] Total number of experimental units
\item[${\mathbf n}$\hfill] Exact design ${\mathbf n}=(n_1, \ldots, n_m)^T$ given $n$, s.t., $\sum_{i=1}^m n_i = n$
\item[$n_i$\hfill] Number of experimental units assigned to or associated with ${\mathbf x}_i$~, $i=1, \ldots, m$, note that $\sum_{i=1}^m n_i = n$
\item[$p$\hfill] Number of parameters for a generalized linear model
\item[${\mathbf q}({\mathbf x}_i)$\hfill] Vector of predictor functions, $(q_1({\mathbf x}_i), \ldots, q_p({\mathbf x}_i))^T$, taking values at ${\mathbf x}_i$  
\item[$q_j({\mathbf x}_i)$\hfill] The $j$th predictor function taking values at the $i$th covariate vector or experimental setting ${\mathbf x}_i$~, $j=1, \ldots, p$ 
\item[$r_j$\hfill] Right end of the range of numerical levels of the $j$th factor, $j=1, \ldots, d$ 
\item[$s$\hfill] Number of continuous factors, $s\in \{0, 1, \ldots, d\}$
\item[$s(\eta_i)$\hfill] Function defined by $s(\eta_i) = {\rm Var}(Y_i)$ for each $i$
\item[$S_m$\hfill] Collection of all approximate allocations for $m$ experimental settings or design points, $S_m = \{{\mathbf w} = (w_1, \ldots, w_m)^T $ $\in \mathbb{R}^m \mid w_i\geq 0, \sum_{i=1}^m w_i = 1\}$
\item[$S_m^+$\hfill] Collection of all feasible approximate allocations of $m$ design points, i.e., $S_m^+ = \{{\mathbf w} \in S_m \mid f({\mathbf w}) > 0\}$
\item[${\mathbf w}$\hfill] Approximate allocation for a design with support points ${\mathbf x}_1, \ldots, {\mathbf x}_m$, that is, ${\mathbf w} = (w_1, \ldots, w_m)^T$, such that, $w_i\geq 0$ and $\sum_{i=1}^m w_i = 1$
\item[${\mathbf W}$\hfill] ${\mathbf W}={\rm diag}\{w_1\nu_1,\ldots, w_m\nu_m\} \in \mathbb{R}^{p\times p}$
\item[$w_i$\hfill] Proportion of experimental units assigned to or associated with ${\mathbf x}_i$~, $i=1, \ldots, m$, $\sum_{i=1}^m w_i = 1$
\item[${\mathbf w_i(z)}$\hfill] ${\mathbf w_i(z)} = \left(\frac{1-z}{1-w_i}w_1,\ldots,\frac{1-z}{1-w_i}w_{i-1},z, \frac{1-z}{1-w_i}w_{i+1},\ldots, \frac{1-z}{1-w_i}w_{m}\right)^T\ = \frac{1-z}{1-w_i}{\mathbf w} + \frac{z-w_i}{1-w_i}{\mathbf e_{i}}$~, for $z\in [0,1]$
\item[${\mathbf W}_{\boldsymbol{\xi}}$\hfill] Weight matrix of $\boldsymbol{\xi}$, i.e., ${\mathbf W}_{\boldsymbol{\xi}} = {\rm diag}\{w_1 \nu(\boldsymbol{\beta}^T {\mathbf q}({\mathbf x}_1)), \ldots, $ $w_m \nu(\boldsymbol{\beta}^T {\mathbf q}({\mathbf x}_m))\} \in \mathbb{R}^{m\times m}$
\item[${\mathbf X}$\hfill] $({\mathbf X}_1, \ldots, {\mathbf X}_m)^T = ({\mathbf q}({\mathbf x}_1), \ldots, {\mathbf q}({\mathbf x}_m))^T \in \mathbb{R}^{m\times p}$
\item[${\cal X}\hfill$] Design space or design region of design points, ${\cal X}$ takes the form of $\prod_{j=1}^d I_j$ if all factors are continuous, $\prod_{j=1}^s I_j \times {\cal D}$ with mixed factors, or ${\cal D}$ itself if all factors are discrete  
\item[${\mathbf X}\cdots$\hfill] ${\mathbf X}[i_1,\ldots,i_p]$, $p \times p$ submatrix of ${\mathbf X}$, consisting of the $i_1$th, $\ldots$, $i_p$th rows of ${\mathbf X}$ 
\item[${\mathbf x}_i$\hfill] The $i$th covariate vector or experimental setting, ${\mathbf x}_i = (x_{i1}, \ldots, x_{id})^T \in\mathbb{R}^d$ 
\item[${\mathbf X}_i$\hfill] The $i$th predictor vector associated with ${\mathbf x}_i$~,\\ ${\mathbf X}_i = {\mathbf q}({\mathbf x}_i) = (q_1({\mathbf x}_i), \ldots, q_p({\mathbf x}_i))^T$ $\in \mathbb{R}^p$ 
\item[$x_{ij}$\hfill] Numerical level of the $j$th covariate or factor associated with the $i$th covariate vector or experimental setting ${\mathbf x}_i$~, $j=1, \ldots, d$
\item[${\mathbf X}_{-j}$\hfill] $m\times (p-1)$ submatrix of ${\mathbf X}$ after removing its $j$th column
\item[${\mathbf X}_{\boldsymbol{\xi}}$\hfill] Model matrix of $\boldsymbol{\xi}$, i.e., ${\mathbf X}_{\boldsymbol{\xi}} = ({\mathbf q}({\mathbf x}_1), \ldots, {\mathbf q}({\mathbf x}_m))^T \in \mathbb{R}^{m\times p}$
\item[${\mathbf X}_{\boldsymbol{\xi}, -j}$\hfill] $m\times (p-1)$ submatrix of ${\mathbf X}_{\boldsymbol{\xi}}$ after removing its $j$th column
\item[$Y$\hfill] Univariate response $Y\in \mathbb{R}$
\item[$Y_i$\hfill] The $i$th response variable associated with ${\mathbf x}_i$
\item[$\boldsymbol{\beta}$\hfill] Parameter vector for a GLM, $\boldsymbol{\beta} = (\beta_1, \ldots, \beta_p)^T \in \mathbb{R}^p$ 
\item[$\beta_j$\hfill] The $j$th model parameter associated with the $j$th predictor function $q_j(\cdot)$, $j=1, \ldots, p$
\item[$\delta$\hfill] Merging threshold for Algorithm~\ref{algo:A_opt_Forlion}, e.g., $\delta=0.1$
\item[$\epsilon$\hfill] Converging threshold for Algorithm~\ref{algo:A_opt_Forlion}, e.g., $\epsilon=10^{-6}$
\item[$\eta_i$\hfill] The $i$th linear predictor associated with ${\mathbf x}_i$~,\\ $\eta_i = g(\mu_i) = {\mathbf X}_i^T \boldsymbol\beta = \sum_{j=1}^p \beta_j q_j({\mathbf x}_i)$
\item[$\theta$\hfill] Single parameter $\theta \in \mathbb{R}$ for an exponential family distribution
\item[$\mu_i$\hfill] Expectation of $Y_i$~, $\mu_i = E(Y_i)$
\item[$\nu$\hfill] Function defined by $\nu =  \left(\left(g^{-1}\right)'\right)^2/s$, such that, $\nu_i = \nu(\eta_i) = \nu\left({\mathbf X}_i^T\boldsymbol\beta\right)$
\item[$\nu_i$\hfill] $\nu_i = \nu(\eta_i) = \nu\left({\mathbf X}_i^T\boldsymbol\beta\right) \in \mathbb{R}$
\item[$\boldsymbol{\xi}$\hfill] A feasible design on ${\cal X}$, i.e., $\boldsymbol{\xi} = \{({\mathbf x}_i, w_i), i=1, \ldots, m\}$ for some $m\geq 1$, ${\mathbf x}_i \in {\cal X}$, and $w_i\geq 0$, such that, $\sum_{i=1}^m w_i = 1$
\item[$\boldsymbol{\Xi}$\hfill] Collection of feasible designs on ${\cal X}$, $\boldsymbol{\Xi} = \{\boldsymbol{\xi} = \{({\mathbf x}_i, w_i),$ $ i=1, \ldots, m\} | m\geq 1, {\mathbf x}_i \in {\cal X}, w_i\geq 0, \sum_{i=1}^m w_i = 1\}$
\item[$\boldsymbol{\Xi}({\cal X})$\hfill] Collection of general designs on ${\cal X}$, also the collection of probability measures on ${\cal X}$, a $\boldsymbol{\xi}\in \boldsymbol{\Xi}({\cal X})$ may contain infinitely many design points
\item[$\varphi({\mathbf x}, \boldsymbol{\xi})$\hfill] Sensitivity function at $\boldsymbol{\xi}$ along the direction of ${\mathbf x}$, i.e.,\\ $\varphi({\mathbf x}, \boldsymbol{\xi}) = \nu\left(\boldsymbol\beta^T {\mathbf q}({\mathbf x})\right) \cdot {\mathbf q}({\mathbf x})^T({\mathbf X}^T_{\boldsymbol\xi} {\mathbf W}_{\boldsymbol\xi} {\mathbf X}_{\boldsymbol\xi})^{-2}{\mathbf q}({\mathbf x})$
\end{list}

\section{Commonly used exponential family distributions}\label{sec:tables}

We list some commonly used exponential family distributions in Table~\ref{tab:exponential_list}.

\medskip\noindent
\begin{table}[ht]     
\caption{Some commonly used exponential family distributions.}
\label{tab:exponential_list}
\centering
\resizebox{0.9\textwidth}{!}{    \begin{tabular}{|ccccc|}\hline
Model & Range of $Y$ & $\log f(y; \theta)$ & $\theta$ & Constant \\ \hline
${\rm Bernoulli}(\theta)$     & $\{0,1\}$ & $y\log\frac{\theta}{1-\theta} + \log(1-\theta)$ & $E(Y)$ & -\\
${\rm Binomial}(n, \theta)$ & $\{0, 1, \ldots, n\}$ & $y\log\frac{\theta}{1-\theta} + n\log(1-\theta) + \log\binom{n}{y}$ & $\frac{E(Y)}{n}$ & $n>0$\\ 
${\rm Poisson}(\theta)$ & $\{0, 1, 2, \ldots\}$ & $y\log\theta - \theta - \log y!$ & $E(Y)$ & -\\
${\rm Gamma}(k, \frac{\theta}{k})$ & $\{y\in \mathbb{R}|y>0\}$ & $y\frac{-k}{\theta} - k\log\frac{\theta}{k} + \log\frac{y^{k-1}}{\Gamma(k)}$ & $E(Y)$ & $k>0$\\
${\rm IG}(\theta, \lambda)$ & $\{y\in \mathbb{R}|y>0\}$ & $y\frac{-\lambda}{2\theta^2} + \frac{\lambda}{\theta} - \frac{\lambda}{2y} + \frac{1}{2}\log\frac{\lambda}{2\pi y^3}$ & $E(Y)$ & $\lambda>0$\\
$N(\theta, \sigma^2)$ & $\mathbb{R}$ & $y\frac{\theta}{\sigma^2} - \frac{\theta^2}{2\sigma^2} - \frac{y^2}{2\sigma^2} - \frac{1}{2}\log(2\pi\sigma^2)$ & $E(Y)$ & $\sigma^2 > 0$\\
\hline
\end{tabular}}
\end{table}

\section{Lift-one, ForLion, and Round-off algorithms}
\label{sec:algorithms}

In this section, we provide the Lift-one A-optimality algorithm (Algorithm~\ref{algo:A_opt_lift_one}) for finding A-optimal allocations under a GLM with discrete factors (see Section~\ref{sec:lift_one_discrete_factor}), the ForLion A-optimality algorithm for finding A-optimal designs under a GLM with mixed factors (see Section~\ref{sec:forlion_mixed_factor}), and a rounding-off algorithm (Algorithm~\ref{algo:exact_A}) converting an approximate design to an exact design under A-criterion.

\begin{algorithm}\label{algo:A_opt_lift_one} \quad {\bf Lift-one A-optimality algorithm} (for GLMs)
\begin{itemize}
 \item[$1^\circ$] Start with arbitrary ${\mathbf w}_0=(w_1, \ldots, w_m)^T$ satisfying $0 < w_i < 1$, $i=1, \ldots, m$, and $\sum_{i=1}^m w_i=1$. We recommend two options: {\it (i)} a uniform initial design with $w_i = 1/m$; {\it (ii)} a random initial design with $w_i = U_i/\sum_{j=1}^m U_j$~, where $U_1, \ldots, U_m$ are iid from the standard exponential distribution, namely $\exp(1)$. 
 Compute $h({\mathbf w}_0)$.
 \item[$2^\circ$] Set up a random order of $i$ going through $\{1,2,\ldots,m\}$.
 \item[$3^\circ$] For each $i$, determine $h_i(z)$ as in \eqref{eq:h_i(x)} in the Supplementary Material (Section~\ref{sec:additional_theorems}).
 \item[$4^\circ$] Determine $z_*$ maximizing $h_i(z)$ with $0\leq z\leq 1$ based on Theorem~\ref{thm:max_h_i(x)} in the Supplementary Material (Section~\ref{sec:additional_theorems}). Let ${\mathbf w}_*^{(i)} = {\mathbf w}_i(z_*) \in S_m$~. Note that $h({\mathbf w}_*^{(i)}) = h_i(z_*) \geq h_i(w_i) = h({\mathbf w}_0)$.
 \item[$5^\circ$] Replace ${\mathbf w}_0$ with ${\mathbf w}_*^{(i)}$, and $h\left({\mathbf w}_0\right)$ with $h({\mathbf w}_*^{(i)})$.
 \item[$6^\circ$] Repeat $2^\circ\sim 5^\circ$ until convergence, that is, $h({\mathbf w}_0)=h({\mathbf w}_*^{(i)})$ for each $i=1,\ldots, m$.
\end{itemize}
\end{algorithm}

\medskip
Lemma~\ref{lem:lift_one_initial} in the Supplementary Material (Section~\ref{sec:lemmas_remark}) guarantees the initial allocation ${\mathbf w}_0$ chosen by Algorithm~\ref{algo:A_opt_lift_one} always satisfies $f({\mathbf w}_0)>0$ and thus $h({\mathbf w}_0)>0$, as long as there exists a ${\mathbf w} \in S_m$ satisfying $f({\mathbf w})>0$.

\medskip
\begin{algorithm}\label{algo:A_opt_Forlion} \quad {\bf ForLion A-optimality algorithm} (for GLMs)
\begin{itemize}
 \item[$0^\circ$] Set up a merging threshold $\delta > 0$ and a converging threshold $\epsilon > 0$ (e.g., $\delta = 0.1$, $\epsilon = 10^{-6}$, see Example~\ref{ex:2_paramrters_logit_continuous} in the Supplementary Material, Section~\ref{sec:examples_supp} for relevant discussions). 
 \item[$1^\circ$] Construct an initial design $\boldsymbol{\xi}_0 = \{ ({\mathbf x}^{(0)}_i, w^{(0)}_i), i = 1,\ldots, m_0\} \in \boldsymbol{\Xi}$ satisfying $f(\boldsymbol\xi_0) > 0$ and $\|{\mathbf x}^{(0)}_i - {\mathbf x}^{(0)}_j\| \geq \delta$ for any $i \neq j$. For example, one may randomly choose ${\mathbf x}^{(0)}_i$ from $\prod^d_{j=1}\{l_j, r_j\}$ or $\prod^d_{j=1} I_j$ one by one, so that the next point is at least $\delta$ away from all previous points, until some $m_0$ such that $|\sum^{m_0}_{i=1} {\mathbf F}_{{\mathbf x}^{(0)}_i}| > 0$. The initial weights $w^{(0)}_i$ are chosen either uniformly or randomly, as in Step~$1^\circ$ of Algorithm~\ref{algo:A_opt_lift_one}.
\item[$2^\circ$] Merging step: Given ${\boldsymbol\xi}_t = \{({\mathbf x}_i^{(t)}, w_i^{(t)}), i=1, \ldots, m_t\} \in \boldsymbol{\Xi}$ at the $t$th iteration, check all pairwise Euclidean distances between ${\mathbf x}_i^{(t)}$'s. If $\|{\mathbf x}_i^{(t)} - {\mathbf x}_j^{(t)}\| < \delta$ for some $i\neq j$, tentatively merge the two points into a new point $(w_i^{(t)}{\mathbf x}_i^{(t)} + w_j^{(t)}{\mathbf x}_j^{(t)})/(w_i^{(t)} + w_j^{(t)})$ with weight $w_i^{(t)} + w_j^{(t)}$, and denote the resulting design as $\boldsymbol{\xi}_{\rm merge}$~. If $|{\mathbf F}(\boldsymbol{\xi}_{\rm merge})|>0$, replace ${\boldsymbol\xi}_t$ with $\boldsymbol{\xi}_{\rm merge}$~, and $m_t$ with $m_t-1$; otherwise, keep ${\boldsymbol\xi}_t$ invariant. Repeat the procedure unless the least pairwise distance is no less than $\delta$ or any further merging leads to a degenerate ${\mathbf F}(\boldsymbol{\xi}_{\rm merge})$.
\item[$3^\circ$] Lift-one step: Given $\boldsymbol{\xi}_t \in \boldsymbol{\Xi}$ with design points ${\mathbf x}^{(t)}_1 , \ldots, {\mathbf x}^{(t)}_{m_t}$~, run Algorithm~\ref{algo:A_opt_lift_one} with converging threshold $\epsilon$ to find the converged allocation $w^\ast_1, \ldots, w^\ast_{m_t}$~. Update $\boldsymbol{\xi}_t$ with the allocations $w^\ast_i, i=1, \ldots, m_t$~.
\item[$4^\circ$] Deleting step: Update $\boldsymbol\xi_t$ by removing all ${\mathbf x}^{(t)}_i$’s with $w^{(t)}_i = 0$.
\item[$5^\circ$] New point step: Given $\boldsymbol\xi_t$~, find ${\mathbf x}^\ast = (x^\ast_1, \ldots, x^\ast_d)^T \in \mathcal{X}$ that maximizes $\varphi({\mathbf x}, \boldsymbol\xi_t) =  \nu(\boldsymbol\beta^T {\mathbf q}({\mathbf x})) \cdot {\mathbf q}({\mathbf x})^T({\mathbf X}^T_{\boldsymbol\xi_t} {\mathbf W}_{\boldsymbol\xi_t} {\mathbf X}_{\boldsymbol\xi_t})^{-2}{\mathbf q}({\mathbf x})$. Recall that the first $s$ factors are continuous. If $1 \leq s < d$ and ${\cal X} = \prod_{j=1}^s I_j \times {\cal D}$, we denote ${\mathbf x}_{(1)} = (x_1, \ldots, x_s)^T$, ${\mathbf x}_{(2)} = (x_{s+1}, \ldots, x_d)^T$, and thus ${\mathbf x} = ({\mathbf x}_{(1)}^T, {\mathbf x}_{(2)}^T)^T$. For each ${\mathbf x}_{(2)} \in {\cal D}$, we use the ``L-BFGS-B'' quasi-Newton method (see Remark~\ref{rem:partial_gradient} in the Supplementary Material, Section~\ref{sec:lemmas_remark}) to find ${\mathbf x}^\ast_{(1)} = {\rm argmax}_{{\mathbf x}_{(1)} \in \prod^s_{j=1}{[l_j, r_j]}} \ \varphi(({\mathbf x}_{(1)}^T,$ ${\mathbf x}_{(2)}^T)^T, \boldsymbol\xi_t)$, which depends on ${\mathbf x}_{(2)}$~. Then we obtain ${\mathbf x}^\ast$ by finding  ${\mathbf x}^\ast_{(2)} = {\rm argmax}_{{\mathbf x}_{(2)} \in {\cal D}}\ \varphi((({\mathbf x}_{(1)}^\ast)^T, {\mathbf x}_{(2)}^T)^T, \boldsymbol\xi_t)$. If $s = d$ and ${\cal X} = \prod_{j=1}^d [l_j, r_j]$, we find ${\mathbf x}^\ast = {\rm argmax}_{{\mathbf x} \in \mathcal{X}} \ \varphi({\mathbf x}, \boldsymbol\xi_t)$ by using the ``L-BFGS-B'' quasi-Newton method directly.
\item[$6^\circ$] If $\varphi({\mathbf x}^\ast, \boldsymbol\xi_t) \leq {\rm tr}({\mathbf F}(\boldsymbol{\xi}_t)^{-1})$, go to Step~$7^\circ$. Otherwise, we let ${\boldsymbol\xi}_{t+1} = (1-\alpha_t){\boldsymbol\xi}_t \bigcup \{({\mathbf x}^*, $ $\alpha_t)\}$, replace $t$ by $t+1$, and go back to Step~$2^\circ$, where $\alpha_t$ is given by Theorem~\ref{thm:alphat} in the Supplementary Material (Section~\ref{sec:additional_theorems}).
\item[$7^\circ$] Report $\boldsymbol\xi_t$ as an A-optimal design.
\end{itemize}
\end{algorithm}

\medskip
Given an approximate design ${\mathbf w} = (w_1, \ldots, w_m)^T\in S_m$ obtained by Algorithm~\ref{algo:A_opt_lift_one}, or $\boldsymbol{\xi} = \{({\mathbf x}_i, w_i), i=1, \ldots, m\} \in \boldsymbol{\Xi}$ obtained by Algorithm~\ref{algo:A_opt_Forlion}, we need a rounding algorithm to trim it to an exact design ${\mathbf n} = (n_1, \ldots, n_m)^T$ consisting of nonnegative integers $n_i$ given $n$, s.t., $\sum_{i=1}^m n_i=n$, which indicates $n_i$ experimental units are assigned to the experimental setting ${\mathbf x}_i$~.

An efficient rounding procedure has been proposed by \cite{pukelsheim1992}, which suggested rounding $(n-m/2)w_i$ to its next integer (see also function {\tt od\_PUK} in \cite{harman2025package}). We adopt in this paper the idea of the constrained round-off algorithm by \citeauthor{huang2025constrained}~(2025, Algorithm~2) that generates feasible exact allocations for D-optimal designs under constraints. More specifically, we recommend Algorithm~\ref{algo:exact_A} below to convert an A-optimal approximate allocation $\mathbf{w} = (w_1, \ldots, w_m)^T$ into an exact allocation $\mathbf{n} = (n_1, \ldots, n_m)^T$. Different from the efficient rounding procedure by \cite{pukelsheim1992}, Algorithm~\ref{algo:exact_A} may remove some design points whose weights are too low (see the ESD experiment revisited by \cite{lin2025ew}). For our examples when {\tt od\_PUK} is applicable, Algorithm~\ref{algo:exact_A} generates the same exact allocations (see Example~\ref{ex:RMSE}).

\begin{algorithm}\label{algo:exact_A} \quad {\bf Round-off algorithm for obtaining an exact allocation from an A-optimal approximate allocation ${\mathbf w} = (w_1, \dots, w_m)^T$} 
\begin{itemize}
 \item[$1^\circ$] First let $n_i = \lfloor nw_i \rfloor$, the largest integer no more than $nw_i$~, $i = 1, \ldots, m$, and $k = n - \sum_{i=1}^{m} n_i$ be the leftovers. Denote $I = \{ i \in \{1, \ldots, m\} \mid w_i > 0 \}$.
 \item[$2^\circ$] While $k > 0$, do
     \begin{enumerate}
        \item[2.1] for each $i \in I$, calculate $d_i = h((n-k+1)^{-1} (n_1, \ldots, n_{i-1}, n_i + 1, n_{i+1}, \ldots, n_m))$;
        
        \item[2.2] pick up any $i_* \in \arg\max_{i \in I} d_i$~;
        
        \item[2.3] let $n_{i_*} \leftarrow n_{i_*} + 1$ and $k \leftarrow k - 1$.
    \end{enumerate}
 \item[$3^\circ$] Output $\mathbf{n} = (n_1, \ldots, n_m)^T$.
\end{itemize}
\end{algorithm}

\section{Additional theorems and technical details}\label{sec:additional_theorems}

In this section, we present further discussion and technical consequences arising directly from Theorem~\ref{thm:m=p}. We also establish two key analytical foundations for our proposed algorithms: Theorem~\ref{thm:max_h_i(x)}, which underpins Algorithm~\ref{algo:A_opt_lift_one} (see Sections~\ref{sec:lift_one_discrete_factor} \& \ref{sec:algorithms}), and Theorem~\ref{thm:alphat}, which supports Algorithm~\ref{algo:A_opt_Forlion} (see Sections~\ref{sec:forlion_mixed_factor} \& \ref{sec:algorithms}).

We first consider a $2^k$ full factorial design, $m=p=2^k$, and
\[
{\mathbf X} = 
\left[
\begin{array}{rr} 
1 & -1\\ 
1 & 1
\end{array} 
\right]
\bigotimes\cdots \bigotimes
\left[
\begin{array}{rr} 
1 & -1\\ 
1 & 1
\end{array} 
\right]
\]
in the lexicographical order of factor levels \citep{cheng2016theory}, where ``$\otimes\cdots \otimes$'' stands for the Kronecker product (see, e.g., Section~11.1 in \cite{seber2008}) of $k$ matrices. In this case, $c_i$—the $i$th diagonal element of $({\mathbf X}{\mathbf X}^T)^{-1}$ required in Theorem~\ref{thm:m=p}—equals $2^{-k}$ for all $i$, leading to the following corollary.

\begin{corollary}\label{cor:Aopt_2^k}
For a $2^k$ full factorial design under a GLM, an allocation ${\mathbf w}_* = (w_1^*, \ldots, $ $w_m^*)^T$ is A-optimal if and only if $w^*_i \propto \nu_i^{-1/2}$. Furthermore, a uniform design with $w^*_i=2^{-k}$ is A-optimal if and only if $\nu_i\equiv$ constant.
\end{corollary}

\begin{example}\label{ex:linear_model_2^k_full_factorial}
{\bf $2^k$ full factorial design under a linear model}\quad     
For linear models, $Y_i \sim N(\theta, \sigma^2)$ with fixed $\sigma^2>0$ (see Table~\ref{tab:exponential_list}), and $\nu_i\equiv \sigma^{-2}>0$ (see Table~5 in the Supplementary Material of \cite{huang2025constrained}). According to Corollary~\ref{cor:Aopt_2^k}, the uniform design with $w^*_i=2^{-k}$ is the only A-optimal allocation.
\hfill{$\Box$}
\end{example}

\begin{example}\label{ex:zero_factor_effect_2^k_full_factorial}
{\bf $2^k$ full factorial design with zero factor effects}\quad     
Consider a $2^k$ full factorial design under a GLM. Note that $q_1(\cdot)\equiv 1$ in this case. Suppose all effects (both main-effects and interactions) are zero, that is, $\beta_2=\cdots=\beta_p=0$. Then $\eta_i\equiv\beta_1$~, the intercept, and $\nu_i \equiv \nu(\beta_1)$. According to Corollary~\ref{cor:Aopt_2^k}, the uniform design with $w^*_i=2^{-k}$ is the only A-optimal allocation. In other words, a uniform allocation under a GLM is a good one under A-optimality if all factor effects are nearly zeros.
\hfill{$\Box$}
\end{example}

The following corollary covers Theorem~4.1 in \cite{chen2023} as a special case.

\begin{corollary}\label{cor:m=p=2}
Consider a GLM with $m=p=2$,  
\[
{\mathbf X} = \left[\begin{array}{c} {\mathbf X}_1^T\\ {\mathbf X}_2^T\end{array}\right]
= \left[\begin{array}{cc} q_1({\mathbf x}_1) & q_2({\mathbf x}_1)\\ q_1({\mathbf x}_2) & q_2({\mathbf x}_2)\end{array}\right]\ 
\]
of full rank, and $\nu_i>0$, $i=1,2$. Then an allocation ${\mathbf w}_* = (w_1^*, w_2^*)^T$ is A-optimal if and only if $w_i^* \propto \{\nu_i [q_1({\mathbf x}_i)^2 + q_2({\mathbf x}_i)^2]\}^{-1/2}$.
\end{corollary}

Two special cases are worth noting. First, for a GLM with $m=p=2$, if $\nu_1=\nu_2$ (see, e.g., Theorem~1 in \cite{yangmin2008}), then we must have $w_i^* \propto [q_1({\mathbf x}_i)^2 + q_2({\mathbf x}_i)^2]^{-1/2}$. Second, we may consider a ``symmetric'' design in which $q_1({\mathbf x}_1) = q_1({\mathbf x}_2) = 1$ and $q_2({\mathbf x}_1) = -q_2({\mathbf x}_2)$. Then $q_1({\mathbf x}_1)^2 + q_2({\mathbf x}_1)^2 = q_1({\mathbf x}_2)^2 + q_2({\mathbf x}_2)^2$, and this common factor cancels, so that $w_i^* \propto \nu_i^{-1/2}$.

Recall from Section~\ref{sec:m>p} that, for $i=1,\ldots,m$ with $0\leq w_i<1$, the allocation ${\mathbf w}_i(z)$ associated with ${\mathbf w}$ yields $f_i(z)=f({\mathbf w}_i(z))$, $f_i^{(-j)}(z)=f_{-j}({\mathbf w}_i(z))$, and $h_i(z)=h({\mathbf w}_i(z))=f_i(z)/\sum_{j=1}^p f_i^{(-j)}(z)$. We provide below (see Theorem~\ref{thm:max_h_i(x)}) the analytic maximizer of $h_i(z)$ over $z\in[0,1]$ used in Step~$4^\circ$ of Algorithm~\ref{algo:A_opt_lift_one}.

According to Lemma~4.1 in \cite{ym2015}, if $0\leq w_i<1$, 
\begin{equation}\label{eq:f_i_a_b}
f_i(z) = a_i z(1-z)^{p-1} + b_i (1-z)^p
\end{equation}
for some constants $a_i\geq 0$ and $b_i\geq 0$. Actually, if $0 < w_i <1$, then $b_i = f_i(0)$, $a_i = [f({\mathbf w})-b_i(1-w_i)^p]/[w_i(1-w_i)^{p-1}]$; if $w_i=0$, then $b_i = f({\mathbf w})$, $a_i =f_i(1/2)\cdot 2^{p}-b_i$; and if $0\leq w_i<1$ and $f({\mathbf w})>0$, then $a_i + b_i >0$. Note that both $a_i$ and $b_i$ here are relevant to $i$, ${\mathbf X}$ and ${\mathbf W}$.

In Lemma~\ref{lem:f(w)>0_a_jb_j_>0} of the Supplementary Material (Section~\ref{sec:lemmas_remark}), we obtain similar results for $f_i^{(-j)}(z)$ with constants $a_{ij}\geq 0$ and $b_{ij}\geq 0$. If $0 < w_i <1$, we let $b_{ij}=f_i^{(-j)}(0)$,
$a_{ij} = [f_{-j}({\mathbf w})-b_{ij}(1-w_i)^{p-1}]/[w_i(1-w_i)^{p-2}]$; if $w_i=0$, then $b_{ij}=f_{-j}({\mathbf w})$, $a_{ij} = f_i^{(-j)}(1/2)\cdot 2^{p-1}-b_{ij}$~, where $f_{-j}({\mathbf w}) = |{\mathbf X}_{-j}^T{\mathbf W}{\mathbf X}_{-j}|$.

Due to \eqref{eq:h_i(x)_1}, \eqref{eq:f_i_a_b}, and Lemma~\ref{lem:f(w)>0_a_jb_j_>0}, it can be verified that 
\begin{equation}\label{eq:h_i(x)}
    h_i(z) = \frac{(b_i-a_i)z^2+(a_i-2b_i)z+b_i}{(A_i - B_i)z + B_i}\ ,
\end{equation}
where $A_i = \sum_{j=1}^p a_{ij}$ and $B_i = \sum_{j=1}^p b_{ij}$~. According to Lemmas~\ref{lem:impossible_case_(4)} and \ref{lem:impossible_case_(6)} in the Supplementary Material (Section~\ref{sec:proofs}), $A_i=0$ implies $a_i=0$, and $B_i=0$ implies $b_i=0$.

\begin{theorem}\label{thm:max_h_i(x)}
Suppose $m > p\geq 2$, $0\leq w_l<1$ for each $l$, and $f({\mathbf w})>0$. Then the solutions for $\max_{0\leq z\leq 1} h_i(z)$ must belong to one of the four cases: (1) If $A_i\neq B_i$, $A_i>0$, $B_i>0$, $a_i>b_i$, $a_iB_i>b_iA_i$, and $b_iA_i < (a_i-b_i)B_i$, then 
\[
\max_{0\leq z\leq 1} h_i(z) = \left(\frac{\sqrt{A_i(a_i-b_i)} - \sqrt{a_iB_i-b_iA_i}}{A_i-B_i}\right)^2
\]
is attained uniquely at $z_* = \frac{t_*-B_i}{A_i-B_i} \in (0,1)$, where $t_*=\sqrt{\frac{A_i(a_iB_i-b_iA_i)}{a_i-b_i}} \in (\min\{A_i,B_i\}, $ $\max\{A_i,$ $ B_i\})$; (2) If $A_i=B_i$ and $a_i>2b_i$, then $\max_{0\leq z\leq 1} h_i(z) = \frac{a_i^2}{4(a_i-b_i)B_i} > 0$ is attained uniquely at $z_* = \frac{a_i-2b_i}{2a_i-2b_i} \in (0,1)$; (3) If $A_i\neq B_i$, $B_i=0$, and $b_i=0$, then $\max_{0\leq z\leq 1} h_i(z) = \frac{a_i}{A_i} > 0$ is attained uniquely at $z_* = 0$; (4) For all other cases, $\max_{0\leq z\leq 1} h_i(z) = \frac{b_i}{B_i}>0$ at the unique $z_*=0$.
\end{theorem} 

The next theorem provides the analytic step size $\alpha_t$ used in Step~$6^\circ$ of the ForLion A-optimality algorithm (Algorithm~\ref{algo:A_opt_Forlion}, see Sections~\ref{sec:forlion_mixed_factor} \& \ref{sec:algorithms}).

\begin{theorem}\label{thm:alphat}
Given ${\boldsymbol\xi}_t = \{({\mathbf x}_i^{(t)}, w_i^{(t)}), i=1, \ldots, m_t\}$ and ${\mathbf x}^* \in {\mathcal X}$, if we consider ${\boldsymbol\xi}_{t+1}$ in the form of $(1-\alpha) {\boldsymbol\xi}_t \bigcup \{({\mathbf x}^*, \alpha)\}$ with $\alpha \in [0,1]$, then $\alpha_t = {\rm argmax}_{\alpha \in [0,1]} h({\boldsymbol\xi}_{t+1})$ must belong to one of the three cases: (1) if $A_t\neq B_t$~, $A_t>0$, $B_t>0$, $a_t>b_t$~, $a_tB_t>b_tA_t$~, and $b_tA_t < (a_t-b_t)B_t$~, then 
    \[
    \alpha_t = \frac{\sqrt{\frac{A_t(a_tB_t-b_tA_t)}{a_t-b_t}}-B_t}{A_t-B_t} \in (0, 1)\ ;
    \]
(2) if $A_t=B_t$ and $a_t>2b_t$~, then $\alpha_t = (a_t-2b_t)/(2a_t-2b_t) \in (0,1)$~; (3) for all other cases, $\alpha_t=0$, where $A_t = \sum_{j=1}^p a^{(t)}_j$, $B_t = \sum_{j=1}^p b^{(t)}_j$, $a_t = f_{m_t+1}(1/2) \cdot 2^p - b_t$~, $b_t = f(\boldsymbol\xi_t)$,  
$a^{(t)}_j = f_{m_t+1}^{(-j)}(1/2)\cdot 2^{p-1}-b^{(t)}_j$~, $b^{(t)}_j = f_{-j}(\boldsymbol\xi_t) = |{\mathbf X}_{\boldsymbol\xi_t, -j}^T {\mathbf W}_{\boldsymbol\xi_t} {\mathbf X}_{\boldsymbol\xi_t, -j}|$, $f_{m_t+1}(1/2) = f(\{({\mathbf x}^{(t)}_1, w^{(t)}_1/2), \ldots, ({\mathbf x}^{(t)}_{m_t}, w^{(t)}_{m_t}/2), ({\mathbf x}^*, 1/2)\})$, $f^{(-j)}_{m_t+1}(1/2) = f_{-j}(\{({\mathbf x}^{(t)}_1, w^{(t)}_1/2), \ldots, \\({\mathbf x}^{(t)}_{m_t}, w^{(t)}_{m_t}/2), ({\mathbf x}^*, 1/2)\})$, and ${\mathbf X}_{\boldsymbol\xi_t, -j}$ is the submatrix of ${\mathbf X}_{\boldsymbol\xi_t}$ after removing its $j$th column. 
\end{theorem}

\section{Lemmas and remarks}\label{sec:lemmas_remark}

In this section, we provide Lemmas~\ref{lem:aobjective}, \ref{lem:f_i(w)_rank_X}, \ref{lem:f_-j(w)_>0}, \& \ref{lem:f(w)>0_a_jb_j_>0} for Section~\ref{sec:m>p}, Lemma~\ref{lem:lift_one_initial} for Section~\ref{sec:lift_one_discrete_factor}, Remark~\ref{rem:partial_gradient} for Step~$5^\circ$ of Algorithm~\ref{algo:A_opt_Forlion} in Section~\ref{sec:algorithms}, and Remark~\ref{rem:cross_entropy_loss} for Example~\ref{ex:RMSE} in Section~\ref{sec:examples_supp}. 

\begin{lemma}\label{lem:aobjective}
We let ${\mathbf X}_{-j}$ be the $m\times (p-1)$ submatrix of ${\mathbf X}$ after removing the $j$th column. If $f({\mathbf w}) > 0$, then 
\begin{equation}\label{eq:tr(xwx)}
h({\mathbf w}) 
= \frac{|{\mathbf X}^T{\mathbf W}{\mathbf X}|}{\sum_{j=1}^p |{\mathbf X}_{-j}^T{\mathbf W}{\mathbf X}_{-j}|} = \frac{f({\mathbf w})}{\sum_{j=1}^p f_{-j}({\mathbf w})}\ ,
\end{equation}
where $f_{-j}({\mathbf w}) = |{\mathbf X}_{-j}^T{\mathbf W}{\mathbf X}_{-j}|$ is an order-$(p-1)$ homogeneous polynomial of $w_1, \ldots, w_m$~.
\end{lemma}

\begin{lemma}\label{lem:f_i(w)_rank_X}
Suppose $0<w_i<1$ for all $i=1, \ldots, m$. Then $f({\mathbf w})>0$ if and only if ${\rm rank}({\mathbf X})=p$. In this case, we have $f_{-j}({\mathbf w})>0$ for all $j=1, \ldots, p$.    
\end{lemma}

\begin{lemma}\label{lem:f_-j(w)_>0}
Suppose $0\leq w_i<1$ for all $i=1, \ldots, m$. If $f({\mathbf w})>0$, then $f_{-j}({\mathbf w})>0$ for all $j=1, \ldots, p$.  
\end{lemma}

\begin{lemma}\label{lem:f(w)>0_a_jb_j_>0}
Suppose $0\leq w_i<1$. Then 
\begin{equation}\label{eq:f_i_-j_a_j_b_j}
f_i^{(-j)}(z) = a_{ij} z(1-z)^{p-2} + b_{ij}(1-z)^{p-1}
\end{equation}
for some constants $a_{ij}\geq 0$ and $b_{ij}\geq 0$. 
If $f({\mathbf w})>0$, then $f_i^{(-j)}(z) > 0$ for all $z\in (0,1)$, which implies $a_{ij}+b_{ij}>0$. If $0 < w_i <1$, then $b_{ij}=f_i^{(-j)}(0)$,
$a_{ij} = [f_{-j}({\mathbf w})-b_{ij}(1-w_i)^{p-1}]/[w_i(1-w_i)^{p-2}]$; if $w_i=0$, then $b_{ij}=f_{-j}({\mathbf w})$, $a_{ij} = f_i^{(-j)}(1/2)\cdot 2^{p-1}-b_{ij}$~.
\end{lemma}

\begin{lemma}\label{lem:lift_one_initial}
Suppose there exists a ${\mathbf w}\in S_m$ satisfying $f({\mathbf w})>0$. Then for any ${\mathbf w}_0 = (w_1', \ldots, w_m')^T\in S_m$ satisfying $0<w_i'<1$ for each $i$, $f({\mathbf w}_0)>0$.    
\end{lemma}

\medskip
\begin{remark}\label{rem:partial_gradient} {\bf First-order derivative of sensitivity function}\quad     
Theorem~\ref{thm:max_h_i(x)} for Step~$4^\circ$ of Algorithm~\ref{algo:A_opt_lift_one} (and thus for Step~$3^\circ$ of Algorithm~\ref{algo:A_opt_Forlion}) and Theorem~\ref{thm:alphat} for Step~$6^\circ$ of Algorithm~\ref{algo:A_opt_Forlion} gain notable advantages on computational simplicity by offering analytic solutions. To fully leverage these benefits, we derive an explicit formula for the first-order derivative of the sensitivity function, which is critical for applying the ``L-BFGS-B'' quasi-Newton method \citep{byrd1995} in Step~$5^\circ$. To simplify the notation, we denote ${\mathbf A} = ({\mathbf X}^T_{\boldsymbol{\xi}_t} {\mathbf W}_{\boldsymbol{\xi}_t} {\mathbf X}_{\boldsymbol{\xi}_t})^{-2}$, which is a known positive definite matrix given $\boldsymbol{\xi_t}$ and $\boldsymbol\beta$. Then the first-order derivative is
\begin{eqnarray*}
\frac{\partial \varphi({\mathbf x}, \boldsymbol{\xi_t})}{\partial {\mathbf x}_{(1)}} &=& \nu'\left(\boldsymbol\beta^T {\mathbf q}({\mathbf x})\right)\cdot{\mathbf q}({\mathbf x})^T{\mathbf A}{\mathbf q}({\mathbf x})\cdot\left[\frac{\partial {\mathbf q}({\mathbf x})}{\partial {\mathbf x}^T_{(1)}}\right]^T\boldsymbol\beta \\
&+& 2\cdot\nu\left(\boldsymbol\beta^T {\mathbf q}({\mathbf x})\right)\cdot\left[\frac{\partial {\mathbf q}({\mathbf x})}{\partial {\mathbf x}^T_{(1)}}\right]^T{\mathbf A}{\mathbf q}({\mathbf x})\ \in \ \mathbb{R}^{s}\ ,
\end{eqnarray*}
where $\partial {\mathbf q}({\mathbf x})/\partial {\mathbf x}^T_{(1)} \in \mathbb{R}^{p\times s}$, ${\mathbf x} = ({\mathbf x}_{(1)}^T, {\mathbf x}_{(2)}^T)^T$ if $1\leq s<d$, and ${\mathbf x} = {\mathbf x}_{(1)}$ if $s=d$.
\hfill{$\Box$}
\end{remark}

\medskip
\begin{remark}\label{rem:cross_entropy_loss} {\bf Cross-entropy loss for evaluating prediction error}\quad    
To compare the prediction accuracy of fitted models based on different samplers in Example~\ref{ex:RMSE} (see Section~\ref{sec:examples_supp} in the Supplementary Material), we denote the whole dataset by ${\cal C} = \{({\mathbf x}_i, y_{ij})\mid i=1, \ldots, 6; j=1, \ldots, N_i\}$. For each sampling procedure, we let $\Lambda \subset {\cal C}$ be the collection of sampled data with sample size $|\Lambda|=n$. We obtain $\hat{\boldsymbol{\beta}}$ from $\Lambda$, i.e., the training data, and use the fitted model to predict the responses of the remaining $N-n$ volunteers in ${\cal C}\setminus \Lambda$, i.e., the testing data. To quantify the prediction errors, we adopt the cross-entropy loss \citep{hastie2009elements}:
\begin{eqnarray*}
{\rm CE}(\Lambda) &=& -\frac{1}{N-n}\sum_{({\mathbf x}, y)\in {\cal C}\setminus \Lambda} \log \hat{p}_y({\mathbf x})\\
&=& -\frac{1}{N-n} \sum_{i=1}^6 \left[\#\{j \mid y_{ij}=1, ({\mathbf x}_i, y_{ij}) \in {\cal C}\setminus \Lambda\} \cdot \log\hat{p}_i\right.\\
& & +\left. \#\{j \mid y_{ij}=0, ({\mathbf x}_i, y_{ij}) \in {\cal C}\setminus \Lambda\}\cdot \log(1-\hat{p}_i)\right]\ ,
\end{eqnarray*}
where $\hat{p}_y({\mathbf x}) = \hat{P}(Y=y\mid {\mathbf x})$ and $\hat{p}_i = \hat{P}(Y=1\mid {\mathbf x}_i)$ are  estimated probabilities based on the fitted model.
Since the simulation is repeated independently for $100$ times, we obtain $100$ CE values for each sampler. The mean CE values over 100 simulations are reported in Example~\ref{ex:RMSE}.
\hfill{$\Box$}
\end{remark}

\section{Proofs}\label{sec:proofs}

\medskip\noindent
{\bf Proof of Theorem~\ref{thm:m=p}:}
In this case, we must have $w_i>0$ for each $i$ to make ${\mathbf X}^T{\mathbf W}{\mathbf X}$ invertible. Then
\begin{eqnarray*}
h({\mathbf w}) &=& \left[{\rm tr}\left(({\mathbf X}^T{\mathbf W}{\mathbf X})^{-1}\right)\right]^{-1} = \left[{\rm tr}\left({\mathbf X}^{-1}{\mathbf W}^{-1}({\mathbf X}^T)^{-1}\right)\right]^{-1}\\
&=& \left[{\rm tr}\left({\mathbf W}^{-1}({\mathbf X}^T)^{-1}{\mathbf X}^{-1}\right)\right]^{-1} = \left[{\rm tr}\left({\mathbf W}^{-1}({\mathbf X}{\mathbf X}^T)^{-1}\right)\right]^{-1}\\
&=& \left(\sum_{i=1}^m \frac{c_i}{w_i\nu_i}\right)^{-1}\ .
\end{eqnarray*}
Since ${\rm rank}({\mathbf X}) = p$, then $({\mathbf X}{\mathbf X}^T)^{-1}$ is positive definite and thus $c_i>0$ for all $i$. It can be verified that ${\mathbf w}_* = (w_1^*, \ldots, w_m^*)^T$ maximizes $h({\mathbf w})$ if and only if 
\[
\frac{c_1}{(w_1^*)^2\nu_1} = \cdots = \frac{c_m}{(w_m^*)^2\nu_m}\ ,
\]
or equivalently, $w_i^* \propto \sqrt{c_i/\nu_i}$~, which leads to the conclusion.
\hfill{$\Box$}

\medskip\noindent
{\bf Proof of Lemma~\ref{lem:aobjective}:} Rewrite the design matrix ${\mathbf X}=[\boldsymbol\alpha_1, \ldots, \boldsymbol\alpha_p]$, where $\boldsymbol\alpha_i$ is the $i$th column of ${\mathbf X}$, and an $m$-dimensional vector. Then ${\mathbf X}^T{\mathbf W}{\mathbf X} = (a_{ij})_{p\times p}$ with entry $a_{ij} = \boldsymbol\alpha_i^T {\mathbf W} \boldsymbol\alpha_j$~. Based on the Cramer's rule in linear algebra, $({\mathbf X}^T{\mathbf W}{\mathbf X})^{-1} = |{\mathbf X}^T{\mathbf W}{\mathbf X}|^{-1}{\mathbf A}^*$ where ${\mathbf A}^*=(a^*_{ij})_{p\times p}$ is the adjugate matrix of ${\mathbf X}^T{\mathbf W}{\mathbf X}$ with $a^*_{ij} = (-1)^{j+i} |{\mathbf X}_{-j}^T {\mathbf W} {\mathbf X}_{-i}|$. Therefore the $j$th diagonal entry of $({\mathbf X}^T{\mathbf W}{\mathbf X})^{-1}$ is $|{\mathbf X}_{-j}^T{\mathbf W}{\mathbf X}_{-j}|/|{\mathbf X}^T{\mathbf W}{\mathbf X}|$ and the trace of $({\mathbf X}^T{\mathbf W}{\mathbf X})^{-1}$ is $\sum_{j=1}^p |{\mathbf X}_{-j}^T{\mathbf W}{\mathbf X}_{-j}|/|{\mathbf X}^T{\mathbf W}{\mathbf X}|$, which leads to the conclusion.

As for $f_{-j}({\mathbf w}) = |{\mathbf X}_{-j}^T{\mathbf W}{\mathbf X}_{-j}|$, following \eqref{eq:|xwx|}, 
\[
f_{-j}({\mathbf w}) =\sum_{1\leq i_1<\cdots<i_{p-1}\leq m} |{\mathbf X}_{-j}[i_1,\ldots,i_{p-1}]|^2 \cdot w_{i_1}\nu_{i_1}\cdots w_{i_{p-1}}\nu_{i_{p-1}}
\]
is an order-$(p-1)$ homogeneous polynomial of $w_1, \ldots, w_m$~.
\hfill{$\Box$}

\medskip\noindent
{\bf Proof of Lemma~\ref{lem:f_i(w)_rank_X}:} 
According to Lemma~3.1 in \cite{ym2015}, we have $f({\mathbf w})=|{\mathbf X}^T{\mathbf W}{\mathbf X}|=\sum_{1\leq i_1<\cdots<i_p\leq m} |{\mathbf X}[i_1,\ldots,i_p]|^2 \cdot w_{i_1}\nu_{i_1}\cdots w_{i_p}\nu_{i_p}$~,
where ${\mathbf X}[i_1,\ldots,i_p]$ is the $p \times p$ submatrix consisting of the $i_1\mbox{th}$, $\ldots$, $i_p\mbox{th}$ rows of the design matrix ${\mathbf X}$, $\nu_i > 0$, and $w_i\geq 0$ for each $i$. Since we further assume $w_i>0$ for each $i$, then $f(w)>0$ if and only if there exists a set of $1\leq i_1<\cdots<i_p\leq m$ satisfying $|{\mathbf X}[i_1,\ldots,i_p]|^2 > 0$. Since ${\mathbf X}$ is an $m\times p$ matrix with $p\leq m$, such a condition is equivalent to that ${\mathbf X}$ is of full column rank $p$.

Similarly, $f_{-j}({\mathbf w})>0$ if and only if ${\mathbf X}_{-j}$ is of full column rank $p-1$, which is guaranteed by ${\rm rank}({\mathbf X})=p$.
\hfill{$\Box$}

\medskip\noindent
{\bf Proof of Lemma~\ref{lem:f_-j(w)_>0}:} 
Because $w_i$ could be zero, the conclusion of Lemma~\ref{lem:f_i(w)_rank_X} does not apply here. 

Since $w_i\geq 0$ and $\nu_i>0$ for all $i$, ${\mathbf X}^T{\mathbf W}{\mathbf X}$ is a $p\times p$ positive semi-definite matrix. If $f({\mathbf w}) = |{\mathbf X}^T{\mathbf W}{\mathbf X}|>0$, then ${\mathbf X}^T{\mathbf W}{\mathbf X}$ is positive definite, which implies that $({\mathbf X}^T{\mathbf W}{\mathbf X})^{-1}$ is positive definite as well.

According to the proof of Lemma~\ref{lem:aobjective}, we know that the $j$th diagonal entry of $({\mathbf X}^T{\mathbf W}{\mathbf X})^{-1}$ is $|{\mathbf X}_{-j}^T{\mathbf W}{\mathbf X}_{-j}|/|{\mathbf X}^T{\mathbf W}{\mathbf X}|$, which must be positive. Therefore, $f_{-j}({\mathbf w}) = |{\mathbf X}_{-j}^T{\mathbf W}{\mathbf X}_{-j}| > 0$ for each $j$. 
\hfill{$\Box$}

\medskip\noindent
{\bf Proof of Lemma~\ref{lem:f(w)>0_a_jb_j_>0}:} 
According to Lemma~4.1 in \cite{ym2015}, given 
\[
f_i(z) = f\left(\frac{1-z}{1-w_i}w_1,\ldots,\frac{1-z}{1-w_i}w_{i-1},z, \frac{1-z}{1-w_i}w_{i+1},\ldots, \frac{1-z}{1-w_i}w_{m}\right)
\]
with $f({\mathbf w}) = |{\mathbf X}^T{\mathbf W}{\mathbf X}|$ as an order-$p$ homogeneous polynomial of $z$, we have $f_i(z) = a_i z(1-z)^{p-1} + b_i(1-z)^p$ for some constants $a_i\geq 0$ and $b_i\geq 0$. In our case, 
\[
f_i^{(-j)}(z) = f_{-j}\left(\frac{1-z}{1-w_i}w_1,\ldots,\frac{1-z}{1-w_i}w_{i-1},z, \frac{1-z}{1-w_i}w_{i+1},\ldots, \frac{1-z}{1-w_i}w_{m}\right)
\]
with $f_{-j}({\mathbf w}) = |{\mathbf X}_{-j}^T{\mathbf W}{\mathbf X}_{-j}|$ as an order-$(p-1)$ homogeneous polynomial of $z$. In other words, ${\mathbf X}$ for $f_i(z)$ is replaced with ${\mathbf X}_{-j}$ for $f_i^{(-j)}(z)$, which is the only change.

As a direct conclusion, $f_i^{(-j)}(z) = a_{ij} z(1-z)^{p-2} + b_{ij}(1-z)^{p-1}$ for some constants $a_{ij}\geq 0$ and $b_{ij}\geq 0$. In particular, letting $z=w_i$~, $0$, or $\frac{1}{2}$~, we obtain $f_{-j}({\mathbf w}) = f_i^{(-j)}(w_i) = a_{ij} w_i(1-w_i)^{p-2} + b_{ij}(1-w_i)^{p-1}$, $f_i^{(-j)}(0)=b_{ij}$ and $f_i^{(-j)}\left(\frac{1}{2}\right)=(a_{ij}+b_{ij})(\frac{1}{2})^{p-1}$, respectively. Solving for $a_{ij}$ and $b_{ij}$ for  two cases, $w_i=0$ and $0<w_i<1$, respectively, we obtain the desired results.
\hfill{$\Box$}

\medskip
\noindent
{\bf Proof of Theorem~\ref{thm:max_h_i(x)}:}  
Under the conditions $m\geq p\geq 2$, $0\leq w_i<1$ for each $i$, and $f({\mathbf w})>0$, we denote the index set $I_{\bf w} = \{i\in \{1, \ldots, m\}\mid w_i>0\}$. Then $|I_{\bf w}| = \#\{i\mid w_i>0\} \geq p$,
\begin{eqnarray*}
    f({\mathbf w}) &=& \sum_{\{i_1, \ldots, i_p\} \subseteq I_{\bf w}} |{\mathbf X}[i_1,\ldots,i_p]|^2 \cdot w_{i_1}\nu_{i_1}\cdots w_{i_p}\nu_{i_p}\ ,\\
    f_{-j}({\mathbf w}) &=& \sum_{\{i_1, \ldots, i_{p-1}\} \subseteq I_{\bf w}} |{\mathbf X}_{-j}[i_1,\ldots,i_{p-1}]|^2 \cdot w_{i_1}\nu_{i_1}\cdots w_{i_{p-1}}\nu_{i_{p-1}}\ .
\end{eqnarray*}
Fixing $i\in \{1, \ldots, m\}$, it can be verified that
\begin{equation}\label{eq:a}
    a_i = \frac{\nu_i}{(1-w_i)^{p-1}} \sum_{\{i_1, \ldots, i_{p-1}\} \subseteq I_{\bf w}\setminus\{i\}} |{\mathbf X}[i_1,\ldots,i_{p-1},i]|^2 w_{i_1}\nu_{i_1}\cdots w_{i_{p-1}}\nu_{i_{p-1}}\ ,
\end{equation}
\begin{equation}\label{eq:b}
b_i = \frac{1}{(1-w_i)^p} \sum_{\{i_1, \ldots, i_p\} \subseteq I_{\bf w}\setminus\{i\}} |{\mathbf X}[i_1,\ldots,i_p]|^2 w_{i_1}\nu_{i_1}\cdots w_{i_p}\nu_{i_p}\ .
\end{equation}
Note that if $w_i=0$, then $i\notin I_{\bf w}$ and $I_{\bf w}\setminus\{i\} = I_{\bf w}$~.

Similarly, given $j\in \{1, \ldots, p\}$, it can be verified that
\begin{equation}\label{eq:a_j}
a_{ij} = \frac{\nu_i}{(1-w_i)^{p-2}} \sum_{\{i_1, \ldots, i_{p-2}\} \subseteq I_{\bf w}\setminus\{i\}} |{\mathbf X}_{-j}[i_1,\ldots,i_{p-2},i]|^2 w_{i_1}\nu_{i_1}\cdots w_{i_{p-2}}\nu_{i_{p-2}}
\end{equation}
if $p\geq 3$; $a_{ij} = \nu_i |{\mathbf X}_{-j}[i]|^2$ if $p=2$; and
\begin{equation}\label{eq:b_j}
    b_{ij} = \frac{1}{(1-w_i)^{p-1}} \sum_{\{i_1, \ldots, i_{p-1}\} \subseteq I_{\bf w}\setminus\{i\}} |{\mathbf X}_{-j}[i_1,\ldots,i_{p-1}]|^2 w_{i_1}\nu_{i_1}\cdots w_{i_{p-1}}\nu_{i_{p-1}}\ .
\end{equation}

We need to show two more lemmas to eliminate some cases for Theorem~\ref{thm:max_h_i(x)}.

\begin{lemma}\label{lem:impossible_case_(4)}
If $A_i=0$, then $a_i=0$.
\end{lemma}

\medskip\noindent
{\bf Proof of Lemma~\ref{lem:impossible_case_(4)}:}
Given that $A_i = \sum_{j=1}^p a_{ij} = 0$ and $a_{ij}\geq 0$ due to Lemma~\ref{lem:f(w)>0_a_jb_j_>0}, we must have $a_{ij}=0$, for all $j=1,\ldots,p$. 

If $p\geq 3$, according to \eqref{eq:a_j}, we must have $|{\mathbf X}_{-j}[i_1,\ldots,i_{p-2},i]| = 0$ for all $\{i_1, \ldots, i_{p-2}\}$ $\subseteq I_{\bf w}\setminus\{i\}$ and all $j=1, \ldots, p$. If ${\rm rank}({\mathbf X}[i_1,\ldots,i_{p-2},i]) = p-1$, then we must have $|{\mathbf X}_{-j}[i_1,\ldots,i_{p-2},i]| \neq 0$ for some $j=1, \ldots, p$. Contradiction! Therefore, ${\rm rank}({\mathbf X}[i_1,\ldots,$ $i_{p-2},i]) < p-1$, for all $\{i_1, \ldots, i_{p-2}\} \subseteq I_{\bf w}\setminus\{i\}$. Then ${\rm rank}({\mathbf X}[i_1,\ldots,i_{p-1},i]) < p$, for all $\{i_1, \ldots, i_{p-1}\} \subseteq I_{\bf w}\setminus\{i\}$, which implies $a_i=0$ according to \eqref{eq:a}.

If $p=2$, $a_{ij} = \nu_i |{\mathbf X}_{-j}[i]|^2 =0$, for $j=1,2$, implies $x_{i1}=x_{i2}=0$, or simply ${\mathbf x}_i = (0,0)^T$. Then ${\rm rank}({\mathbf X}[i_1,i]) < 2$, for all $\{i_1\} \subseteq I_{\bf w}\setminus\{i\}$, which implies $a_i=0$ according to \eqref{eq:a} with $p=2$.
\hfill{$\Box$}

\begin{lemma}\label{lem:impossible_case_(6)}
If $B_i=0$, then $b_i=0$.
\end{lemma}

\medskip\noindent
{\bf Proof of Lemma~\ref{lem:impossible_case_(6)}:}
Given that $B_i = \sum_{j=1}^p b_{ij} = 0$ and $b_{ij}\geq 0$ due to Lemma~\ref{lem:f(w)>0_a_jb_j_>0}, we must have $b_{ij}=0$, for all $j=1,\ldots,p$. 

According to \eqref{eq:b_j}, we must have $|{\mathbf X}_{-j}[i_1,\ldots,i_{p-1}]| = 0$ for all $\{i_1, \ldots,$ $i_{p-1}\} \subseteq I_{\bf w}\setminus\{i\}$ and all $j=1, \ldots, p$. If ${\rm rank}({\mathbf X}[i_1,\ldots,i_{p-1}]) = p-1$, then we must have $|{\mathbf X}_{-j}[i_1,\ldots,i_{p-1}]| \neq 0$ for some $j=1, \ldots, p$. Contradiction! Therefore, ${\rm rank}({\mathbf X}[i_1,\ldots,$ $i_{p-1}]) < p-1$, for all $\{i_1, \ldots, i_{p-2}\} \subseteq I_{\bf w}\setminus\{i\}$. Then ${\rm rank}({\mathbf X}[i_1,\ldots,i_p]) < p$, for all $\{i_1, \ldots, i_p\} \subseteq I_{\bf w}\setminus\{i\}$, which implies $b_i=0$ according to \eqref{eq:b}.
\hfill{$\Box$}

\medskip
\noindent
Now we are ready to continue the proof of   Theorem~\ref{thm:max_h_i(x)}: 

\medskip
\noindent
{\it Case (i):} $A_i =B_i$ and $a_i=b_i$~. According to Lemma~\ref{lem:f(w)>0_a_jb_j_>0}, we must have $a_i=b_i>0$ and $A_i =B_i > 0$. In this case, $h_i(z)=\frac{b_i(1-z)}{B_i}$, which is a monotonic function. Then $\max_{0\leq z\leq 1} h_i(z) = \frac{b_i}{B_i} > 0$ attained uniquely at $z_*=0$.

\medskip
\noindent
{\it Case (ii):} $A_i =B_i$ but $a_i\neq b_i$~. According to Lemma~\ref{lem:f(w)>0_a_jb_j_>0}, we still have $A_i =B_i > 0$. In this case,
\begin{eqnarray*}
    h_i(z) &=& \frac{b_i-a_i}{B_i} \left(z-\frac{b_i}{b_i-a_i}\right) (z-1)\\
    &=& \frac{b_i-a_i}{B_i} \left(z - \frac{2b_i-a_i}{2b_i-2a_i}\right)^2 + \frac{a_i^2}{4(a_i-b_i) B_i}
\end{eqnarray*}
is a quadratic function. It can be verified that if $a_i>2b_i$~, then 
\[
\max_{z\in [0,1]} h_i(z) = \frac{a_i^2}{4(a_i-b_i) B_i} > 0\ ,
\] 
which is attained at the unique $z_* = \frac{a_i-2b_i}{2a_i-2b_i} \in (0, 1)$; otherwise, 
\[
\max_{z\in [0,1]} h_i(z) = \frac{b_i}{B_i} > 0\ ,
\]
which is attained at the unique $z_*=0$, the same as in Case (i).

\medskip
\noindent
{\it Case (iii):} $A_i \neq B_i$ and $A_i = 0$. In this case, we must have $B_i > 0$ and then 
\[
h_i(z) = \frac{1}{B_i}\left[b_i+(a_i-b_i)z\right]\ .
\]
According to Lemma~\ref{lem:impossible_case_(4)}, we must have $a_i=0$ and thus $b_i>0$. 
Since $a_i<b_i$~, then $\max_{z\in [0,1]} h_i(z) = \frac{b_i}{B_i} > 0$ attained at the unique $z_*=0$.

\medskip
\noindent
{\it Case (iv):} $A_i \neq B_i$ and $B_i = 0$. In this case, we must have $A_i > 0$ and then 
\[
h_i(z) = \frac{1}{A_i}\left[(b_i-a_i)z + \frac{b_i}{z} + a_i - 2b_i\right]\ .
\]
According to Lemma~\ref{lem:impossible_case_(6)}, we must have $b_i=0$ and thus $a_i>0$.
Then $h_i(z) = \frac{a_i}{A_i}(1-z)$, and $\max_{z\in [0,1]} h_i(z) = \frac{a_i}{A_i} > 0$ attained at the unique $z_*=0$.

\medskip
\noindent
{\it Case (v):} $A_i \neq B_i$, $A_i>0$ and $B_i>0$. In this case,
\begin{eqnarray*}
\left(A_i - B_i\right)^2\cdot   h_i(z) &=& (b_i-a_i) \cdot t + A_i\left(b_iA_i-a_iB_i\right)\cdot \frac{1}{t}\\
  & & +\ (a_i-2b_i)A_i + a_iB_i\\
  &\triangleq& s(t)  \ ,
\end{eqnarray*}
where $t=\left(A_i - B_i\right)z + B_i$ takes values between $A_i$ and $B_i$~, which is always positive.

If $(b_i-a_i)(b_iA_i-a_iB_i)\le0$, that is, the coefficients of $t$ and $1/t$ have different signs, then $s(t)$ is a monotone function of $t\geq 0$, which implies that $h_i(z)$ is a monotone function of $z\in [0,1]$. Since $b_i=0$ implies $a_i>0$ and $(b_i-a_i)(b_iA_i-a_iB_i) = a_i^2B_i>0$, we must have $b_i>0$. Then $h_i(0) = b_i/B_i > 0 = h_i(1)$, and $\max_{z\in [0,1]} h_i(z) = \frac{b_i}{B_i} > 0$ attained at the unique $z_*=0$. 

If $(b_i-a_i)(b_iA_i-a_iB_i)>0$, then the coefficients of $t$ and $1/t$ have the same sign. If $b_i>a_i$ and $b_iA_i>a_iB_i$, it can be verified that $s(t)$ strictly decreases before $t_* = \left[\frac{A_i(b_iA_i-a_iB_i)}{b_i-a_i}\right]^{1/2}$ and strictly increases after $t_*$~. In this case, we always have $\max_{z\in [0,1]} h_i(z) = \frac{b_i}{B_i} > 0$ attained at the unique $z_*=0$. If $b_i<a_i$ and $b_iA_i<a_iB_i$, it can be verified that $s(t)$ strictly increases before $t_*$ and strictly decreases after $t_*$~. We can also verify that if $b_i<a_i$ and $(a_i-b_i)B_i\leq b_iA_i < a_iB_i$~, which implies $b_i>0$, we have $\max_{z\in [0,1]} h_i(z) = \frac{b_i}{B_i} > 0$ attained at the unique $z_*=0$; if $b_i<a_i$ and $b_iA_i < (a_i-b_i)B_i$~, we have 
\[
\max_{z\in [0,1]} h_i(z) = \left(\frac{\sqrt{A_i(a_i-b_i)} - \sqrt{a_iB_i-b_iA_i}}{A_i-B_i}\right)^2
\] 
attained at the unique $z_*=\frac{t_*-B_i}{A_i-B_i} \in (0,1)$.

Summarizing the results, we obtain the four cases in Theorem~\ref{thm:max_h_i(x)}.
\hfill{$\Box$}

\medskip\noindent
{\bf Proof of Theorem~\ref{thm:for_lift_one}:}\quad
First of all, if $S_m^+ = \emptyset$, that is, $f({\mathbf w})=0$ and thus $h({\mathbf w})=0$ for all ${\mathbf w}\in S_m$~, then all ${\mathbf w}\in S_m$ are A-optimal, and the set of A-optimal allocations is $S_m$ itself and thus convex.

Now we assume that $S_m^+ \neq \emptyset$, that is, there exists a ${\mathbf w}\in S_m$~, such that $f({\mathbf w})>0$. We claim that Assumptions (A1), (A2), (B1) $\sim$ (B4) in \cite{fedorov2014} are all satisfied and thus their Theorem~2.2 can be applied to our case.

Actually, in our case ${\cal X} = \{{\mathbf x}_1, \ldots, {\mathbf x}_m\}$ is a collection of finite points in $\mathbb{R}^d$ and must be compact. Then Assumption~(A1) in \cite{fedorov2014} is satisfied.

Since ${\cal X}$ contains only a finite number of points, then the Fisher information matrix ${\mathbf F}_{\mathbf x} \in \mathbb{R}^{p\times p}$ is continuous with respect to ${\mathbf x} \in {\cal X}$, which satisfies Assumption~(A2) in \cite{fedorov2014}. 
Actually, suppose $\{ {\mathbf x}_n, n\geq 1\} \subseteq {\cal X}$ satisfying $\lim_{n\rightarrow \infty} {\mathbf x}_n = {\mathbf x}_0 \in {\cal X}$, we must have ${\mathbf x}_n \equiv {\mathbf x}_0$ for all large enough $n$. Therefore, $\lim_{n\rightarrow \infty} {\mathbf F}_{{\mathbf x}_n} = {\mathbf F}_{{\mathbf x}_0}$ in this case.

According to Section~2.4.2 in \cite{fedorov2014}, their Assumptions~(B1), (B2), and (B4) are always satisfied for A-optimality. We only need to verify their Assumption~(B3). To simplify the notation, we denote ${\mathbf F}({\mathbf w})  = {\mathbf X}^T {\mathbf W}  {\mathbf X}$, the Fisher information matrix (up to a multiplicative constant $n$) associated with ${\mathbf w}$. Since there exists a ${\mathbf w} \in S_m$ satisfying $f({\mathbf w}) = |{\mathbf F}({\mathbf w})| > 0$, according to Lemma~\ref{lem:f_-j(w)_>0}, $h({\mathbf w})$ is a finite positive number, which qualifies Assumption~(B3) in \cite{fedorov2014}.

Then {\it (i)} and {\it (ii)} are direct conclusions of Theorem~2.2 in \cite{fedorov2014}. As for {\it (iii)}, the only thing left is to show that their $\psi({\mathbf x}_i, {\mathbf w}_*) \geq 0$ for each $i=1, \ldots, m$.

We denote $\bar{\mathbf w}_i = (0, \ldots, 0, 1, 0, \ldots, 0)^T\in S_m$ with its $i$th coordinate being $1$. Since $p\geq 2$ and ${\mathbf F}(\bar{\mathbf w}_i) = {\mathbf F}_{{\mathbf x}_i} = \nu(\boldsymbol{\beta}^T {\mathbf q}({\mathbf x}_i)) {\mathbf q}({\mathbf x}_i) {\mathbf q}({\mathbf x}_i)^T \in \mathbb{R}^{p\times p}$ has only rank $1$, then $f(\bar{\mathbf w}_i) = |F(\bar{\mathbf w}_i)| = 0$. Therefore, $f({\mathbf w}_*) > 0$ implies $0\leq w^*_i<1$ for each $i$. It can be verified that associated with ${\mathbf w}_*$~, 
\[
h_i(z) = h\left({\mathbf w}_* + \alpha (\bar{\mathbf w}_i - {\mathbf w}_*)\right)
\]
with $\alpha = (z-w^*_i)/(1-w_i^*)$, and $h_i(w_i^*) = h({\mathbf w}_*)$. Then 
\[
h'_i(z) = \nabla h\left({\mathbf w}_* + \alpha (\bar{\mathbf w}_i - {\mathbf w}_*)\right)^T \frac{\bar{\mathbf w}_i - {\mathbf w}_*}{1-w_i^*}
\]
implies
\begin{equation}\label{eq:hi'(wi*)}
h'_i(w_i^*) = \frac{1}{1-w_i^*} \cdot \nabla h({\mathbf w}_*)^T (\bar{\mathbf w}_i - {\mathbf w}_*)\ ,
\end{equation}
where $\nabla h({\mathbf w}_*) = (\partial h/\partial w_i)_i \in \mathbb{R}^m$ is the gradient of $h({\mathbf w})$ at ${\mathbf w} = {\mathbf w}_*$~.  

It can be verified that, in this case, 
\[
{\mathbf F}({\mathbf w}_* + \alpha (\bar{\mathbf w}_i - {\mathbf w}_*) ) = {\mathbf F}({\mathbf w}_*) + \alpha [{\mathbf F}(\bar{\mathbf w}_i) - {\mathbf F}({\mathbf w}_*)]\ .
\]
According to Section~2.5 in \cite{fedorov2014}, their $\Psi({\mathbf F}({\mathbf w})) = {\rm tr}({\mathbf F}({\mathbf w})^{-1}) = h^{-1}({\mathbf w})$, 
\begin{eqnarray*}
    G(\alpha) &=& \Psi\left({\mathbf F}({\mathbf w}_*) + \alpha [{\mathbf F}(\bar{\mathbf w}_i) - {\mathbf F}({\mathbf w}_*)]\right)\\
    &=& \Psi\left({\mathbf F}({\mathbf w}_* + \alpha (\bar{\mathbf w}_i - {\mathbf w}_*) )\right)\\
    &=& h^{-1}({\mathbf w}_* + \alpha (\bar{\mathbf w}_i - {\mathbf w}_*) )\ .
\end{eqnarray*}
Then $G'(\alpha) = -h^{-2}({\mathbf w}_* + \alpha (\bar{\mathbf w}_i - {\mathbf w}_*)) \cdot \nabla h({\mathbf w}_* + \alpha (\bar{\mathbf w}_i - {\mathbf w}_*))^T (\bar{\mathbf w}_i - {\mathbf w}_*)$, 
\begin{eqnarray*}
\psi({\mathbf x}_i, {\mathbf w}_*) &=& \lim_{\alpha\rightarrow 0} G'(\alpha) \\
&=&  -h^{-2}({\mathbf w}_*) \cdot \nabla h({\mathbf w}_*)^T (\bar{\mathbf w}_i - {\mathbf w}_*)\\
&=& -h^{-2}({\mathbf w}_*) \cdot (1-w_i^*) h'_i(w_i^*)
\end{eqnarray*}
due to \eqref{eq:hi'(wi*)}. Since $f({\mathbf w}_*)>0$, then $h({\mathbf w}_*)>0$ due to Lemma~\ref{lem:f_-j(w)_>0}. For each $i=1, \ldots, m$, if $w_i^* \in (0,1)$, then $h'_i(w_i^*) = 0$ since $w_i^*$ maximizes $h_i(z)$ for $z\in [0,1]$; if $w_i^*=0$, then we must have $h'_i(w_i^*)\leq 0$ due to the same reason. In both cases, we have $\psi({\mathbf x}_i, {\mathbf w}_*) \geq 0$ for each $i$. 
Due to Theorem~2.2 in \cite{fedorov2014}, we must have ${\mathbf w}_*$ is A-optimal among $S_m$~.

On the other hand, if ${\mathbf w}_*$ is A-optimal among $S_m$~, we claim that for each $i$, $w_i^*$ maximizes $h_i(z), z\in [0,1]$ associated with ${\mathbf w}_*$~. Actually, if there exists a $z_*\in [0,1]$ such that $h_i(z_*)>h_i(w_i^*)$, then $h_i(z_*) = h({\mathbf w}_* + \alpha_*(\bar{\mathbf w}_i - {\mathbf w}_*)) > h({\mathbf w}_*) = h_i(w_i^*)$, where $\alpha_* = (z_*-w_i^*)/(1-w_i^*)$, which violates the optimality of ${\mathbf w}_*$~. 
\hfill{$\Box$}

\medskip\noindent
{\bf Proof of Lemma~\ref{lem:lift_one_initial}:}\quad
We denote such a ${\mathbf w}=(w_1, \ldots, w_m)^T$. According to Lemma~3.1 in \cite{ym2015}, 
\[
f({\mathbf w})=\sum_{1\leq i_1<\cdots<i_p\leq m} |{\mathbf X}[i_1,\ldots,i_p]|^2 \cdot w_{i_1}\nu_{i_1}\cdots w_{i_p}\nu_{i_p}
\]
is a homogeneous polynomial of $w_1, \ldots, w_m$ up to order $p$. Since $w_i\geq 0$ and $\nu_i\geq 0$ for all $i$, $f({\mathbf w})>0$ implies the existence of $1\leq i_1 < \cdots < i_p \leq m$, such that, $|{\mathbf X}[i_1,\ldots,i_p]|^2 \cdot w_{i_1}\nu_{i_1}\cdots w_{i_p}\nu_{i_p} > 0$. Therefore, $|{\mathbf X}[i_1,\ldots,i_p]|^2 \cdot w'_{i_1}\nu_{i_1}\cdots w'_{i_p}\nu_{i_p} > 0$ for the same set of $\{i_1, \ldots, i_p\}$, since $w'_i>0$ for each $i$. Therefore, $f({\mathbf w}_0) \geq |{\mathbf X}[i_1,\ldots,i_p]|^2 \cdot w'_{i_1}\nu_{i_1}\cdots w'_{i_p}\nu_{i_p} >0$.
\hfill{$\Box$}

\medskip\noindent
{\bf Proof of Corollary~\ref{cor:lift_one}:}\quad
Suppose ${\mathbf w}_0$ is the initial allocation in Algorithm~\ref{algo:A_opt_lift_one}, according to Lemma~\ref{lem:lift_one_initial}, we must have $f({\mathbf w}_0)>0$. Therefore, $h({\mathbf w}_0)>0$ due to Lemma~\ref{lem:f_-j(w)_>0}.

Suppose ${\mathbf w}_* = (w_1^*, \ldots, w_m^*)^T$ is a converged allocation in Algorithm~\ref{algo:A_opt_lift_one}, then we must have $h({\mathbf w}_*) \geq h({\mathbf w}_0) > 0$ due to Step~$4^\circ$. Since ${\mathbf w}_*$ is a converged allocation, $w_i^*$ maximizes $h_i(z)$ associated with ${\mathbf w}_*$ for each $i$ due to Step~$6^\circ$. According to Theorem~\ref{thm:for_lift_one}, ${\mathbf w}_*$ must be A-optimal among $S_m$~. 

On the other hand, if ${\mathbf w}_*$ is A-optimal among $S_m$~, we must have $h({\mathbf w}_*) \geq h({\mathbf w}_0) > 0$. According to Theorem~\ref{thm:for_lift_one}, $w_i^*$ maximizes $h_i(z)$ associated with ${\mathbf w}_*$ for each $i$. That is,  ${\mathbf w}_* = (w_1^*, \ldots, w_m^*)^T$ is a converged allocation in Algorithm~\ref{algo:A_opt_lift_one}.
\hfill{$\Box$}

\medskip\noindent
{\bf Proof of Theorem~\ref{thm:GLM_A_optimality}:}\quad 
First of all, since ${\cal X} \subset \mathbb{R}^d$ is compact, then Assumption~(A1) in \cite{fedorov2014} is satisfied.

Secondly, we claim that the Fisher information matrix ${\mathbf F}_{\mathbf x} \in \mathbb{R}^{p\times p}$ is continuous with respect to ${\mathbf x} \in {\cal X}$, which satisfies Assumption~(A2) in \cite{fedorov2014}.

Actually, suppose $\{ {\mathbf x}_n, n\geq 1\} \subseteq {\cal X}$ satisfying $\lim_{n\rightarrow \infty} {\mathbf x}_n = {\mathbf x}_0 \in {\cal X}$. If $s=d$, that is, all factors are continuous, then we must have $\lim_{n\rightarrow \infty} q_j({\mathbf x}_n) = q_j({\mathbf x}_0)$ for each $j=1, \ldots, p$, since $q_j$ is continuous with respect to all continuous factors of ${\mathbf x}\in {\cal X}$. Then 
\[
{\mathbf F}_{{\mathbf x}_n} = \nu(\boldsymbol{\beta}^T {\mathbf q}({\mathbf x}_n)) {\mathbf q}({\mathbf x}_n) {\mathbf q}({\mathbf x}_n)^T\ \longrightarrow\  \nu(\boldsymbol{\beta}^T {\mathbf q}({\mathbf x}_0)) {\mathbf q}({\mathbf x}_0) {\mathbf q}({\mathbf x}_0)^T = {\mathbf F}_{{\mathbf x}_0}\ ,
\]
as $n$ goes to $\infty$. 

If $1\leq s \leq d-1$, that is, the last $d-s$ factors are discrete, ${\cal X} = \prod_{j=1}^s I_j \times {\cal D}$ with ${\cal D}$ containing only a finite number of distinct points. We denote ${\mathbf x}_n = (x_{n1}, \ldots, x_{nd})^T$, ${\mathbf x}_{n}^{(1)} = (x_{n1}, \ldots, x_{ns})^T$ for the continuous factors, and ${\mathbf x}_{n}^{(2)} = (x_{n,s+1}, \ldots, x_{nd})^T$ for the discrete factors, $n=0, 1, \ldots$~. Then $\lim_{n\rightarrow\infty} {\mathbf x}_n = {\mathbf x}_0$ implies ${\mathbf x}_n^{(2)} \equiv {\mathbf x}_0^{(2)}$ for all large enough $n$. Since $q_j$ is continuous with respect to all continuous factors of ${\mathbf x}\in {\cal X}$, then $\lim_{n\rightarrow\infty} q_j({\mathbf x}_n) = q_j({\mathbf x}_0)$ for each $j=1, \ldots, p$. Thus we also have $\lim_{n\rightarrow \infty} {\mathbf F}_{{\mathbf x}_n} = {\mathbf F}_{{\mathbf x}_0}$ in this case.

If $s=0$, that is, all $d$ factors are discrete and ${\cal X} = {\cal D}$ contains only a finite number of design points. Then we must have ${\mathbf x}_n \equiv {\mathbf x}_0$ for all large enough $n$. Therefore, $\lim_{n\rightarrow\infty} q_j({\mathbf x}_n) = q_j({\mathbf x}_0)$ for each $j$ and $\lim_{n\rightarrow \infty} {\mathbf F}_{{\mathbf x}_n} = {\mathbf F}_{{\mathbf x}_0}$ in this case as well.

According to Section~2.4.2 in \cite{fedorov2014}, their Assumptions~(B1), (B2), and (B4) are always satisfied for A-optimality. We only need to verify their Assumption~(B3). Actually, since there exists a $\boldsymbol{\xi}\in \boldsymbol{\Xi}({\cal X})$ satisfying $|{\mathbf F}(\boldsymbol{\xi})| > 0$, according to Theorem~2.1 in \cite{fedorov2014}, there always exists a $\boldsymbol{\xi}_0\in \boldsymbol{\Xi}$, such that ${\mathbf F}(\boldsymbol{\xi}_0) = {\mathbf F}(\boldsymbol{\xi})$. According to Lemma~\ref{lem:f_-j(w)_>0}, $f(\boldsymbol{\xi}_0)>0$ also implies $h(\boldsymbol{\xi}_0)$ is a finite positive number, which qualifies Assumption~(B3) in \cite{fedorov2014}.

Then {\it (i)} and {\it (ii)} are direct conclusions of Theorem~2.2 in \cite{fedorov2014}. As for {\it (iii)}, the only thing left is that given a feasible design $\boldsymbol{\xi}$ with $f(\boldsymbol{\xi}) = |{\mathbf F}(\boldsymbol{\xi})| > 0$, the sensitivity function \citep{fedorov2014, huang2024forlion} at $\boldsymbol{\xi}$ along the direction of ${\mathbf x}$ is
\begin{eqnarray*}
\varphi({\mathbf x}, \boldsymbol{\xi}) &=& {\rm tr}({\mathbf F}_{\mathbf x} {\mathbf F}(\boldsymbol\xi)^{-2})\nonumber\\
&=& {\rm tr}\big(\nu\left(\boldsymbol\beta^T {\mathbf q}({\mathbf x})\right) {\mathbf q}({\mathbf x}) {\mathbf q}({\mathbf x})^T({\mathbf X}^T_{\boldsymbol\xi} {\mathbf W}_{\boldsymbol\xi} {\mathbf X}_{\boldsymbol\xi})^{-2}\big)\\
&=& \nu\left(\boldsymbol\beta^T {\mathbf q}({\mathbf x})\right)\cdot {\mathbf q}({\mathbf x})^T({\mathbf X}^T_{\boldsymbol\xi} {\mathbf W}_{\boldsymbol\xi} {\mathbf X}_{\boldsymbol\xi})^{-2}{\mathbf q}({\mathbf x})
\end{eqnarray*}
for each ${\mathbf x} \in {\cal X}$. Then {\it (iii)} is obtained from Theorem~2.2 in \cite{fedorov2014} as well.
\hfill{$\Box$}

\medskip\noindent
{\bf Proof of Theorem~\ref{thm:alphat}:}\quad
We denote ${\mathbf w} = (w_1^{(t)}, \ldots, w_{m_t}^{(t)}, 0)^T \in S_{m_t+1}$~. Associated with ${\mathbf w}$ and $\{{\mathbf x}_1^{(t)}, \ldots, {\mathbf x}_{m_t}^{(t)}, {\mathbf x}^*\}$, ${\mathbf W} \in \mathbb{R}^{(m_t+1)\times (m_t+1)}$, and ${\mathbf X} = ({\mathbf q}({\mathbf x}_1^{(t)}), \ldots, {\mathbf q}({\mathbf x}_{m_t}^{(t)}), {\mathbf q}({\mathbf x}_*))^T \in \mathbb{R}^{(m_t+1)\times p}$. Since ${\mathbf w}$'s last coordinate is $0$, 
\begin{eqnarray*}
f({\mathbf w}) &=& |{\mathbf X}^T {\mathbf W} {\mathbf X}| = \left|{\mathbf X}_{\boldsymbol{\xi}_t}^T {\mathbf W}_{\boldsymbol{\xi}_t} {\mathbf X}_{\boldsymbol{\xi}_t}\right| = |{\mathbf F}(\boldsymbol{\xi}_t)| = f(\boldsymbol{\xi}_t)\ ,\\
f_{-j}({\mathbf w}) &=& |{\mathbf X}_{-j}^T {\mathbf W} {\mathbf X}_{-j}| = \left|{\mathbf X}_{\boldsymbol{\xi}_t, -j}^T {\mathbf W}_{\boldsymbol{\xi}_t} {\mathbf X}_{\boldsymbol{\xi}_t,-j}\right| = f_{-j}(\boldsymbol{\xi}_t)\ .
\end{eqnarray*}
With $i=m_t+1$ and $w_{m_t+1}=0$, we have for $\alpha\in [0,1]$,
\begin{eqnarray*}
    f_{m_t+1}(\alpha) &=& f\left((1-\alpha)w_1^{(t)}, \ldots, (1-\alpha)w_{m_t}^{(t)}, \alpha)\right)\ ,\\
    f_{m_t+1}^{(-j)}(\alpha) &=& f_{-j}\left((1-\alpha)w_1^{(t)}, \ldots, (1-\alpha)w_{m_t}^{(t)}, \alpha)\right)\ ,\\
    h_{m_t+1}(\alpha) &=& h\left((1-\alpha)w_1^{(t)}, \ldots, (1-\alpha)w_{m_t}^{(t)}, \alpha)\right) = h(\boldsymbol{\xi}_{t+1})\ . 
\end{eqnarray*}
Then maximizing $h(\boldsymbol{\xi}_{t+1})$ for $\alpha\in [0,1]$ is the same as maximizing $h_{m_t+1}(z)$ for $z\in [0,1]$. The desired results are direct conclusions of Theorem~\ref{thm:max_h_i(x)}.  
\hfill{$\Box$}

\section{Additional examples and tables}\label{sec:examples_supp}

{\bf Example~\ref{ex:wavesolder_example}. Wave-soldering defects (continued)}\quad {\rm To assess whether the conclusions of Example~\ref{ex:wavesolder_example} depend on the main-effects model chosen for illustration purposes, we repeat the comparison under the Poisson regression model with log link presented by \citet{hamada1997glmquality} after removing a potential outlier. The model contains six main effects (A, B, C, D, E, and G) and two interactions (AC and BD), with fitted parameters  provided in their Table~7. 

For comparison purposes, we keep all other settings unchanged, including the sixteen candidate runs and the simulation protocol. As shown in Figure~\ref{fig:rmse_ws_hamada}, the A-optimal design again attains the smallest mean RMSE at all sample sizes except $n=500$, where the original design is marginally better. The improvement of A-optimal design over the original design is not so significant, because under this model the A-efficiency of the original design is already as high as $88.3\%$.
\hfill{$\Box$}

\begin{figure}[htbp]
  \centering
  \includegraphics[width=0.5\textwidth]{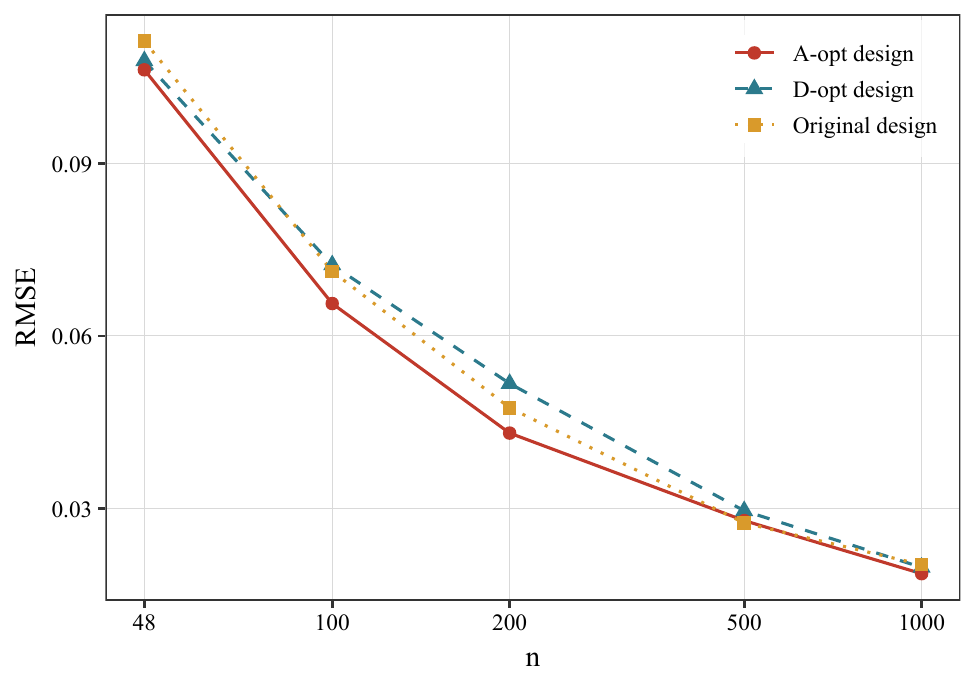}
  \caption{Mean RMSE versus sample size $n$, averaged over 100 simulated data sets, under the model of \citet{hamada1997glmquality}, with the horizontal axis on a logarithmic scale.}
  \label{fig:rmse_ws_hamada}
\end{figure}

\begin{example}\label{ex:RMSE} 
{\bf Paid research study}\quad {\rm
Inspired by Example 6 in \cite{huang2025constrained}, we use a paid research study scenario to illustrate the difference between A-optimal and D-optimal allocations. Suppose there are $N = 5,000$ eligible volunteers, grouped into $m=6$ categories according to gender ($x_{i1}=0$ for female, $1$ for male) and age ($x_{i2}=0$ for $18-25$, $1$ for $26-64$, and $2$ for $65$ or above). Then the covariate setting ${\mathbf x}_i = (x_{i1}, x_{i2})^T$ is $(0, 0), (0, 1), (0, 2), (1, 0), (1, 1)$, or $(1, 2)$, with corresponding population sizes
$(N_1,\ldots,N_6)=(500,400,100,2000,1500,500)$, for $i=1, \ldots, 6$. The responses of participants are either $0$ for no treatment effect, or $1$ for effective treatment. Following \cite{huang2025constrained}, we assume a logistic regression model (see also Example~\ref{ex:main_effects}) as follows: 
\begin{equation}\label{eq:paid_research_study}
{\rm logit}(P(Y_{ij} = 1|x_{i1}, x_{i2})) = \beta_0 + \beta_1 x_{i1} + \beta_{21} \mathds{1}_{\{x_{i2} = 1\}} + \beta_{22} \mathds{1}_{\{x_{i2} = 2\}}\ ,
\end{equation}
where $i = 1, \ldots, 6$, $j = 1, \ldots , N_i$~, $Y_{ij} \in \{0,1\}$ is the response of the $j$th volunteer in the $i$th category, $\mathds{1}$ stands for an indicator function, and $\boldsymbol\beta = (\beta_0, \beta_1, \beta_{21}, \beta_{22})^T = (0, 3, 3, 3)^T$. 

Suppose the budget allows $n = 200$ participants and stratified sampling is adopted. Given a predetermined integer-valued allocation ${\mathbf n} = (n_1, \ldots, n_6)^T$, such that $\sum_{i=1}^6 n_i = n$, known as an {\it exact allocation}, we randomly choose $n_i$ out of $N_i$ volunteers in the $i$th category. The proportionally stratified allocation is ${\mathbf n}_p = (20, 16, 4, 80, 60, 20)^T$, a uniformly stratified allocation is ${\mathbf n}_u = (34, 34, 33, 33, 33, 33)^T$, and the (locally) D-optimal allocation is ${\mathbf w}_D = (0.25, 0.25, 0.25, 0.25, 0, 0)^T$ or ${\mathbf n}_D = (50, 50, 50, 50, 0, 0)^T$. By using Algorithms~\ref{algo:A_opt_lift_one} and \ref{algo:exact_A} (a rounding-off algorithm in the Supplementary Material, Section~\ref{sec:algorithms}), we obtain the (locally) A-optimal allocation ${\mathbf w}_A = (0.2208, 0.2597, 0.2597, 0.2597, 0,$ $0)^T$ or ${\mathbf n}_A = (44, 52, 52, 52, 0, 0)^T$. Both the D- and A-optimal allocations assign no units to strata $(1,1)$ and $(1,2)$. These strata have $\eta_5 = \eta_6 = 6$ and thus $\nu_5 = \nu_6  \approx 0.0025$, so they carry little information. Under the main-effects model, their response probabilities can still be inferred from the other four strata. Such allocations therefore rely on the main-effects assumption. If interactions may be present, the saturated model in Example~\ref{ex:saturated} can be used, under which every stratum is sampled.

To compare the accuracy of parameter estimates based on different allocations or samplers, we use the root mean squared error (RMSE), $[\sum_{i \in I} (\hat{\beta}_i$ $- \beta_i)^2/|I|]^{1/2}$, given an index set $I$ of parameters of interest. For each of $100$ simulations, responses for all $5,000$ volunteers are generated from~\eqref{eq:paid_research_study}, and samples of size $n=200$ are drawn using the simple random sample without replacement (SRSWOR, see \cite{lohr2022sampling}), proportionally stratified sampler (${\mathbf n}_p$), uniformly stratified sampler (${\mathbf n}_u$), D-optimal sampler (${\mathbf n}_D$), and A-optimal sampler (${\mathbf n}_A$). Parameter estimates $\hat{\boldsymbol{\beta}} = (\hat\beta_0, \hat\beta_1, \hat\beta_{21}, \hat\beta_{22})^T$ and corresponding RMSEs are computed. Figure~\ref{fig:rmse_allx0} summarizes the RMSEs for the coefficients of interest ($I = \{1, 21, 22\}$) across the samplers. Their averages and standard deviations (sd) for all coefficients are reported in Table~\ref{RMSE0333}, which also includes the RMSEs based on the full data (i.e., $5,000$ volunteers) for readers' reference.

\begin{figure}[htbp]
    \centering
    \begin{minipage}[t]{0.45\textwidth}
        \centering
        \includegraphics[width=\textwidth]{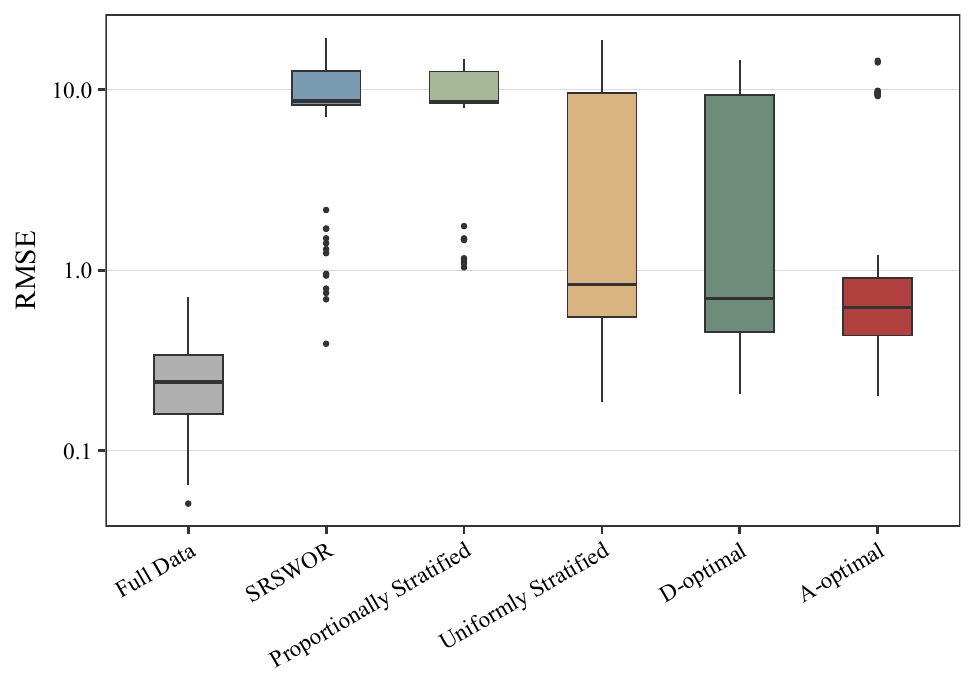}
        \caption{RMSE for all coefficients except $\beta_0$ over 100 simulations in Example~\ref{ex:RMSE}, on a logarithmic scale.}
        \label{fig:rmse_allx0}
    \end{minipage}
    \hfill
    \begin{minipage}[t]{0.45\textwidth}
        \centering
        \includegraphics[width=\textwidth]{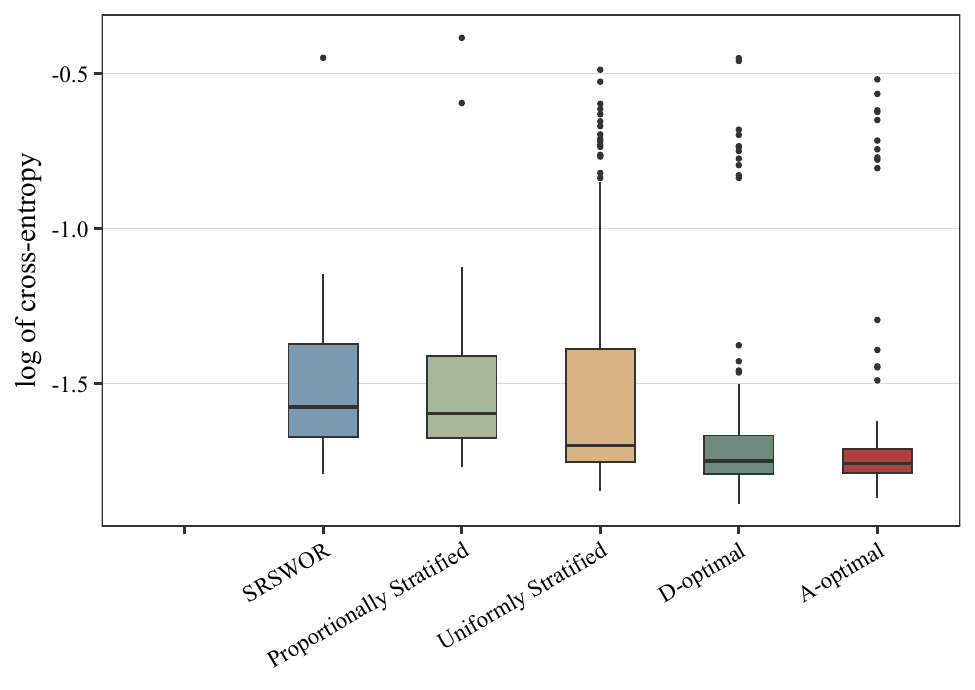}
        \caption{Boxplots of cross-entropy loss over 100 simulations for Example~\ref{ex:RMSE}.}
        \label{fig:log_of_cross_entropy}
    \end{minipage}
\end{figure}

As shown in Figure~\ref{fig:rmse_allx0} (with the underlying per-coefficient averages and standard deviations reported in Table~\ref{RMSE0333}), the proportionally stratified sampler and SRSWOR are the least accurate ones, the uniformly stratified sampler is better, and so is the D-optimal sampler, and the A-optimal sampler seems the best. Compared with the D-optimal sampler, the A-optimal sampler achieves smaller RMSE for each $\beta_i$ except $\beta_0$~, as detailed in that table, which is typically of less interest in practice. 

Actually, across various sets of parameter values (results not shown here), A-optimal allocations tend to outperform the corresponding D-optimal ones in terms of RMSE. The reason is that an A-optimal allocation minimizes $\mathrm{tr}(({\mathbf X}^T{\mathbf W}{\mathbf X})^{-1})$, which is asymptotically the sum of the variances of the estimated parameters (up to a constant). 
Since the maximum likelihood estimator $\hat{\boldsymbol\beta}$ is asymptotically unbiased, in large samples the quantity $\mathrm{tr}(({\mathbf X}^T{\mathbf W}{\mathbf X})^{-1})$ approximates the expected squared RMSE of $\hat{\boldsymbol\beta}$ (up to a constant). 
Therefore, A-optimal designs are expected to perform well in terms of RMSE when the sample size is large.

Prediction accuracy is further evaluated using cross-entropy (CE) loss on predicting the responses of unsampled individuals (see Remark~\ref{rem:cross_entropy_loss} in Section~\ref{sec:lemmas_remark}). Over $100$ simulations, the A-optimal sampler achieves the lowest mean CE of $0.2088$ with the least variation (see Figure~\ref{fig:log_of_cross_entropy}), the D-optimal sampler is the second lowest with a mean CE of $0.2128$. Both substantially outperform SRSWOR ($0.2237$), the proportionally stratified sampler ($0.2233$), and the uniformly stratified sampler ($0.2484$).
}\hfill{$\Box$}
\end{example}

\begin{example}\label{ex:saturated} {\bf Paid research study with a saturated model}\quad 
{\rm We revisit the paid research study of Example~\ref{ex:RMSE} under a saturated model, that is, whose number of parameters equals the number of distinct design points. Actually, by adding the interactions $x_{i1}\mathds{1}_{\{x_{i2}=1\}}$ and $x_{i1}\mathds{1}_{\{x_{i2}=2\}}$ to model~\eqref{eq:paid_research_study}, we consider the model as follows: ${\rm logit}\big(P(Y_{ij}=1\mid x_{i1},x_{i2})\big)=\beta_0+\beta_1 x_{i1}+\beta_{21}\mathds{1}_{\{x_{i2}=1\}}+\beta_{22}\mathds{1}_{\{x_{i2}=2\}}+\beta_{31}\,x_{i1}\mathds{1}_{\{x_{i2}=1\}}+\beta_{32}\,x_{i1}\mathds{1}_{\{x_{i2}=2\}}$, with six parameters over six categories. For illustration purposes, we keep $(\beta_0,\beta_1,\beta_{21},\beta_{22})=(0,3,3,3)$ as in Example~\ref{ex:RMSE} and vary the values of $\beta_{31}$ and $\beta_{32}$ from $-9$ to $-3$ (see Figure~\ref{fig:supp_saturated}).

\begin{figure}[t]
\centering
\includegraphics[width=0.7\linewidth]{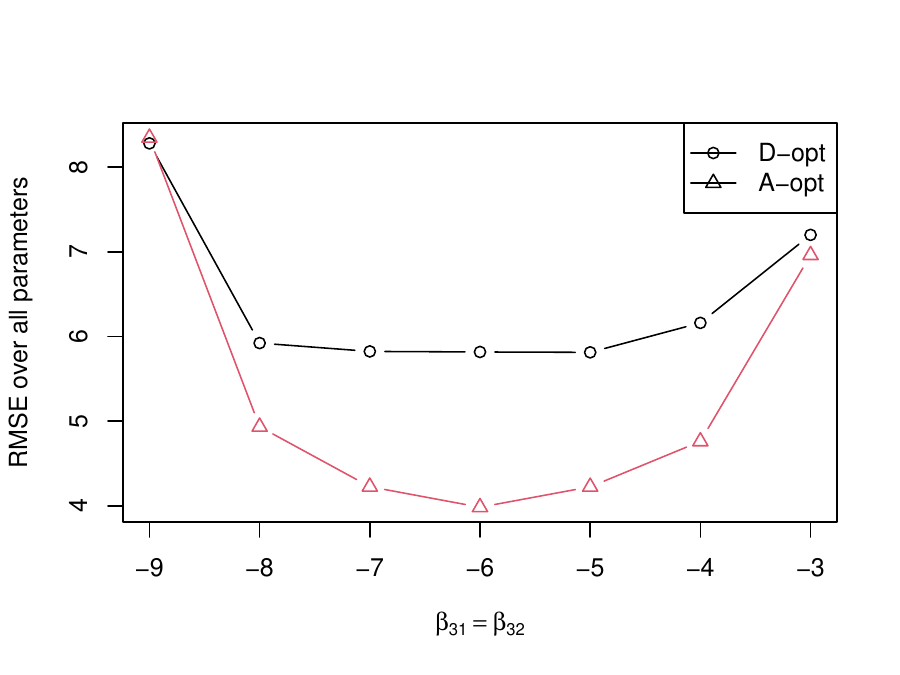}
\caption{RMSE over all parameters under the interaction model versus $\beta_{31}=\beta_{32}\in[-9,-3]$, for the locally D- and A-optimal allocations.}
\label{fig:supp_saturated}
\end{figure}

Under this saturated model, the D-optimal design does not depend on $\boldsymbol\beta$, while the A-optimal design does. Averaged over $100$ simulations, the A-optimal design achieves a lower RMSE over all parameters across the range $[-9,-3]$ of $\beta_{31}=\beta_{32}$ (see Figure~\ref{fig:supp_saturated}). The reduction reaches about $30\%$ at $\beta_{31}=\beta_{32} = -6$. 
}\hfill{$\Box$}
\end{example}

\begin{example}\label{ex:compare_A_D} {\bf Printed circuit board}\quad 
{\rm In their Example~5.1,  \cite{ym2015} modified an experiment on printed circuit boards (PCB) discussed by \cite{jeng2008modeling}. They considered a binary response indicating whether an open circuit fault occurs during the inner layer manufacturing, and two discrete experimental factors, namely Preheat (factor A, two levels) and Lamination Temperature (factor B, three levels). Following Example~3.2 in \cite{ym2015}, we denote the model matrix by
\begin{equation}\label{eq:PCB_X}
{\mathbf X} = \begin{bmatrix}
1 & 1 & 1 & 1 \\
1 & 1 & 0 & -2 \\
1 & 1 & -1 & 1 \\
1 & -1 & 1 & 1 \\
1 & -1 & 0 & -2 \\
1 & -1 & -1 & 1 \\
\end{bmatrix}\ ,
\end{equation}
whose $i$th row is regarded as $(1,{\mathbf x}_i^T)$ with ${\mathbf x}_i = (x_{i1}, x_{i2}, x_{i3})^T$, and whose four columns represent the intercept ($\beta_0$), factor A ($\beta_1$), the linear component of factor B ($\beta_{21}$), and the quadratic component of factor B ($\beta_{22}$). In other words, the original two discrete factors are now coded as six experimental settings ${\mathbf x}_i$~, $i=1, \ldots, 6$ listed in ${\mathbf X}$ (see also, e.g., \cite{joseph2004}). Under a GLM with Bernoulli distribution and logit link (i.e., a logistic regression model) with assumed parameter values $\boldsymbol{\beta} = (\beta_0, \beta_1, \beta_{21}, \beta_{22})^T = (-2.5, 0.15, 0.70, 0.10)^T$, \cite{ym2015} recommended a D-optimal allocation ${\mathbf w}_D = (0.216, 0.186, 0.198, 0.206, 0.115, $ $0.080)^T$ for ${\mathbf X}$.

Since in this case the model matrix \eqref{eq:PCB_X} is built from a custom set of design points that does not correspond to a full factorial or product space, the R package \texttt{OptimalDesign} \citep{harman2025package} is not applicable, which assumes that the design space is a Cartesian product of factor levels.
On the contrary, our Algorithm~\ref{algo:A_opt_lift_one} can incorporate an arbitrary design matrix  allowing for non-product-space structures. By applying Algorithm~\ref{algo:A_opt_lift_one}, we obtain an A-optimal allocation ${\mathbf w}_A = (0.1458, 0.1407, 0.2261,$ $0.1510, 0.1385, 0.1980)^T$ associated with ${\mathbf X}$.

\begin{table}[ht]
\caption{Average (sd) of RMSE over 100 simulations for Example~\ref{ex:compare_A_D}}\label{tab:YM2015_RMSE_100}
\resizebox{\textwidth}{!}{\begin{tabular}{@{\extracolsep\fill}lccccc}
\toprule
Sampler & $\beta_0$ & all except $\beta_0$ & $\beta_1$ & $\beta_{21}$ & $\beta_{22}$ \\
\midrule
Full Data  & 2.503(0.026) & 0.319(0.216) & 0.018(0.012) & 0.024(0.020) & 0.013(0.010) \\
SRSWOR  & 2.506(0.080) & 0.340(0.203) & 0.062(0.043) & 0.085(0.061) & 0.036(0.025) \\
Uniformly/Proportionally  & 2.518(0.077) & 0.337(0.194) & 0.052(0.039) & 0.073(0.057) & 0.050(0.036) \\
D-optimal  & 2.520(0.090) & 0.339(0.204) & 0.051(0.045) & 0.075(0.068) & 0.048(0.033) \\
A-optimal  & 2.514(0.075) & 0.338(0.208) & 0.052(0.043) & 0.065(0.050) & 0.036(0.027) \\
\bottomrule
\end{tabular}}
\end{table}

\begin{table}[ht]
\caption{$p$-values of two-sided pairwise $t$-tests between a sampler's 100 RMSEs and A-optimal ones for Example~\ref{ex:compare_A_D}}\label{tab:p_values_1}
\centering
\resizebox{0.7\textwidth}{!}
{\begin{tabular}{@{\extracolsep\fill}lccccc}
\toprule
Sampler & $\beta_0$ & all except $\beta_0$ & $\beta_1$ & $\beta_{21}$ & $\beta_{22}$ \\
\midrule
SRSWOR  & 1.0000 & 1.0000 & 0.7600 & 0.1200 & 1.0000 \\
Uniformly/Proportionally & 1.0000 & 1.0000 & 1.0000 & 1.0000 & 0.0024 \\
D-optimal  & 1.0000 & 1.0000 & 1.0000 & 1.0000 & 0.0136 \\
\bottomrule
\end{tabular}}
\end{table}

Similarly to Example~\ref{ex:RMSE}, we use ${\rm RMSE} = [\sum_{i \in I} (\hat{\beta}_i$ $- \beta_i)^2/|I|]^{1/2}$ to compare the estimation accuracy of model parameters based on different samplers.
In this case, we simulate $N_i=4,800$ responses for each ${\mathbf x}_i$ and treat the total $N = 6N_i = 28,800$ observations as the target population or whole dataset, denoted by ${\cal C} = \{({\mathbf x}_i, y_{ij})\mid i=1, \ldots, 6; j=1, \ldots, N_i\}$. Suppose we aim to sample $n=2,880$ observations out of ${\cal C}$. By applying Algorithm~\ref{algo:exact_A}, we convert each given approximate allocation ${\mathbf w} = (w_1, \ldots, w_6)^T \in S_6$ to an exact allocation ${\mathbf n} = (n_1, \ldots, n_6)^T$, such that $\sum_{i=1}^6 n_i = n$. In this case, both the proportionally and uniformly stratified exact allocations are $n_u = (480, 480, 480, 480, 480, 480)^T$. The D-optimal exact allocation is $n_D = (621, 534, 569, 593, 332, 231)^T$, and A-optimal exact allocation is $n_A = (420, 405, 651, 435, 399, 570)^T$. Given an exact allocation ${\mathbf n} = (n_1, \ldots, n_6)^T$, we pick up a simple random sample of  size $n_i$ from $N_i$ observations associated with ${\mathbf x}_i$~, $i=1, \ldots, 6$, and denote the collection of sampled observations by $\Lambda \subset {\cal C}$ with $|\Lambda|=n$. Similarly to Example~\ref{ex:RMSE}, we estimate $\hat{\boldsymbol{\beta}} = (\hat{\beta}_0, \hat{\beta}_1, \hat{\beta}_{21}, \hat{\beta}_{22})^T$ and compute the RMSE for each of 100 independent simulations. 

The average and standard deviation (sd) of RMSEs over 100 simulations are listed in Table~\ref{tab:YM2015_RMSE_100}. In terms of the RMSEs for $\beta_1$~, the uniformly (and proportionally) stratified sampler, the D-optimal and A-optimal samples have comparable accuracy and all better than SRSWOR (not significantly according to Table~\ref{tab:p_values_1} though). In terms of estimations for $\beta_{22}$~, SRSWOR and the A-optimal allocation seem to yield lower RMSE values than the uniformly (and proportionally) stratified sampler and the D-optimal sampler (significantly according to Table~\ref{tab:p_values_1}). Overall, we conclude that the A-optimal sampler provides more accurate estimates on non-intercept parameters in this case. The $p$-values of two-sided pairwise $t$-tests listed in Table~\ref{tab:p_values_1} confirm our conclusion.
}\hfill{$\Box$}
\end{example}

\begin{example}\label{ex:2_paramrters_logit_continuous} {\bf Logistic model with one continuous factor}\quad {\rm  
In Example~1 of \cite{yangmin2008}, a two-parameter logistic regression model with one continuous factor $x_i \in \mathbb{R}$ is considered, such that, $\eta_i = \alpha + \beta x_i$ and $g(\mu_i) = \log(\mu_i/(1 - \mu_i))$ as in \eqref{eq:glm}. 
Given $\alpha = -2$ and $\beta = 0.5$, \cite{yangmin2008} obtained theoretically a (locally) A-optimal design $\boldsymbol{\xi}_o = \{(x_1^o = 0.2579, w_1^o = 0.8832),\ (x_2^o = 7.7421, w_2^o = 0.1168)\}$. By applying Algorithm~\ref{algo:A_opt_Forlion} with $\delta = 0.3$ and $\epsilon = 10^{-12}$, we obtain an A-optimal design $\boldsymbol{\xi}_a = \{(x_1^a = 0.2542, w_1^a = 0.8833), (x_2^a = 7.7459, w_2^a = 0.1167)\}$, which is fairly close to $\boldsymbol{\xi}_o$~. Its relative efficiency is $h(\boldsymbol{\xi}_a)/h(\boldsymbol{\xi}_o) = 99.9997\%$.

To further illustrate the effects of the merging threshold $\delta$ and the converging threshold $\epsilon$ of Algorithm~\ref{algo:A_opt_Forlion}, we list in Table~\ref{tab:relative_difference_tolerance} the performance of Algorithm~\ref{algo:A_opt_Forlion} across different levels of $\delta$ and $\epsilon$. As shown, the computational time is influenced more by the choice of $\delta$, with smaller $\delta$ leading to substantially longer running time. When $\delta$ is not too small (e.g., $\delta = 0.3$ or $0.4$), the time cost increases as $\epsilon$ decreases. Most configurations result in $2$ support points, while a $\delta$ as small as $0.1$ causes unnecessary support points with a small minimum distance ($0.134$), which implies that $\delta$ is too small. The relative efficiency remains high across all settings. Overall, in this case we recommend $\delta = 0.3$ and $\epsilon = 10^{-6}$ to balance the relative efficiency and computational cost.

\begin{table}[ht]
\centering
\caption{Effects of tuning parameters of ForLion algorithm for Example~\ref{ex:2_paramrters_logit_continuous}}\label{tab:relative_difference_tolerance}
{\renewcommand{\arraystretch}{0.6}
\resizebox{0.8\textwidth}{!}
{\begin{tabular*}{1.1\textwidth}{@{\extracolsep\fill}lrrrrrrrr}
\toprule%
& \multicolumn{4}{@{}c@{}}{Time cost (in seconds)} & \multicolumn{4}{@{}c@{}}{Minimum distance between points} \\ \cmidrule(l{2pt}r{2pt}){2-5}\cmidrule(l{2pt}r{2pt}){6-9}%
$\delta \backslash \epsilon$ & $10^{-6}$ & $10^{-8}$ & $10^{-10}$ & $10^{-12}$ & $10^{-6}$ & $10^{-8}$ & $10^{-10}$ & $10^{-12}$ \\
\midrule
0.1 & 167.20 & 44.76 & 121.58 & 79.02 & 0.134 & 7.508 & 7.452 & 7.472 \\
0.2 & 12.61  & 18.81 & 14.78  & 15.95 & 7.493 & 7.492 & 7.440 & 7.440 \\
0.3 & 10.16  & 12.66 & 12.03  & 13.56 & 7.492 & 7.492 & 7.492 & 7.492 \\
0.4 & 9.90   & 12.67 & 12.23  & 13.41 & 7.492 & 7.492 & 7.492 & 7.492 \\
\bottomrule
& \multicolumn{4}{@{}c@{}}{Number of support points} & \multicolumn{4}{@{}c@{}}{Relative efficiency w.r.t. $\boldsymbol{\xi}_o$ (\%)} \\ \cmidrule(l{2pt}r{2pt}){2-5}\cmidrule(l{2pt}r{2pt}){6-9}%
$\delta \backslash\epsilon$ & $10^{-6}$ & $10^{-8}$ & $10^{-10}$ & $10^{-12}$ & $10^{-6}$ & $10^{-8}$ & $10^{-10}$ & $10^{-12}$ \\
\midrule
0.1 & 4 & 2 & 2 & 2 & 99.9690 & 99.9993 & 99.9980 & 99.9997 \\
0.2 & 2 & 2 & 2 & 2 & 99.9997 & 99.9997 & 99.9963 & 99.9963 \\
0.3 & 2 & 2 & 2 & 2 & 99.9997 & 99.9997 & 99.9997 & 99.9997 \\
0.4 & 2 & 2 & 2 & 2 & 99.9997 & 99.9997 & 99.9997 & 99.9997 \\
\bottomrule
\end{tabular*}}
}\end{table}

\begin{table}[ht]
\centering
\caption{A-optimal designs with constrained design spaces for Example~\ref{ex:2_paramrters_logit_continuous}}\label{tab_constrained}
{\renewcommand{\arraystretch}{0.6}
\resizebox{0.9\textwidth}{!}{\begin{tabular}{llr}
\toprule
Design space  & A-optimal design by Algorithm~\ref{algo:A_opt_Forlion} & Relative efficiency w.r.t. $\boldsymbol{\xi}_o$ \\
\midrule
$\left[0, 7\right]$ & $\{(x_1 = 0.1721,\ w_1 = 0.8894), (x_2 = 7,\ w_2 = 0.1106)\}$ & 0.9967    \\
$\left[0, 5\right]$ & $\{(x_1 = 0,\ w_1 = 0.8841), (x_2 = 5,\ w_2 = 0.1159)\}$ & 0.9520 \\
$\left[0, 3\right]$ & $\{(x_1 = 0,\ w_1 = 0.8255), (x_2 = 3,\ w_2 = 0.1745)\}$ & 0.7769 \\
$\left[0, 1\right]$ & $\{(x_1 = 0,\ w_1 = 0.6276), (x_2 = 1,\ w_2 = 0.3724)\}$ & 0.2495 \\
\bottomrule
\end{tabular}}
}\end{table}

In \cite{yangmin2008}, $\boldsymbol{\xi}_o$ is guaranteed to be A-optimal if the design region is unconstrained, i.e., the whole real line. In practice, however, a feasible design region is often bounded due to various physical, experimental, or operational limitations. As such, the levels of continuous or discrete factors are typically confined to  predefined bounded regions. For those scenarios, we can still use our Algorithm~\ref{algo:A_opt_Forlion} to find the corresponding A-optimal designs. For illustration purpose, we let the design space in this case change from $[0, 7]$ to $[0, 1]$ gradually. As shown in Table~\ref{tab_constrained}, by using Algorithm~\ref{algo:A_opt_Forlion} with $\delta = 0.3$ and $\epsilon = 10^{-6}$, the support points move toward the boundary of the regions, and the relative efficiency with respect to $\boldsymbol{\xi}_o$ decreases as the design space becomes more and more restricted.
}\hfill{$\Box$}
\end{example}

\begin{example}\label{ex:3_paramrters_gamma} {\bf Gamma model with two factors}\quad {\rm  
\cite{gaffke2019} considered a Gamma regression model with two continuous factors $x_{i1}$ and $x_{i2}$~, such that $\eta_i = \beta_0 + \beta_1 x_{i1} + \beta_2 x_{i2}$ and $g(\mu_i) = {\mu_i}^{-1}$ (see model~\eqref{eq:glm}).
Assuming $\beta_0=1$, $\beta_1=\beta_2=\gamma$, they theoretically justified that the (locally) A-optimal design on ${\cal X} = [0,1]^2$ is supported only on the four vertices of ${\cal X}$, namely ${\mathbf x}_i = (0,0)^T, (1,0)^T, (0,1)^T$ or $(1,1)^T$ for $i=1,2,3,4$, respectively. Then they used the multiplicative algorithm (see also Example~\ref{ex:main_effects}) to obtain an A-optimal allocation for the four vertices, denoted by ${\mathbf w}_\gamma$~.

\begin{table}[ht]
\centering
\caption{A-optimal designs by Algorithm~\ref{algo:A_opt_Forlion} for Example~\ref{ex:3_paramrters_gamma}}\label{tab:gamma_3parameters}
\resizebox{0.8\textwidth}{!}{%
\begin{tabular}{lccccc}
\toprule
$\gamma$  & $w_1$ & $w_2$ & $w_3$ & $w_4$ & Relative efficiency w.r.t. ${\mathbf w}_\gamma$\\
\midrule
-0.45 & 0.1136 & 0.3984 & 0.3983 & 0.0897 &  100.0000\%  \\
0 & 0.3560 & 0.2257 & 0.2250 & 0.1933  & 100.0100\% \\
1 & 0.2690 & 0.3003 & 0.3001 & 0.1307 & 100.0006\% \\
2 & 0.2208 & 0.3805 & 0.3806 & 0.0182 &  100.0000\% \\
\bottomrule
\end{tabular}
}\end{table}

To show how our algorithms work for this experimental design problem, we first apply Algorithm~\ref{algo:A_opt_Forlion} to the continuous design problem with ${\cal X}=[0,1]^2$. Remarkably, for each $\gamma$ listed in Table~\ref{tab:gamma_3parameters} (see also Table~1 in \cite{gaffke2019}), our ForLion algorithm ends with the same four vertices as the support points, along with the weights $w_i$ provided in Table~\ref{tab:gamma_3parameters}. Compared with ${\mathbf w}_\gamma$ listed in Table~1 of \cite{gaffke2019}, our designs can be slightly better according to the relative efficiencies. }\hfill{$\Box$}
\end{example}

For the rest of this section, we provide more tables supporting the examples earlier, including 
Table~\ref{tab:2^k_main_tab} for Example~\ref{ex:main_effects} in Section~\ref{sec:lift_one_discrete_factor}, 
Table~\ref{tab:rmse_ws} for Example~\ref{ex:wavesolder_example} in Section~\ref{sec:examples}, 
Tables~\ref{tab:rmse_potato} \& \ref{tab:potato_rmse_pairwise} for Example~\ref{ex:potato_example}  in Section~\ref{sec:examples},
and Table~\ref{RMSE0333} for Example~\ref{ex:RMSE} in this section.

\begin{table}[ht]
    \centering
    \caption{A-optimal designs for $2^k$ main-effects model in Example~\ref{ex:main_effects}}\label{tab:2^k_main_tab}  
{    \renewcommand{\arraystretch}{0.6}
    \resizebox{0.9\textwidth}{!}{\begin{tabular}{crrrrrrrr}
        \toprule
    \multicolumn{9}{c}{\bf\Large Total computation time in seconds across 100 simulations}\\
    \midrule
        Experiment & Algorithm 1 & Nelder-Mead & Quasi-Newton & Conjugate Gradient & Simulated Annealing & REX & VDM & MUL \\
        \midrule
        $2^2$ & 0.22 & 0.64 & 0.27 & 2.36 & 48.26 & 0.56 & 156.45 & 0.97 \\
        $2^3$ & 0.72 & 4.01 & 0.61 & 3.64 & 54.35 & 1.99 & 596.78 & 1.04 \\
        $2^4$ & 3.70 & 11.35 & 1.67 & 6.09 & 68.33 & 7.83 & 2536.88 & 3.37 \\
        $2^5$ & 22.34 & 20.16 & 9.50 & 12.94 & 98.75 & 112.81 & 3608.08 & 59.16 \\
        $2^6$ & 83.87 & 25.78 & 6.96 & 9.68 & 130.32 & 130.02 & 4901.03 & 148.53 \\
        $2^7$ & 405.85 & 81.81 & 39.14 & NA & 516.36 & 650.24 & 4887.25 & 1022.8 \\
        \bottomrule
    \multicolumn{9}{c}{\bf\Large Average number of support points over 100 simulations}\\
    \midrule
        Experiment & Algorithm 1 & Nelder-Mead & Quasi-Newton & Conjugate Gradient & Simulated Annealing & REX & VDM & MUL \\
        \midrule
        $2^2$ & 3.40 & 4.00 & 4.00 & 4.00 & 4.00 & 3.40 & 3.40 & 4.00 \\
        $2^3$ & 5.84 & 8.00 & 8.00 & 8.00 & 8.00 & 5.82 & 5.89 & 7.95 \\
        $2^4$ & 9.59 & 16.00 & 16.00 & 16.00 & 16.00 & 9.42 & 10.09 & 15.60 \\
        $2^5$ & 14.02 & 32.00 & 32.00 & 32.00 & 32.00 & 13.57 & 16.49 & 27.07 \\
        $2^6$ & 19.91 & 64.00 & 64.00 & 64.00 & 64.00 & 18.76 & 26.50 & 48.37 \\
        $2^7$ & 28.17 & 128.00 & 128.00 & NA & 128.00 & 25.79 & 41.96 & 91.14 \\
        \bottomrule
        \multicolumn{9}{c}{\bf\Large Average relative efficiency over 100 simulations}\\
    \midrule
        Experiment & Algorithm 1 & Nelder-Mead & Quasi-Newton & Conjugate Gradient & Simulated Annealing & REX & VDM & MUL \\
        \midrule
        $2^2$ & 100.00 & 100.00 & 100.00 & 100.00 & 83.99 & 100.00 & 100.00 & 100.00 \\
        $2^3$ & 100.00 & 99.96 & 100.00 & 99.83 & 77.42 & 100.00 & 100.00 & 100.00 \\
        $2^4$ & 100.00 & 99.28 & 100.00 & 97.66 & 72.54 & 100.00 & 100.00 & 100.00 \\
        $2^5$ & 100.00 & 94.37 & 99.99 & 90.68 & 66.86 & 100.00 & 100.00 & 100.00 \\
        $2^6$ & 100.00 & 75.98 & 99.97 & 83.89 & 61.92 & 100.01 & 100.01 & 100.01 \\
        $2^7$ & 100.00 & 60.47 & 98.54 & NA & 58.13 & 100.01 & 100.01 & 100.01 \\
        \bottomrule
    \end{tabular}}
\caption*{
\footnotesize{Note: ``NA'' indicates that the algorithm fails under this setting.}}
}\end{table}

\begin{table}[htbp]
\centering
\caption{Average (sd) of RMSEs over 100 simulations for Example~\ref{ex:wavesolder_example}.}
\label{tab:rmse_ws}
\begin{tabular}{ccccc}
\hline
$n$ & A-optimal & D-optimal & Original \\
\hline
$48$   & $0.058\;(0.043)$ & $0.063\;(0.025)$ & $0.064\;(0.020)$ \\
$100$  & $0.035\;(0.010)$ & $0.041^{*}\;(0.012)$     & $0.043^{*}\;(0.013)$ \\
$200$  & $0.025\;(0.008)$ & $0.029^{*}\;(0.011)$ & $0.029^{*}\;(0.009)$ \\
$500$  & $0.015\;(0.004)$ & $0.017^{*}\;(0.006)$ & $0.018^{*}\;(0.006)$ \\
$1000$ & $0.011\;(0.003)$ & $0.012^{*}\;(0.004)$     & $0.012^{*}\;(0.004)$ \\
\hline
\end{tabular}
\caption*{
\footnotesize{Note: ``*'' marks a significant difference in RMSE between the A-optimal design and the marked design at that $n$ (two-sided paired $t$-test, $p<0.05$), with the A-optimal design having the smaller average RMSE.}}
\end{table}

\begin{table}[htbp]
\centering
\caption{Average (sd) of RMSEs over 100 simulations for Example~\ref{ex:potato_example}.}
\label{tab:rmse_potato}
\begin{tabular}{ccccc}
\hline
 & A-optimal & A-optimal & I-optimal & D-optimal \\
$n$ & Algorithm~\ref{algo:A_opt_Forlion} & \cite{li2021} & \cite{li2021} & \cite{li2021} \\
\hline
$200$  & $0.334$ $(0.137)$ & $0.345$ $(0.141)$ & $0.381$ $(0.172)$ & $0.396$ $(0.206)$ \\
$500$  & $0.207$ $(0.073)$ & $0.206$ $(0.070)$ & $0.216$ $(0.079)$ & $0.245$ $(0.103)$ \\
$1000$ & $0.143$ $(0.051)$ & $0.145$ $(0.054)$ & $0.150$ $(0.061)$ & $0.174$ $(0.074)$ \\
$1500$ & $0.114$ $(0.039)$ & $0.112$ $(0.040)$ & $0.122$ $(0.046)$ & $0.142$ $(0.063)$ \\
$2000$ & $0.095$ $(0.031)$ & $0.098$ $(0.034)$ & $0.102$ $(0.041)$ & $0.125$ $(0.054)$ \\
\hline
\end{tabular}
\end{table}

\begin{table}[htbp]
\centering
\caption{Mean paired difference (with $p$-value) in RMSEs for Example~\ref{ex:potato_example}.}
\label{tab:potato_rmse_pairwise}
\begin{tabular}{cccc}
\hline
$n$ & vs.\ A-optimal & vs.\ I-optimal & vs.\ D-optimal \\
\hline
$200$  & $-0.0108\ (0.608)$ & $-0.0476\ (0.027)$ & $-0.0625\ (0.009)$   \\
$500$  & $+0.0016\ (0.888)$ & $-0.0088\ (0.399)$ & $-0.0382\ (0.002)$   \\
$1000$ & $-0.0026\ (0.746)$ & $-0.0068\ (0.385)$ & $-0.0314\ (<0.001)$  \\
$1500$ & $+0.0016\ (0.800)$ & $-0.0089\ (0.136)$ & $-0.0286\ (<0.001)$  \\
$2000$ & $-0.0034\ (0.506)$ & $-0.0070\ (0.162)$ & $-0.0307\ (<0.001)$  \\
\hline
\end{tabular}
\caption*{
\footnotesize{Note: Each cell reports the mean paired difference in RMSE (Algorithm~\ref{algo:A_opt_Forlion}'s minus the corresponding design's of \cite{li2021}) with the two-sided paired $t$-test $p$-value in parentheses.}}
\end{table}

\begin{table}[ht]
\caption{Average (sd) of RMSEs over 100 simulations for Example~\ref{ex:RMSE}}\label{RMSE0333}
{\renewcommand{\arraystretch}{0.6}
\resizebox{\textwidth}{!}{\begin{tabular}{@{\extracolsep\fill}lrrrrr}
\toprule
Sampler & $\beta_0\hspace{1cm}$ & all except $\beta_0$ & $\beta_1\hspace{1cm}$ & $\beta_{21}\hspace{9mm}$ & $\beta_{22}\hspace{9mm}$ \\
\midrule 
Full Data  & 0.073(0.051) & 0.267(0.147) & 0.106(0.078) & 0.194(0.144) & 0.338(0.298) \\
SRSWOR  & 0.362(0.331) & 9.133(3.853) & 0.730(2.092) & 7.026(7.753) & 12.229(5.599) \\
Proportionally Stratified  & 0.378(0.289) & 9.155(3.386) & 0.909(2.526) & 5.768(7.464) & 12.759(5.235) \\
Uniformly Stratified  & 0.287(0.251) & 4.813(5.153) & 3.567(6.393) & 3.709(6.536) & 2.736(5.654) \\
D-optimal  & 0.204(0.186) & 3.104(4.232) & 2.365(5.127) & 1.924(4.621) & 1.978(4.683) \\
A-optimal   & 0.213(0.165) & 2.281(3.740) & 2.109(4.916)  & 1.458(3.889) & 1.132(3.241) \\
\bottomrule
\end{tabular}}
}\end{table}

\end{document}